\documentstyle [epsfig] {article}

\def \sm {Standard Model }

\def \susyq {supersymmetric }

\def \sugra {supergravity }

\def\l {\lambda }

\def \ud {{1 \over 2} }

\def \bea {\begin{equation} }
\def \eea {\end{equation} }
\def \cc {coupling constant }

\def \Eslash {E \kern-.5em\slash }
\def \pslash {p \kern-.5em\slash }
\def \kslash {k \kern-.5em\slash }
\newcommand{\rpv}{\mbox{$\not \hspace{-0.10cm} R_p$ }}

\if@twoside\oddsidemargin25mm
\else\oddsidemargin25mm\evensidemargin25mm\marginparwidth25mm\fi
\begin{document}

\large

\title{\bf Single superpartner production at Tevatron Run II}
\author{F. D\'eliot$^1$, G. Moreau$^2$, C. Royon$^{1,3,4}$ \\ \\
{\it  1: Service de Physique des Particules, DAPNIA }\\
{ \it  CEA/Saclay 91191 Gif-sur-Yvette Cedex, France } \\
{\it  2: Service de Physique Th\'eorique }\\
{ \it  CEA/Saclay 91191 Gif-sur-Yvette Cedex, France }\\
{\it 3: Brookhaven National Laboratory, Upton, New York, 11973;}\\
{\it 4: University of Texas, Arlington, Texas, 76019}}
\maketitle

\begin{abstract}
We study the single productions of supersymmetric particles at Tevatron Run
II
which occur in the $2 \to 2-body$ processes involving R-parity violating
couplings
of type $\l'_{ijk} L_i Q_j D_k^c$.
We focus on the single gaugino productions which receive contributions from
the resonant slepton productions.
We first calculate the amplitudes of the single gaugino productions.
Then we perform analyses of the single gaugino productions based on the
three
charged leptons and like sign dilepton signatures. These analyses allow to
probe
supersymmetric particles masses beyond the present experimental limits, and
many of the $\l'_{ijk}$ coupling constants down to values smaller than the
low-energy bounds. Finally, we show that the studies of
the single gaugino productions offer the opportunity to reconstruct
the $\tilde \chi^0_1$, $\tilde \chi^{\pm}_1$, $\tilde \nu_L$ and
$\tilde l^{\pm}_L$ masses with a good accuracy in a model independent way.
\end{abstract}

\section{Introduction}
\label{intro}

In the Minimal Supersymmetric Standard Model (MSSM),
the \susyq (SUSY) particles must
be produced in pairs. The phase space 
is largely suppressed in pair production of SUSY particles
due to the important masses of the superpartners.
The R-parity violating (\rpv) extension of the MSSM contains the following
additional terms in the superpotential,
which are trilinear in the quarks and leptons superfields,
\begin{eqnarray}
W_{\rpv}=\sum_{i,j,k} \bigg (\ud \l _{ijk} L_iL_j E^c_k+
\l ' _{ijk} L_i Q_j D^c_k+ \ud \l '' _{ijk} U_i^cD_j^cD_k^c   \bigg ),
\label{super}
\end{eqnarray}
where $i,j,k$ are flavour indices.
These \rpv couplings offer the opportunity to produce the scalar
\susyq particles as resonances \cite{Dim1,Dreinoss}. Although the \rpv
coupling constants are severely constrained by the
low-energy experimental bounds \cite{Drein,Bhatt,GDR}, the resonant
superpartner
production reaches high cross sections both at leptonic
\cite{Han} and hadronic \cite{Dim2} colliders.

The resonant production of SUSY particle has another interest:
since its cross section is proportional
to a power $2$ of the relevant \rpv coupling,
this reaction would allow an easier determination of the
\rpv couplings than the pair production provided the \rpv coupling 
is large enough.
As a matter of fact in the pair production study,
the sensitivity on the \rpv couplings is
mainly provided by the displaced vertex analysis of the
Lightest Supersymmetric Particle (LSP) decay
which is difficult experimentally, especially at hadronic colliders.

Neither the Grand Unified Theories (GUT), the string theories nor
the study of the discrete gauge symmetries give a strong theoretical
argument in favor of the R-parity violating or
R-parity conserving scenarios \cite{Drein}. Hence,
the resonant production of SUSY particle through \rpv couplings
is an attractive possibility which must be considered in the
phenomenology of supersymmetry.

The hadronic colliders have an advantage in detecting
new particles resonance. Indeed, due to the wide energy
distribution of the
colliding partons, the resonance can be probed in a wide range of the
new particle mass. This is in contrast with the leptonic colliders
for which the center of mass energy must be fine-tuned in order to
discover new narrow width resonances.

At hadronic colliders, either a slepton or a squark can be produced at
the resonance respectively through a $\l'$ or a $\l''$ coupling constant.
In the hypothesis of a single dominant \rpv
coupling constant, the resonant scalar particle
can decay through the same \rpv coupling as in the production,
leading to a two quark final state for the hard process
\cite{Bin,Dat,Oak,Rizz,Chiap}.
In the case where both $\l'$ and $\l$ couplings
are non-vanishing, the slepton produced via $\l'$ can
decay through $\l$
giving rise to the same final state as in Drell-Yan process,
namely two leptons
\cite{Rizz,Kal,SonPart}. However, for reasonable values of the
\rpv coupling constants, the decays
of the resonant scalar particle via gauge interactions are
typically dominant if kinematically allowed \cite{Han,lola}.\\
The main decay of the resonant scalar particle
through gauge interactions is the decay
into its \sm partner plus a gaugino. Indeed,
in the case where the resonant scalar particle is a squark,
it is produced
through $\l''$ interactions so that it
must be a Right squark $\tilde q_R$ and thus it
cannot decay into the $W^{\pm}$-boson,
which is the only other possible
decay channel via gauge interactions. Besides,
in the case where the resonant scalar particle is a slepton,
it is a Left slepton produced via a $\l'$ coupling
but it cannot generally decay as
$\tilde l^{\pm}_L \to W^{\pm} \tilde \nu_L$ or as
$\tilde \nu_L \to W^{\pm} \tilde l^{\mp}_L$.
The reason is that in most of the SUSY models,
as for example the supergravity or the gauge
mediated models, the mass difference
between the Left charged slepton and the Left sneutrino
is due to the D-terms so that it is fixed by the relation
$m^2_{\tilde l^{\pm}_L}-m^2_{\tilde \nu_L}=\cos 2 \beta M_W^2$
\cite{Iban}
and thus it does not exceed the $W^{\pm}$-boson mass.
Nevertheless, we note that in the large $\tan \beta$ scenario, a
resonant
scalar particle of the third generation
can generally decay into the $W^{\pm}$-boson due to the large mixing in the
third family sfermion sector. For instance, in the SUGRA model with
a large $\tan \beta$ a tau-sneutrino produced at the resonance
can decay as $\tilde \nu_{\tau} \to W^{\pm} \tilde \tau^{\mp}_1$,
$\tilde \tau^{\mp}_1$ being the lightest stau.\\
The resonant scalar particle production
at hadronic colliders leads thus mainly
to the single gaugino production,
in case where the decay of the relevant scalar particle
into gaugino is kinematically allowed.
In this paper, we study the single gaugino productions
at Tevatron Run II.
The single gaugino productions at hadronic colliders
were first studied in \cite{Dreinoss,Dim2}.
Later, studies on the single neutralino \cite{Rich}
and single chargino \cite{greg}
productions at Tevatron have been performed.
The single neutralino \cite{Rich2} and single chargino
\cite{Gia1} productions have also been considered
in the context of physics at LHC.
In the present article, we also study the single superpartner
productions at Tevatron Run II which occur via $2 \to 2-body$
processes and do not receive
contributions from resonant SUSY particle productions.

The singly produced superpartner
initiates a cascade decay ended typically by
the \rpv decay of the LSP. In case of a single dominant
$\l''$ coupling constant, the LSP decays into quarks so that
this cascade decay leads to multijet final states having a
large QCD background \cite{Dim2,Bin}. Nevertheless, if some leptonic decays,
as for instance $\tilde \chi^{\pm} \to l^{\pm} \nu \tilde \chi^0$,
$\tilde \chi^{\pm}$ being the chargino and $\tilde \chi^0$ the neutralino,
enter the chain reaction, clearer leptonic signatures can be investigated
\cite{Berg}.
In contrast, in the hypothesis of a single dominant $\l'$ coupling constant,
the LSP decay into charged leptons naturally favors leptonic signatures
\cite{Dreinoss}. We will thus study
the single superpartner production reaction
at Tevatron Run II within the scenario of a single dominant $\l'_{ijk}$
coupling constant.

In section \ref{theoretical}, we define our theoretical framework.
In section \ref{discussion}, we present the values of the cross sections for
the various single superpartner productions
via $\l'_{ijk}$ at Tevatron Run II and we discuss the interesting
multileptonic signatures that these processes can generate.
In section \ref{analysis1}, we analyse the three lepton signature
induced by the single chargino production.
In section \ref{analysis2}, we study the like sign dilepton
final state generated by the single neutralino and chargino productions.

\section{Theoretical framework}
\label{theoretical}

Our framework throughout this paper will be
the so-called minimal \sugra model (mSUGRA)
which assumes the existence of a grand unified gauge theory and
family universal boundary conditions on the supersymmetry breaking
parameters.
We choose the 5 following parameters:
$m_0$ the universal scalars mass at the unification scale $M_X$,
$m_{1/2}$ the universal gauginos mass at $M_X$,
$A=A_{t}=A_{b}=A_{\tau}$ the trilinear Yukawa coupling at $M_X$,
$sign(\mu)$ the sign of the $\mu(t)$ parameter
($t=\log (M_X^2/Q^2)$, $Q$ denoting the running scale)
and $\tan \beta=<H_u>/<H_d>$ where $<H_u>$ and
$<H_d>$ denote the vacuum expectation values of the Higgs fields.
In this model, the
higgsino mixing parameter $\vert \mu \vert$ is determined by the
radiative electroweak symmetry breaking condition.
Note also that the parameters $m_{1/2}$ and $M_2(t)$ ($\tilde W$ wino mass)
are related by the solution of the one loop renormalization group
equations $m_{1/2}=(1-\beta_a t)M_a(t)$ with $\beta_a=g_X^2 b_a/(4 \pi)^2$,
where $\beta_a$ are the beta functions,
$g_X$ is the \cc at $M_X$ and
$b_a=[3,-1,-11]$, $a=[3,2,1]$ corresponding to
the gauge group factors $SU(3)_c,SU(2)_L,SU(1)_Y$.
We shall set the unification scale at $M_X=2 \ 10^{16} GeV$
and the running scale at the $Z^0$-boson mass: $Q=m_{Z^0}$.

We also assume the infrared fixed point hypothesis for the top quark
Yukawa coupling \cite{Pok} that provides a natural explanation of
a large top quark mass $m_{top}$.
In the infrared fixed point approach,
$\tan \beta$ is fixed up to the ambiguity
associated with large or low $\tan \beta$ solutions.
The low solution of $\tan \beta$ is fixed by
the equation $m_{top}=C \sin \beta$,
where $C \approx 190-210 \ GeV$ for $\alpha_s(m_{Z^0})=0.11-0.13$.
For instance, with a top quark mass of
$m_{top}=174.2GeV$ \cite{top}, the low solution is given by $\tan \beta \approx 1.5$.
The second important effect of the infrared fixed point hypothesis is
that the dependence of the electroweak symmetry breaking
constraint on the $A$ parameter becomes
weak so that $\vert \mu \vert$ is a known function of the
$m_0$, $m_{1/2}$ and $\tan \beta$ parameters \cite{Pok}.

Finally, we consider the \rpv extension of the mSUGRA model
characterised by a single dominant \rpv coupling constant of type
$\l'_{ijk}$.

\section{Single superpartner productions via $\l'_{ijk}$ at Tevatron Run II}
\label{discussion}

\subsection{Resonant superpartner production}
\label{resonant}

\begin{figure}[t]
\begin{center}
\leavevmode
\centerline{\psfig{figure=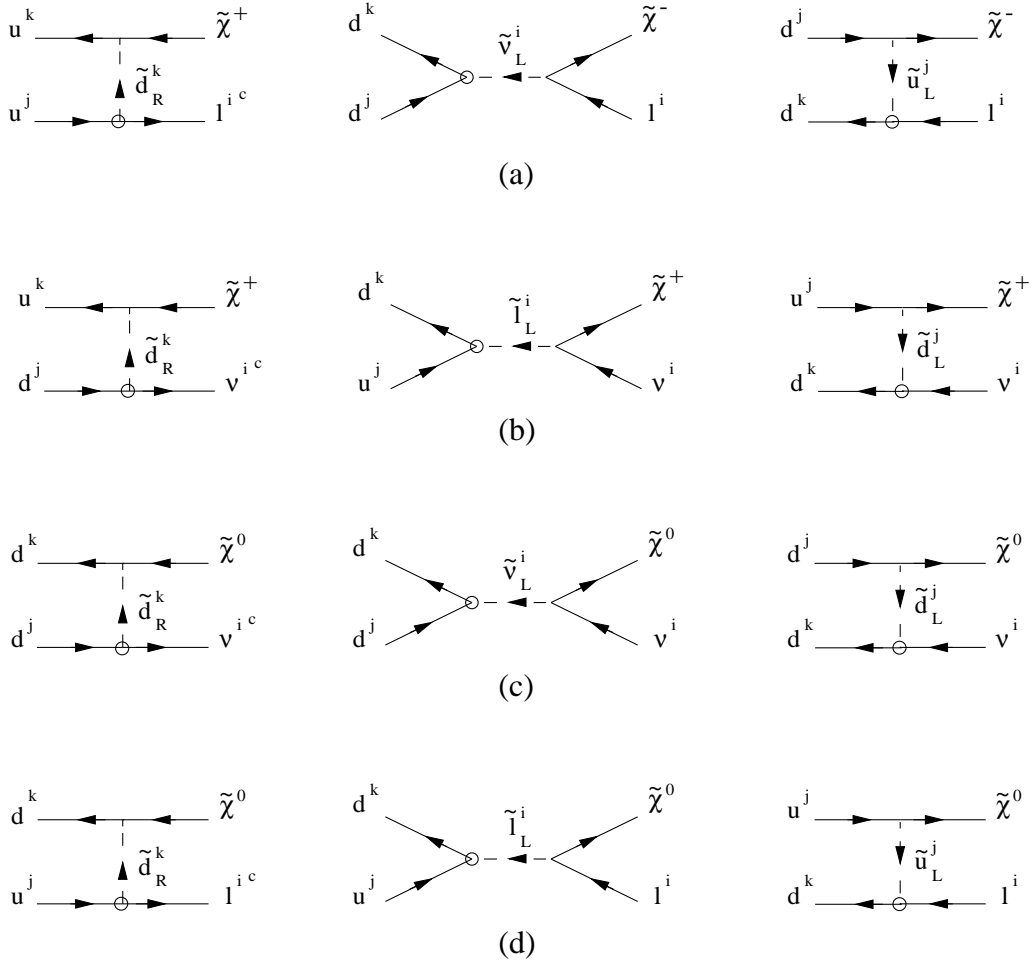,height=5.in}}
\end{center}
\caption{\footnotesize  \it
Feynman diagrams for the 4 single production reactions
involving  $\l'_{ijk}$  at hadronic colliders which
receive a contribution from a resonant \susyq particle production.
The $\l'_{ijk}$ coupling constant is symbolised by a small circle
and the arrows denote the flow of the particle momentum.
\rm \normalsize }
\label{graphes}
\end{figure}

At hadronic colliders, either a sneutrino ($\tilde \nu$) or a charged
slepton ($\tilde l$) can be produced
at the resonance via the $\l'_{ijk}$ coupling.
As explained in Section \ref{intro}, for most of the SUSY models,
the slepton produced at the resonance has two possible gauge
decays, namely a decay into either a chargino or a neutralino.
Therefore, in the scenario of a single dominant
$\l'_{ijk}$ coupling and for most of the SUSY models,
either a chargino or a neutralino
is singly produced together
with either a charged lepton or a neutrino,
through the resonant superpartner production at hadronic colliders.
There are thus four main possible types
of single superpartner production reaction
involving $\l'_{ijk}$ at hadronic colliders
which receive a contribution from resonant SUSY particle production.
The diagrams associated
to these four reactions are drawn in Fig.\ref{graphes}.
As can be seen in this figure, these single superpartner productions receive
also some contributions from both the $t$ and $u$ channels.
Note that all the single superpartner production processes drawn in
Fig.\ref{graphes} have charge conjugated processes.
We have calculated the amplitudes of the processes shown in
Fig.\ref{graphes}
and the results are given in Appendix \ref{formulas}.

\subsubsection{Cross sections}
\label{cross1}

\begin{figure}[t]
\begin{center}
\leavevmode
\centerline{\psfig{figure=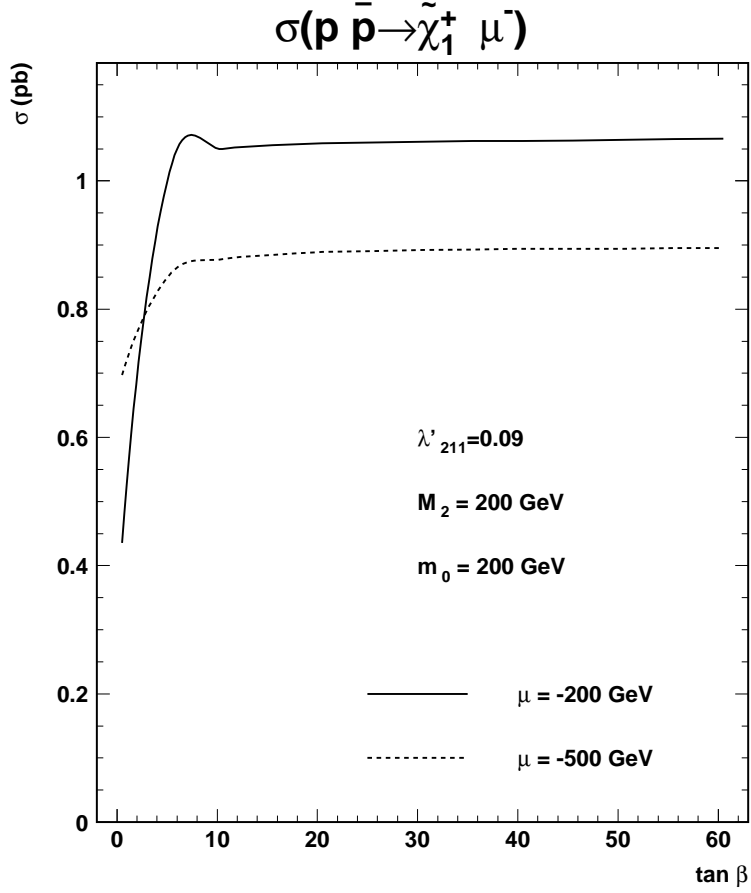,height=5.in}}
\end{center}
\caption{\footnotesize  \it
Cross sections (in $pb$) of the single chargino production
$p \bar p \to \tilde \chi^+_1 \mu^-$ at a center of mass energy
of $2 TeV$ as a function of the $\tan \beta$ parameter
for $\l'_{211}=0.09$, $M_2=200GeV$,
$m_0=200GeV$ and two values of the $\mu$ parameter:
$\mu=-200GeV,-500GeV$.
\rm \normalsize }
\label{XStan}
\end{figure}

In this section, we discuss the dependence of the
single gaugino production cross sections on the various
\susyq parameters. We will not assume here the
radiative electroweak symmetry breaking condition in order to
study the variations of the cross sections with the
higgsino mixing parameter $\mu$.

First, we study the cross section of the single chargino production
$p \bar p \to \tilde \chi^+ l_i^-$ which
occurs through the $\l'_{ijk}$ coupling
(see Fig.\ref{graphes}(a)).
The differences between the $\tilde \chi^+ e^-$, $\tilde \chi^+ \mu^-$
and $\tilde \chi^+ \tau^-$ production
(occuring respectively through the $\l'_{1jk}$, $\l'_{2jk}$ and
$\l'_{3jk}$ couplings with identical $j$ and $k$ indices)
cross sections involve $m_{l_i}$
lepton mass terms (see Appendix \ref{formulas}) and are thus negligible.
The $p \bar p \to \tilde \chi^+ l_i^-$ reaction receives contributions
from the $s$ channel sneutrino exchange and
the $t$ and $u$ channels squark exchanges as shown in Fig.\ref{graphes}.
However,
the $t$ and $u$ channels represent small contributions
to the whole single chargino production cross section 
when the sneutrino exchanged in the
$s$ channel is real, namely for $m_{\tilde \nu_{iL}}>m_{\tilde
\chi^{\pm}}$.
The $t$ and $u$ channels cross sections will be relevant only
when the produced sneutrino is virtual since the $s$ channel
contribution is small.
In this situation the single chargino production
rate is greatly reduced compared
to the case where the exchanged sneutrino is produced as a resonance.
Hence, The $t$ and $u$ channels do not represent important contributions
to the $\tilde \chi^+ l_i^-$ production rate.

\begin{figure}[t]
\begin{center}
\leavevmode
\centerline{\psfig{figure=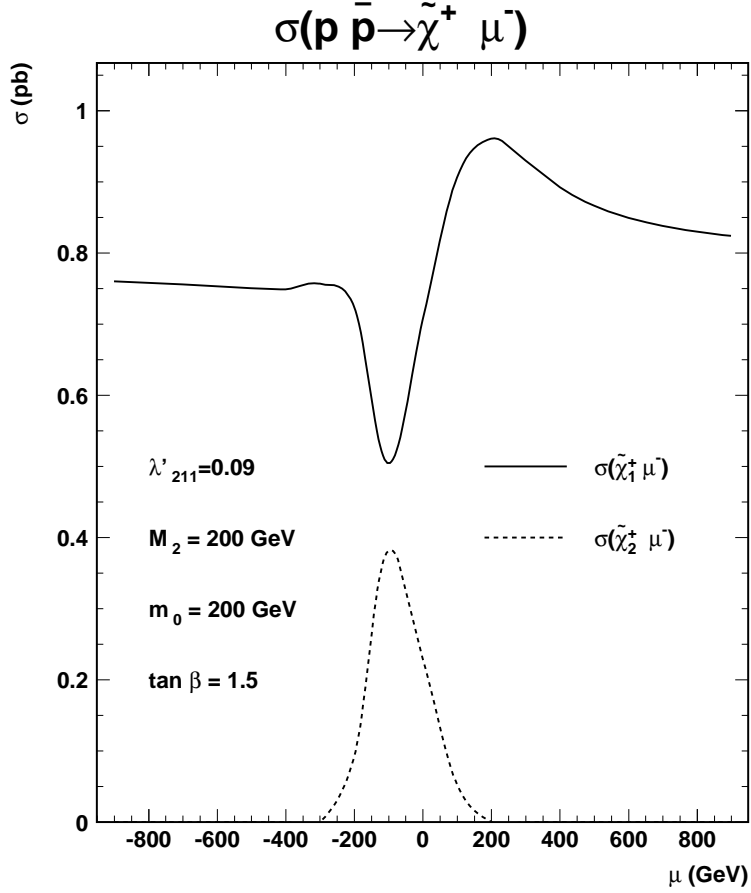,height=5.in}}
\end{center}
\caption{\footnotesize  \it
Cross sections (in $pb$) of the single chargino productions
$p \bar p \to \tilde \chi^+_{1,2} \mu^-$
as a function of the $\mu$ parameter (in $GeV$)
for $\l'_{211}=0.09$, $M_2=200GeV$, $\tan \beta=1.5$
and $m_0=200GeV$ at a center of mass energy
of $2 TeV$.
\rm \normalsize }
\label{XSmu}
\end{figure}

The dependence of the $\tilde \chi^+ l_i^-$ production rate
on the $A$ coupling is weak. Indeed, the rate depends on the $A$
parameter only through the masses of the third generation squarks
eventually exchanged in the $t$ and $u$ channels
(see Fig.\ref{graphes}). Similarly,
the dependences on the $A$ coupling
of the rates of the other single gaugino productions
shown in Fig.\ref{graphes} are weak.
Therefore, in this article we present the results for
$A=0$. Later, we will discuss the effects
of large $A$ couplings on the cascade decays which
are similar to the effects of large $\tan \beta$ values.

{\bf $\tan \beta$ dependence:}
The dependence of the $\tilde \chi^+ l_i^-$ production rate
on $\tan \beta$ is also weak, except for $\tan \beta<10$.
This can be seen in Fig.\ref{XStan} where the cross section
of the $p \bar p \to \tilde \chi^+_1 \mu^-$ reaction occuring
through the $\l'_{211}$ coupling is shown
as a function of the $\tan \beta$ parameter.
The choice of the $\l'_{211}$ coupling is motivated by the fact
that the analysis in Sections \ref{analysis1} and \ref{analysis2}
are explicitly made for this \rpv coupling. In Fig.\ref{XStan},
we have taken the $\l'_{211}$ value equal to
its low-energy experimental bound
for $m_{\tilde d_R}=100GeV$ which is $\l'_{211}<0.09$ \cite{Bhatt}. \\
At this stage, some remarks on the values of the
cross sections presented in this section must be done. First,
the single gaugino production rates must be multiplied
by a factor 2 in order to take into account the charge conjugated
process, which is for example
in the present case $p \bar p \to \tilde \chi^- \mu^+$. Furthermore,
the values of the cross sections for all the single gaugino productions
are obtained using the CTEQ4L structure function \cite{CTEQ4}.
Choosing other parametrizations does not change significantly
the results since proton structure functions in our kinematical domain
in Bjorken $x$ are known and have been already measured.
For instance, with the set of parameters $\l'_{211}=0.09$,
$M_2=100GeV$, $\tan \beta=1.5$, $m_0=300GeV$ and $\mu=-500GeV$,
the $\tilde \chi^+_1 \mu^-$ production cross section is
$0.503pb$ for the CTEQ4L structure function \cite{CTEQ4},
$0.503pb$ for the BEP structure function \cite{BEP},
$0.480pb$ for the MRS (R2) structure function \cite{MRS} and
$0.485pb$ for the GRV LO structure function \cite{GRV}.

\begin{figure}[t]
\begin{center}
\leavevmode
\centerline{\psfig{figure=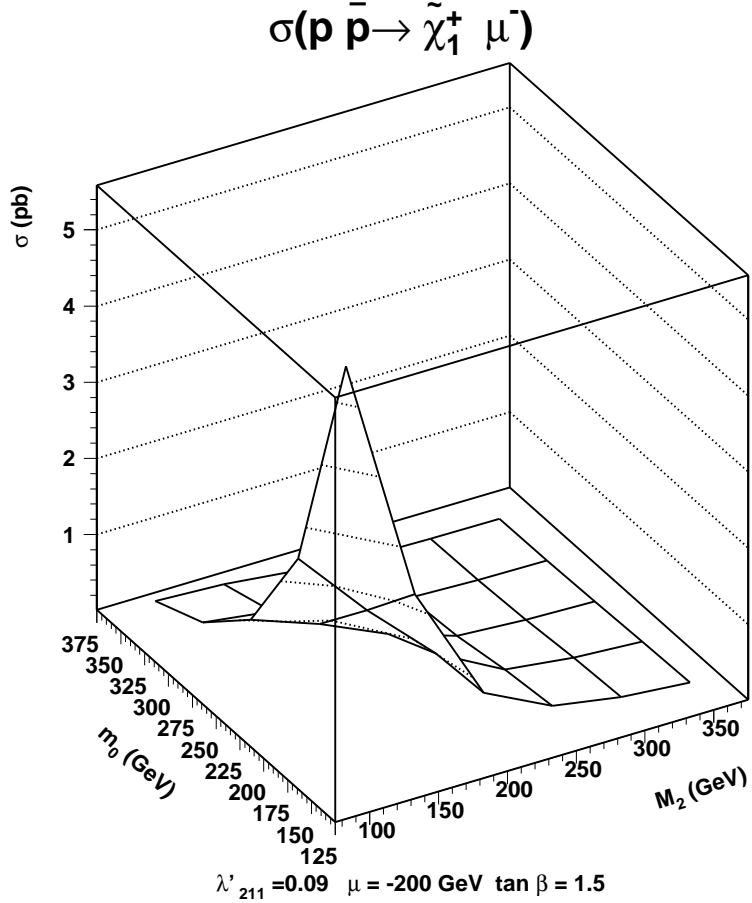,height=5.in}}
\end{center}
\caption{\footnotesize  \it
Cross section (in $pb$) of the single chargino production
$p \bar p \to \tilde \chi^+_1 \mu^-$
as a function of the $m_0$ (in $GeV$)
and $M_2$ (in $GeV$) parameters.
The center of mass energy is $\sqrt s=2 TeV$
and the other parameters are: $\l'_{211}=0.09$,
$\tan \beta=1.5$ and $\mu=-200GeV$.
\rm \normalsize }
\label{XS02}
\end{figure}

{\bf $\mu$ dependence:}
In Fig.\ref{XSmu},
we present the cross sections of the $\tilde \chi_1^+ \mu^-$ and
$\tilde \chi_2^+ \mu^-$ productions as a function of the $\mu$ parameter.
We observe in this figure the weak dependence of the
cross section $\sigma(p \bar p \to \tilde \chi_1^+ \mu^-)$ on $\mu$
for $| \mu | > M_2$.
The reason is the smooth dependence of the $\tilde \chi_1^{\pm}$
mass
on $\mu$ in this domain.
However, the rate strongly decreases in the region $| \mu | < M_2$
in which the
$\tilde \chi_1^{\pm}$ chargino is mainly composed by the higgsino.
Nevertheless, the small $ \vert \mu \vert$ domain  ($\vert \mu \vert$ smaller than
$\sim 100 GeV$ for $\tan \beta=1.41$, $M_2>100 GeV$, $m_0=500 GeV$ and $\l' \neq 0$)
is excluded by the present experimental limits
derived from the LEP data \cite{Mass}. \\
In contrast, the cross section $\sigma(p \bar p \to \tilde \chi_2^+ \mu^-)$
increases in the domain $\vert \mu \vert < M_2$.
The explanation is that
the $\tilde \chi_2^{\pm}$ mass is enhanced as $\vert \mu \vert$ increases.
The region in which $\sigma(p \bar p \to \tilde \chi_2^+ \mu^-)$
becomes important is at small values of $\vert \mu \vert$,
near the LEP limits of \cite{Mass}.
We also remark in Fig.\ref{XSmu} that the single $\tilde \chi_1^+$
production rate values remain
above the single $\tilde \chi_2^+$ production rate values in all
the considered range of $\mu$. 
In this figure, we also notice that the cross section is smaller
when $\mu$ is negative. To be conservative, we will take $\mu <0$
in the following.

{\bf $m_0$ and $M_2$ dependences:}
In fact, the cross section
$\sigma(p \bar p \to \tilde \chi^+ l_i^-)$ depends mainly
on the $m_0$ and $M_2$ parameters.
We present in Fig.\ref{XS02} the rate
of the $\tilde \chi^+_1 \mu^-$ production
as a function of the $m_0$ and $M_2$ parameters.
The rate decreases at high values of $m_0$
since the sneutrino becomes heavier as $m_0$ increases 
and more energetic initial partons are
required in order to produce the resonant sneutrino.
The decrease of the rate at large values of $M_2$ is due to
the increase of the chargino mass and thus
the reduction of the phase space factor.

\begin{figure}[t]
\begin{center}
\leavevmode
\centerline{\psfig{figure=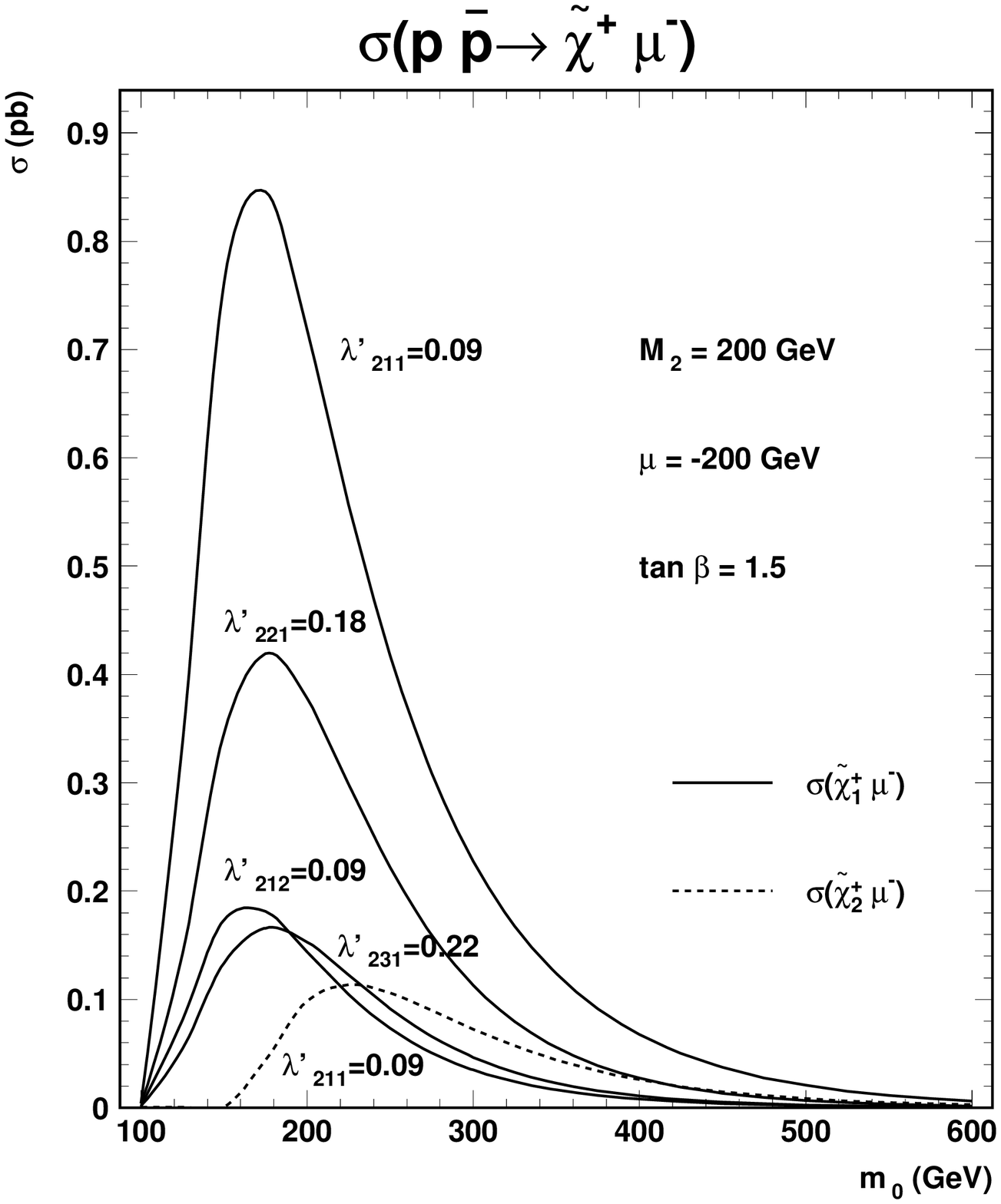,height=5.in}}
\end{center}
\caption{\footnotesize  \it
Cross sections (in $pb$) of the single chargino productions
$p \bar p \to \tilde \chi^+_{1,2} \mu^-$
as a function of the $m_0$ parameter (in $GeV$).
The center of mass energy is taken at $\sqrt s=2 TeV$
and $\l'_{211}=0.09$,
$M_2=200GeV$, $\tan \beta=1.5$ and $\mu=-200GeV$.
The rates of the single $\tilde \chi^+_1$ production via the
\rpv couplings $\l'_{212}=0.09$, $\l'_{221}=0.18$ and $\l'_{231}=0.22$
are also shown. The chosen values of the \rpv couplings correspond
to the low-energy limits \cite{Bhatt} for a squark mass of $100GeV$.
\rm \normalsize }
\label{XScl}
\end{figure}

In Fig.\ref{XScl}, we show the variations of the
$\sigma(p \bar p \to \tilde \chi^+_1 \mu^-)$ cross sections with $m_0$
for fixed values of $M_2$, $\mu$ and $\tan \beta$.
The cross sections corresponding to the $\tilde \chi^+_1 \mu^-$
production through various \rpv couplings of type $\l'_{2jk}$ are
presented. In this figure, we only consider the \rpv couplings giving 
the highest cross sections.
The values of the considered $\l'_{2jk}$ couplings have been taken
at their low-energy limit \cite{Bhatt} for a squark
mass of $100GeV$. The rate of the $\tilde \chi^+_2 \mu^-$ production
through $\l'_{211}$ is also shown in this figure.
We already notice that the cross section is significant for many \rpv
couplings and we will come back on this important statement in the following.\\
The $\sigma(p \bar p \to \tilde \chi^+ \mu^-)$
rates decrease as $m_0$ increases for the same reason as in
Fig.\ref{XS02}.
A decrease of the rates also occurs at small values of $m_0$.
The reason is the following.
When $m_0$ decreases, the $\tilde \nu$ mass is getting closer
to the $\tilde \chi^{\pm}$ masses so that the phase space
factor associated to the decay
$\tilde \nu_{\mu} \to \tilde \chi^{\pm} \mu^{\mp}$ decreases.\\
We also observe that the single $\tilde \chi_2^+$
production rate is much smaller than the single
$\tilde \chi_1^+$ production rate, as in Fig.\ref{XSmu}.\\
The differences between the $\tilde \chi^+_1 \mu^-$
production rates occuring via the various $\l'_{2jk}$ couplings
are explained by the different parton densities.
Indeed, as shown in Fig.\ref{graphes}
the hard process associated to the $\tilde \chi^+_1 \mu^-$
production occuring through the
$\l'_{2jk}$ coupling constant
has a partonic initial state $ \bar q_j q_k$.
The $\tilde \chi^+_1 \mu^-$ production via the $\l'_{211}$ coupling
has first generation quarks in the initial state
which provide the maximum parton density.

\begin{figure}[t]
\begin{center}
\leavevmode
\centerline{\psfig{figure=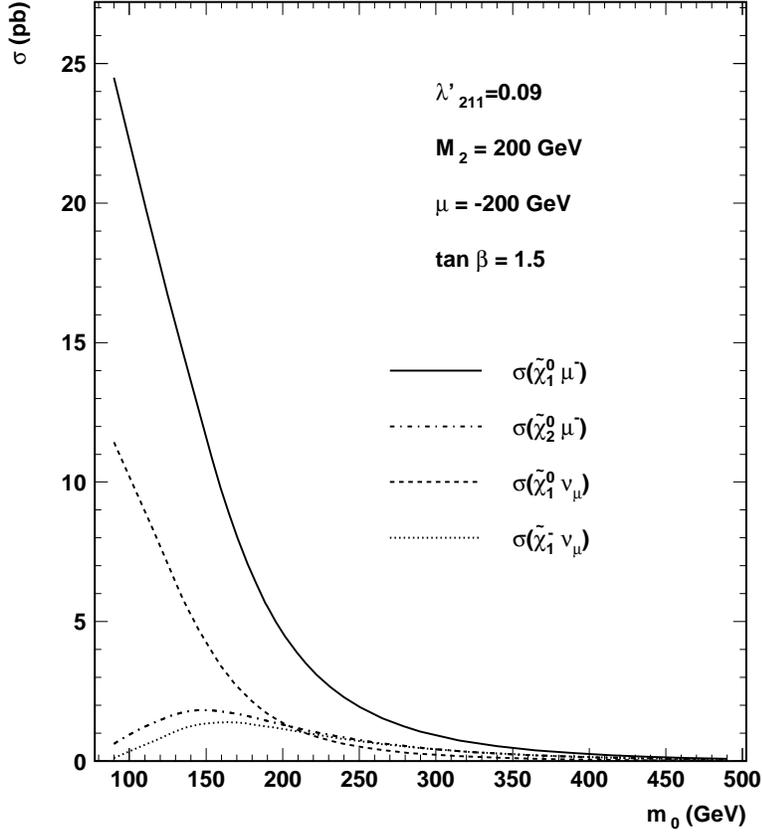,height=5.in}}
\end{center}
\caption{\footnotesize  \it
Cross sections (in $pb$) of the reactions
$p \bar p \to \tilde \chi^-_1 \nu$,
$p \bar p \to \tilde \chi^0_{1,2} \mu^-$
and $p \bar p \to \tilde \chi^0_1 \nu$
as a function of the $m_0$ parameter (in $GeV$).
The center of mass energy is taken at $\sqrt s=2 TeV$
and the considered set of parameters is: $\l'_{211}=0.09$,
$M_2=200GeV$, $\tan \beta=1.5$ and $\mu=-200GeV$.
\rm \normalsize }
\label{allXS}
\end{figure}

We now discuss the rate behaviours for the reactions
$p \bar p \to \tilde \chi^- \nu_{\mu}$, $p \bar p \to \tilde \chi^0 \mu^-$
and $p \bar p \to \tilde \chi^0 \nu_{\mu}$ which
occur via $\l'_{211}$, in the SUSY parameter space.
The dependences of these rates on the $A$, $\tan \beta$, $\mu$ and $M_2$
parameters are typically the same as for the
$\tilde \chi^+ \mu^-$ production rate.
The variations of the
$\tilde \chi^-_1 \nu_{\mu}$, $\tilde \chi^0_{1,2} \mu^-$
and $\tilde \chi^0_1 \nu_{\mu}$ productions cross sections
with the $m_0$ parameter are shown in Fig.\ref{allXS}.
The $\tilde \chi^-_2 \nu_{\mu}$, $\tilde \chi^0_{3,4} \mu^-$ and
$\tilde \chi^0_{3,4} \nu_{\mu}$ production rates are comparatively
negligible
and thus have not been represented.
We observe in this figure that the cross sections decrease
at large $m_0$ values like the $\tilde \chi^+ \mu^-$ production rate.
However, while
the single $\tilde \chi^{\pm}_1$ productions rates decrease at
small $m_0$ values
(see Fig.\ref{XScl} and Fig.\ref{allXS}),
this is not true for the single $\tilde \chi^0_1$ productions
(see Fig.\ref{allXS}). The reason is that in mSUGRA
the $\tilde \chi^0_1$ and $\tilde l_{iL}$
($l_i=l_i^{\pm},\nu_i$) masses are never close enough
to induce a significant decrease of the
cross section associated to the reaction
$p \bar p \to \tilde l_{iL} \to \tilde \chi^0_1 l_i$,
where $l_i=l_i^{\pm},\nu_i$ (see Fig.\ref{graphes}(c)(d)),
caused by a phase space factor reduction. Therefore,
the resonant slepton contribution to the single $\tilde \chi^0_1$ production
is not reduced at small $m_0$ values like the
resonant slepton contribution to the single $\tilde \chi^{\pm}_1$
production.
For the same reason,
the single $\tilde \chi^0_1$ productions have much higher cross sections
than the single $\tilde \chi^{\pm}_1$ productions in most
of the mSUGRA parameter space, as illustrate Fig.\ref{XScl} and
Fig.\ref{allXS}.
We note that in the particular case of a single dominant $\l'_{3jk}$
coupling
constant and of large $\tan \beta$ values, the rate of the reaction
$p \bar p \to \tilde \tau^{\pm}_1 \to \tilde \chi^0_1 \tau^{\pm}$
(see Fig.\ref{graphes}(d)), where $\tilde \tau^{\pm}_1$ is the lightest
tau-slepton,
can be reduced at low $m_0$ values since
then $m_{\tilde \tau^{\pm}_1}$ can be closed to $m_{\tilde \chi^0_1}$ due to
the large
mixing occuring in the staus sector.
By analysing Fig.\ref{XScl} and Fig.\ref{allXS},
we also remark that the $\tilde \chi^- \nu_{\mu}$
($\tilde \chi^0 \mu^-$) production rate is larger than the
$\tilde \chi^+ \mu^-$ ($\tilde \chi^0 \nu_{\mu}$) one.
The explanation is that in $p \bar p$ collisions
the initial states of the resonant charged slepton production
$u_j \bar d_k, \bar u_j d_k$ have higher partonic densities than the
initial states of the resonant sneutrino production
$d_j \bar d_k, \bar d_j d_k$. 
This phenomenon also increases the difference between the rates of the
$\tilde \chi^0_1 \mu^-$ and $\tilde \chi^+_1 \mu^-$ productions
at Tevatron.
\\ Although the single $\tilde \chi^{\pm}_1$ production cross sections
are smaller than the $\tilde \chi^0_1$ ones, it is interesting to
study both of them since they have quite high values.

\subsection{Non-resonant superpartner production}
\label{non-resonant}

At hadronic colliders, the single productions of SUSY particle
via $\l'_{ijk}$ can occur through some $2 \to 2-body$
processes which do not receive contributions
from any resonant superpartner production.
These non-resonant superpartner productions are
(one must also add the charge conjugated processes):

\begin{itemize}
\item The gluino production $\bar u_j d_k \to \tilde g l_i$ via the exchange
of a $\tilde u_{jL}$ ($\tilde d_{kR}$) squark in the $t$ ($u$) channel.
\item The squark production $\bar d_j g \to \tilde d_{kR}^* \nu_i$ via the
exchange
of a $\tilde d_{kR}$ squark ($d_j$ quark) in the $t$ ($s$) channel.
\item The squark production $\bar u_j g \to \tilde d_{kR}^* l_i$ via the
exchange
of a $\tilde d_{kR}$ squark ($u_j$ quark) in the $t$ ($s$) channel.
\item The squark production $d_k g \to \tilde d_{jL} \nu_i$ via the exchange
of a $\tilde d_{jL}$ squark ($d_k$ quark) in the $t$ ($s$) channel.
\item The squark production $d_k g \to \tilde u_{jL} l_i$ via the exchange
of a $\tilde u_{jL}$ squark ($d_k$ quark) in the $t$ ($s$) channel.
\item The sneutrino production $\bar d_j d_k \to Z \tilde \nu_{iL}$ via the
exchange
of a $d_k$ or $d_j$ quark ($\tilde \nu_{iL}$ sneutrino) in the $t$ ($s$)
channel.
\item The charged slepton production $\bar u_j d_k \to Z \tilde l_{iL}$ via
the exchange
of a $d_k$ or $u_j$ quark ($\tilde l_{iL}$ slepton) in the $t$ ($s$)
channel.
\item The sneutrino production $\bar u_j d_k \to W^- \tilde \nu_{iL}$
via the exchange of a $d_j$ quark ($\tilde l_{iL}$ sneutrino) in the $t$
($s$) channel.
\item The charged slepton production $\bar d_j d_k \to W^+ \tilde l_{iL}$
via the exchange of a $u_j$ quark ($\tilde \nu_{iL}$ sneutrino) in the $t$
($s$) channel.
\end{itemize}

The single gluino production cannot reach high cross sections
due to the strong experimental limits on the squarks and gluinos
masses which are typically about $m_{\tilde q},m_{\tilde g}
\stackrel{>}{\sim}200GeV$ \cite{PDG98}. Indeed, the single gluino
production occurs through the exchange of squarks in the $t$
and $u$ channels, as described above, so that the cross section
of this production decreases as the squarks and gluinos
masses increase. For the value
$m_{\tilde q}=m_{\tilde g}=250GeV$ which is close to the
experimental limits, we find the
single gluino production rate
$\sigma(p \bar p \to \tilde g \mu) \approx 10^{-2}pb$
which is consistent with the results of \cite{Dim2}.
The cross sections given in this section are computed at a center of mass
energy of
$\sqrt s=2TeV$ using the version 33.18 of the COMPHEP routine \cite{COMPHEP}
with the CTEQ4m structure function and an \rpv coupling
$\l'_{211}=0.09$.
Similarly, the single squark production cross section
cannot be large: for $m_{\tilde q}=250GeV$,
the rate $\sigma(p \bar p \to \tilde u_L \mu)$
is of order $\sim 10^{-3}pb$.
The production of a slepton together
with a massive gauge boson has a small phase space factor
and does not involve strong interaction couplings.
The cross section of this type of reaction is thus small.
For instance, with a slepton mass of $m_{\tilde
l}=100GeV$ we find the cross section
$\sigma(p \bar p \to Z \tilde \mu_L)$ to be of order $10^{-2}pb$.

As a conclusion, the non-resonant single superpartner productions
have small rates and will not be considered here.
Nevertheless, some of 
these reactions are interesting as their cross section
involves few SUSY parameters, namely only one scalar
superpartner mass and one \rpv coupling constant.

\section{Three lepton signature analysis}
\label{analysis1}

\subsection{Signal}
\label{signal1}

In this section, we study the three lepton signature
at Tevatron Run II
generated by the single chargino production through
$\l'_{ijk}$,
$p \bar p \to \tilde \chi^{\pm} l^{\mp}_i$, followed by
the cascade decay,
$\tilde \chi^{\pm} \to \tilde \chi^0_1 l^{\pm} \nu$,
$\tilde \chi^0_1 \to l_i u_j \bar d_k, \ \bar l_i \bar u_j d_k$
(the indices $i,j,k$ correspond to the indices of $\l'_{ijk}$).
In fact, the whole final state is
3 charged leptons + 2 hard jets + missing energy ($\Eslash$).
The two jets and the missing energy come respectively from
the quarks and the neutrino produced in the cascade decay.
In the mSUGRA model, which predicts the $\tilde \chi^0_1$ as the LSP
in most of the parameter space,
the $p \bar p \to \tilde \chi^{\pm} l_i^{\mp}$ reaction
is the only single gaugino production allowing the three lepton signature
to be generated in a significant way.
Since the $\tilde \chi^{\pm}_1 l^{\mp}_i$
production rate is dominant
compared to the $\tilde \chi^{\pm}_2 l^{\mp}_i$ production rate,
as discussed in Section \ref{cross1}, we only consider
the contribution to the three lepton signature
from the single lightest chargino production.

For $m_{\tilde \nu},m_{\tilde l},m_{\tilde q},
m_{\tilde \chi^0_2}>m_{\tilde \chi^{\pm}_1}$,
the branching ratio $B(\tilde \chi^{\pm}_1 \to \tilde \chi^0_1 l^{\pm} \nu)$
is typically of order $30\%$
and is smaller than for the other possible decay
$\tilde \chi^{\pm}_1 \to \tilde \chi^0_1 \bar q_p q'_p$
because of the color factor.\\
Since in our framework the $\tilde \chi^0_1$ is the LSP,
it can only decay via $\l'_{ijk}$, either as
$\tilde \chi^0_1 \to l_i u_j d_k$ or as $\tilde \chi^0_1 \to \nu_i d_j d_k$,
with a branching ratio $B(\tilde \chi^0_1 \to l_i u_j d_k)$ ranging between
$\sim 40\%$ and $\sim 70\%$.

The three lepton signature
is particularly attractive at hadronic colliders because of
the possibility to reduce the associated \sm background.
In Section \ref{back1} we describe this \sm background
and in Section \ref{cut1} we show how it can be reduced.

\subsection{Standard Model
background of the 3 lepton signature at Tevatron}
\label{back1}

The first source of \sm background for the three leptons final state is
the top quark pair production
$q \bar q \to t \bar t$ or $g g \to t \bar t$. Since the top quark life time
is smaller than its hadronisation time, the top decays and its main channel
is the decay into a $W$ gauge boson and a bottom quark as $t \to b W$.
The $t \bar t$ production can thus give rise to a $3l$
final state if the $W$ bosons and one of the b-quarks undergo leptonic
decays simultaneously. The cross section,
calculated at leading order with PYTHIA \cite{PYTHIA}
using the CTEQ2L structure function, times the branching fraction is
$\sigma (p \bar p \to t \bar t)
\times B^2(W \to l_p \nu_p) \approx  863fb$ ($704fb$) with $p=1,2,3$
at $\sqrt s= 2 TeV$ for a top quark mass of $m_{top}=170 GeV$ ($175 GeV$).

The other major source of \sm background is the
$W^{\pm} Z^0$ production followed
by the leptonic decays of the gauge bosons, namely $W \to l \nu$ and $Z \to
l \bar l$. The value for
the cross section times the branching ratios is $\sigma (p \bar p \to W Z)
\times B(W \to l_p \nu_p) \times B(Z \to l_p \bar l_p) \approx  82fb$
($p=1,2,3$) at leading
order with a center of mass energy of $\sqrt s= 2 TeV$.
\\ The $W^{\pm} Z^0$ production gives also a small contribution to the 3
leptons
background through the decays: $W \to b u_p$ and $Z \to b \bar b$,
$W \to l \nu$ and $Z \to b \bar b$ or
$W \to b u_p$ and $Z \to l \bar l$, if a lepton is produced in each of the b
jets.

Similarly, the $Z^0 Z^0$ production followed by the decays $Z \to l \bar
l$ ($l=e,\mu$), $Z \to \tau \bar \tau$, where one of the $\tau$ decays into
lepton while
the other decays into jet, leads to three leptons in the final state.
Within the same framework as above,
the cross section is of order $\sigma (p \bar p \to Z Z \to 3 l)
\approx  2fb$.
\\ The $Z^0 Z^0$ production can also contribute weakly to the 3 leptons
background via the decays: $Z \to l \bar l$ and $Z \to b \bar b$
or $Z \to b \bar b$ and $Z \to b \bar b$,
since a lepton can be produced in a b jet.

It has been pointed out recently that the $W Z^*$ (throughout this
paper a star indicates a virtual particle) and the $W \gamma^*$ productions
could represent important contributions to the trilepton background
\cite{Lyk,Match}.
The complete list of contributions to the 3 leptons final state
from the $WZ$,$W \gamma^*$ and $ZZ$ productions, including cases where
either one or both of the gauge bosons can be virtual, has been calculated
in
\cite{Pai}. The authors of \cite{Pai} have found that the $WZ$, $W \gamma^*$
and $ZZ$
backgrounds (including virtual boson(s)) at the upgraded Tevatron
have together a cross section
of order $0.5 fb$ after the following cuts have been implemented:
$P_t(l_1)>20GeV$, $P_t(l_2)>15GeV$, $P_t(l_3)>10GeV$;
$\vert \eta (l_1,l_{2,3}) \vert <1.0,2.0$; $ISO_{\delta R=0.4}<2GeV$;
$\Eslash_T >25GeV$;
$81GeV<M_{inv}(l \bar  l)<101GeV$; $12GeV<M_{inv}(l \bar l)$;
$60GeV<m_T(l,\Eslash_T)<85GeV$.
\\ We note that there is at most one hard jet
in the 3 leptons backgrounds generated by the
$WZ$, $W \gamma*$ and $ZZ$ productions (including virtual boson(s)).
Since the number of hard jets is equal to 2 in our signal
(see Section \ref{signal1}), a jet veto can thus
reduce this \sm background with respect to the signal.

Other small sources of \sm background have been estimated in \cite{Barb}:
The productions like $Zb$, $Wt$ or $W t \bar t$.
After applying cuts
on the geometrical acceptance, the transverse momentum and the
isolation,
these backgrounds are expected to be at most of order $10^{-4} pb$
in $p \bar p$ collisions with a center of mass energy of $\sqrt s=2 TeV$.
We have checked that the $Zb$ production gives a
negligible contribution to the 3 lepton signature.

There are finally some non-physics sources of background.
First, the 4 leptons signal, which can be generated by the $Z^0 Z^0$
and $t \bar t$ productions, appears as a 3 leptons
signature if one of the leptons is missed.
Besides, the
processes $p \bar p \to \ Z \ + \ X, \ Drell-Yan \ +  \ X$
would mimic a trilepton signal
if $X$ fakes a lepton. Monte Carlo simulations using simplified detector
simulation, like for example SHW \cite{SHW}
as in the present study (see Section \ref{cut1}),
cannot give a reliable estimate of this
background.
A knowledge of the details of the detector response as well as the jet
fragmentation
is necessary in order to determinate the probability to fake a lepton.
In \cite{Kam}, using standard cuts the background coming from
$p \bar p \to \ Z \ + \ X, \ Drell-Yan \ +  \ X$ has been estimated to be of
order
$2 fb$ at Tevatron with $\sqrt s=2 TeV$. The authors of \cite{Kam}
have also estimated the background from the three-jet events faking
trilepton signals
to be around $10^{-3} fb$.

Hence for the study of the \sm background associated to
the 3 lepton signature at Tevatron Run II,
we consider the $W^{\pm} Z^0$ production and both the physics and non-physics
contributions generated by the $Z^0 Z^0$ and $t \bar t$ productions.

\subsection{Supersymmetric background
of the 3 lepton signature at Tevatron}
\label{susyback1}

If an excess of events is
observed in the three lepton channel at Tevatron,
one would wonder what is the origin of those anomalous events.
One would thus have to consider all of the \susyq productions leading to
the three lepton signature.
In the present context of R-parity violation, multileptonic
final states can be generated by the single chargino production
involving \rpv couplings, but
also by the \susyq particle pair production which involves
only gauge couplings \cite{Atlas,RunII}.
In \rpv models, the superpartner
pair production can even
lead to the trilepton signature \cite{Barg,tat1,tat2}.
As a matter of fact, both of the produced \susyq particles
decay, either directly or through cascade decays, into the LSP
which is the neutralino in our framework. In the hypothesis of a
dominant
$\l'$ coupling constant, each of the 2 produced neutralinos can decay into
a charged lepton and two quarks: at least 
two charged leptons and four jets in the final state are produced.
The third charged lepton can be generated in the cascade decays
as for example at the level of
the chargino decay $\tilde \chi^{\pm} \to \tilde \chi^0 l^{\pm} \nu$.

\begin{table}[t]
\begin{center}
\begin{tabular}{|c|c|c|c|c|c|}
\hline
$m_{1/2} \ \backslash \ m_0$
&                 $100GeV$ & $200GeV$   & $300GeV$  & $400GeV$   & $500GeV$
\\
\hline
$100GeV$ & $5.775$         &   $3.376$        &     $2.849$     &    $2.974$
&   $3.195$        \\
\hline
$200GeV$ & $0.147$         &   $0.122$        &     $0.123$     &    $0.130$
&   $0.138$        \\
\hline
$300GeV$ & $1.8 \ 10^{-2}$ &  $1.3 \ 10^{-2}$ & $1.2 \ 10^{-2}$ & $1.3 \
10^{-2}$  &  $1.3 \ 10^{-2}$ \\
\hline
\end{tabular}
\caption{Cross section (in $pb$) of the sum of all the
superpartners pair productions at Tevatron Run II
as a function of the $m_0$ and $m_{1/2}$ parameters
for $\tan \beta=1.5$, $sign(\mu)<0$ and $\l'_{211}=0.05$
at a center of mass energy of $\sqrt s=2 TeV$.
These rates have been calculated with HERWIG \cite{HERWIG} using
the EHLQ2 structure function.}
\label{xosuper}
\end{center}
\end{table}

In Table \ref{xosuper}, we show
for different mSUGRA points the cross section of
the sum of all superpartner pair productions,
namely the $R_p$ conserving SUSY background
of the 3 lepton signature generated by the
single chargino production. As can be seen in this table, the
summed superpartner pair production rate
decreases as $m_0$ and $m_{1/2}$ increase.
This is due to the increase of the superpartner masses as
the $m_0$ or $m_{1/2}$ parameter increases.
The SUSY background will be important only for low values 
of $m_0$ and $m_{1/2}$ as we will see in the following.

\subsection{Cuts}
\label{cut1}

In order to simulate the single chargino production
$p \bar p \to \tilde \chi^{\pm}_1 l^{\mp}$ at Tevatron,
the matrix elements (see Appendix \ref{formulas}) of this process
have been implemented
in a version of the SUSYGEN event generator \cite{SUSYGEN3}
allowing the generation of $p \bar p$ reactions \cite{priv}.
The \sm background
($W^{\pm} Z^0$, $Z^0 Z^0$ and $t \bar t$ productions)
has been simulated
using the PYTHIA event generator \cite{PYTHIA}
and the SUSY background (all SUSY particles pair productions) using the
HERWIG event generator \cite{HERWIG}.
SUSYGEN, PYTHIA and HERWIG have been interfaced with
the SHW detector simulation package \cite{SHW},
which mimics an average of the CDF and D0 Run II detector performance.

We have developped a series of cuts
in order to enhance the signal-to-background ratio.\\
First, we have selected the events with at least three leptons where
the leptons are either an electron, a muon or a
tau reconstructed from a jet, namely $N_l \geq 3 \ [l=e,\mu,\tau]$.
We have also considered the case where the selected leptons
are only electrons and muons, namely $N_l \geq 3 \ [l=e,\mu]$.

The selection criteria on the jets
was to have a number of jets greater or equal to two, where the jets have a
transverse momentum
higher than $10 GeV$, namely $N_j \geq 2$ with $P_t(j) > 10 GeV$. This jet
veto
reduces the 3 lepton backgrounds coming from the $W^{\pm} Z^0$ and $Z^0
Z^0$
productions. Indeed, the
$W^{\pm} Z^0$ production generates no hard jets and the $Z^0 Z^0$
production generates at most one hard jet. Moreover,
the hard jet produced in the $Z^0 Z^0$ background is generated by
a tau decay (see Section \ref{back1}) and can thus be identified as a tau.

\begin{figure}[t]
\begin{center}
\leavevmode
\centerline{\psfig{figure=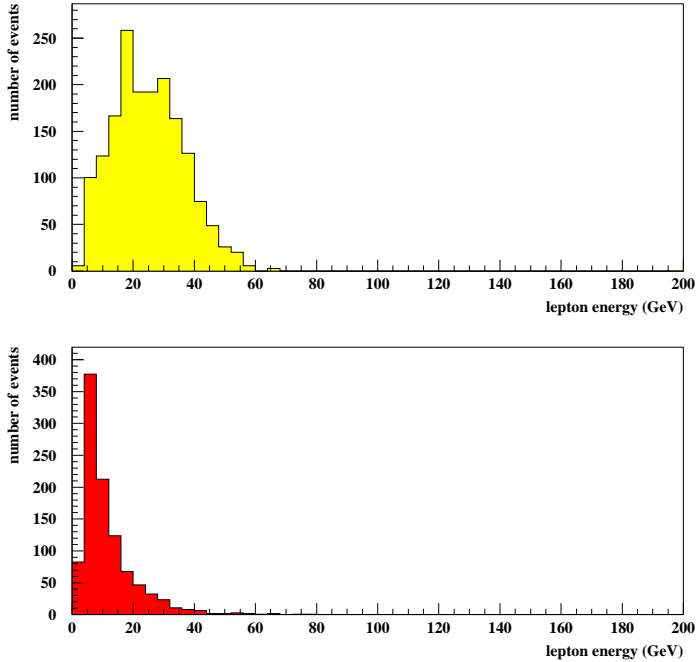,height=4.in}}
\end{center}
\caption{\footnotesize  \it
Distributions of the lowest lepton energy (in $GeV$) among the energies
of the 3 leading leptons (electrons and muons)
in the events containing at least 3 charged leptons and 2 jets
generated by the \sm background (lower curve),
namely the
$W^{\pm} Z^0$, $Z^0 Z^0$ and $t \bar t$ productions,
and the SUSY signal (upper curve), for $\l'_{211}=0.09$,
$M_2=150GeV$, $m_0=200GeV$, $\tan \beta=1.5$
and $sign(\mu)<0$.
The numbers of events correspond to an integrated luminosity of
${\cal L}=10fb^{-1}$.
\rm \normalsize }
\label{distener}
\end{figure}

Besides, some effective cuts concerning the energies of the produced leptons
have been applied. In Fig.\ref{distener},
we show the distributions of the third leading lepton energy
in the 3 lepton events produced by the \sm background
($W^{\pm} Z^0$, $Z^0 Z^0$ and $t \bar t$) and the SUSY signal.
Based on those kinds of distributions, we have chosen the following cut
on the third leading lepton energy: $E(l_3)>10GeV$.
Similarly, we have required that the energies of the 2 leading leptons
verify $E(l_2)>20GeV$ and $E(l_1)>20GeV$.

We will refer to all the selection criteria described above, namely
$N_l \geq 3 \ [l=e,\mu,\tau]$ with $E(l_1)>20GeV$, $E(l_2)>20GeV$,
$E(l_3)>10GeV$,
and $N_j \geq 2$ with $P_t(j) > 10 GeV$,
as cut $1$.

\begin{figure}[t]
\begin{center}
\leavevmode
\centerline{\psfig{figure=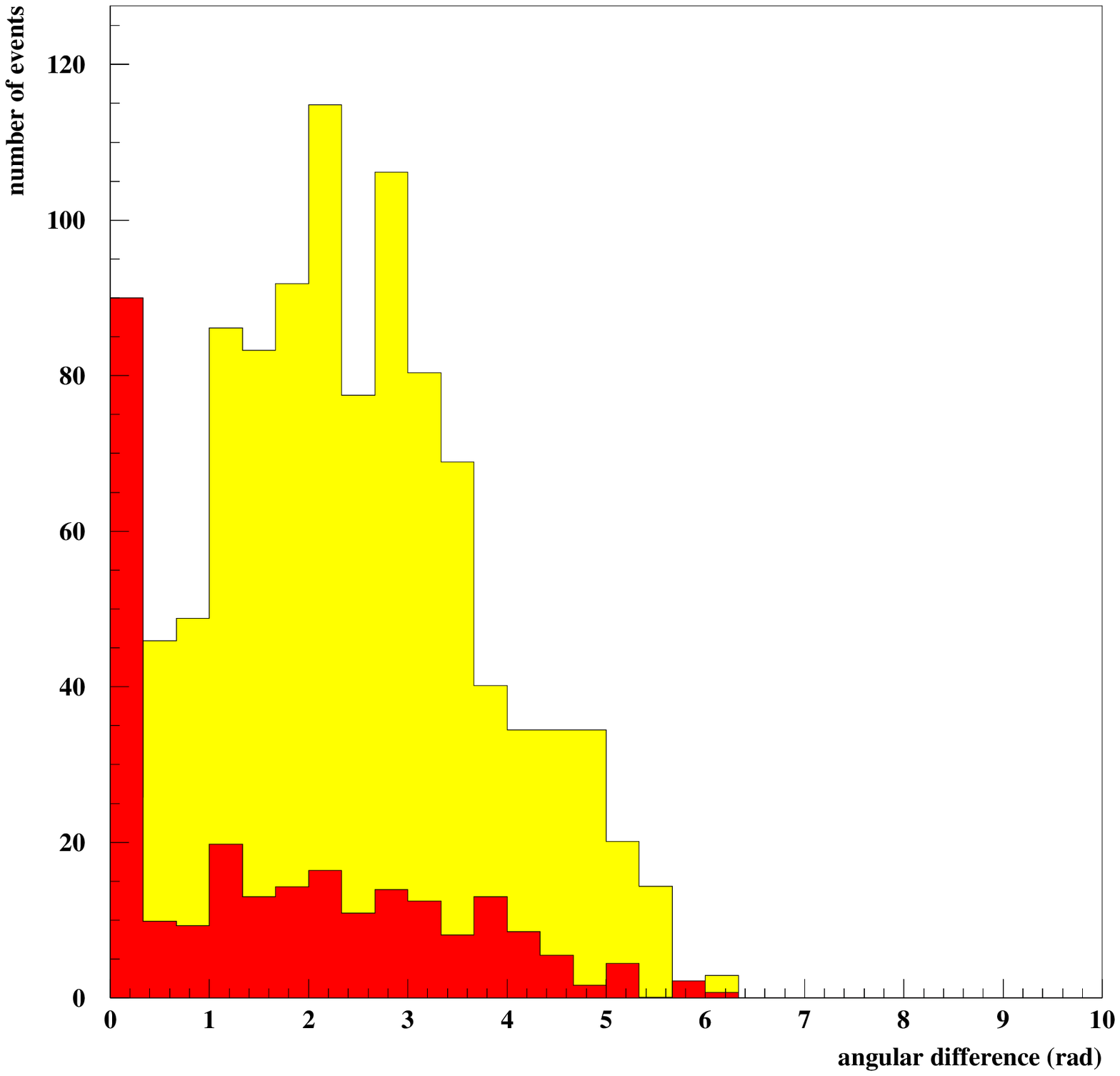,height=4.in}}
\end{center}
\caption{\footnotesize  \it
Distributions of the $\Delta R$ angular difference (in $rad$)
between the third leading lepton (electron or muon) and the second leading
jet in the 3 leptons events selected by applying cut $1$ and
produced by the \sm background (curve in black),
namely the
$W^{\pm} Z^0$, $Z^0 Z^0$ and $t \bar t$ productions, and the SUSY signal
(curve in grey), for $\l'_{211}=0.09$,
$M_2=150GeV$, $m_0=200GeV$, $\tan \beta=1.5$ and $sign(\mu)<0$.
The numbers of events correspond to an integrated luminosity of
${\cal L}=10fb^{-1}$.
\rm \normalsize }
\label{distang}
\end{figure}

Finally, since the leptons originating from the hadron decays
(as in the $t \bar t$ production) are
not well isolated, we have
applied some cuts on the lepton isolation.
We have imposed the isolation cut
$\Delta R=\sqrt{\delta \phi^2+\delta \theta^2}>0.4$ where
$\phi$ is the azimuthal angle and $\theta$ the polar angle between
the 3 most energetic charged leptons and the 2 hardest jets.
Such a cut is for instance motivated
by the distributions shown in Fig.\ref{distang}
of the $\Delta R$ angular difference
between the third leading lepton and the second leading
jet, in the 3 lepton events generated by the SUSY signal
and \sm background. We call cut $\Delta R>0.4$ together with 
cut $1$, cut $2$.

In order to eliminate poorly isolated leptons,
we have also required that $E<2GeV$, where $E$ represents the
summed energies of the jets being close to a muon or an electron,
namely the jets contained in the cone centered on a muon or an electron
and defined by $\Delta R<0.25$.
This cut is not applied for taus candidates as they have hadronic decays.
It is quite efficient (see Fig.\ref{dianmu} for the 2 lepton case) since
we sum over all jet energies in the cone. The \sm background shows more jets 
and less separation between jets and leptons in $(\theta, \phi)$ in 
final state than the single productions \footnote{This cut will have to be fine tuned
with real events since it will depend on the energy flow inside the detector,
the overlap and minimum biased events.}.
We denote cut $E<2GeV$ plus cut $2$ as cut $3$
\footnote{Although it has not been applied, we mention 
another kind of isolation cut which allows
to further reduce the \sm background:
$\delta \phi>70^{\circ}$ between the leading
charged lepton and the 2 hardest jets.}.

\begin{figure}[t]
\begin{center}
\leavevmode
\centerline{\psfig{figure=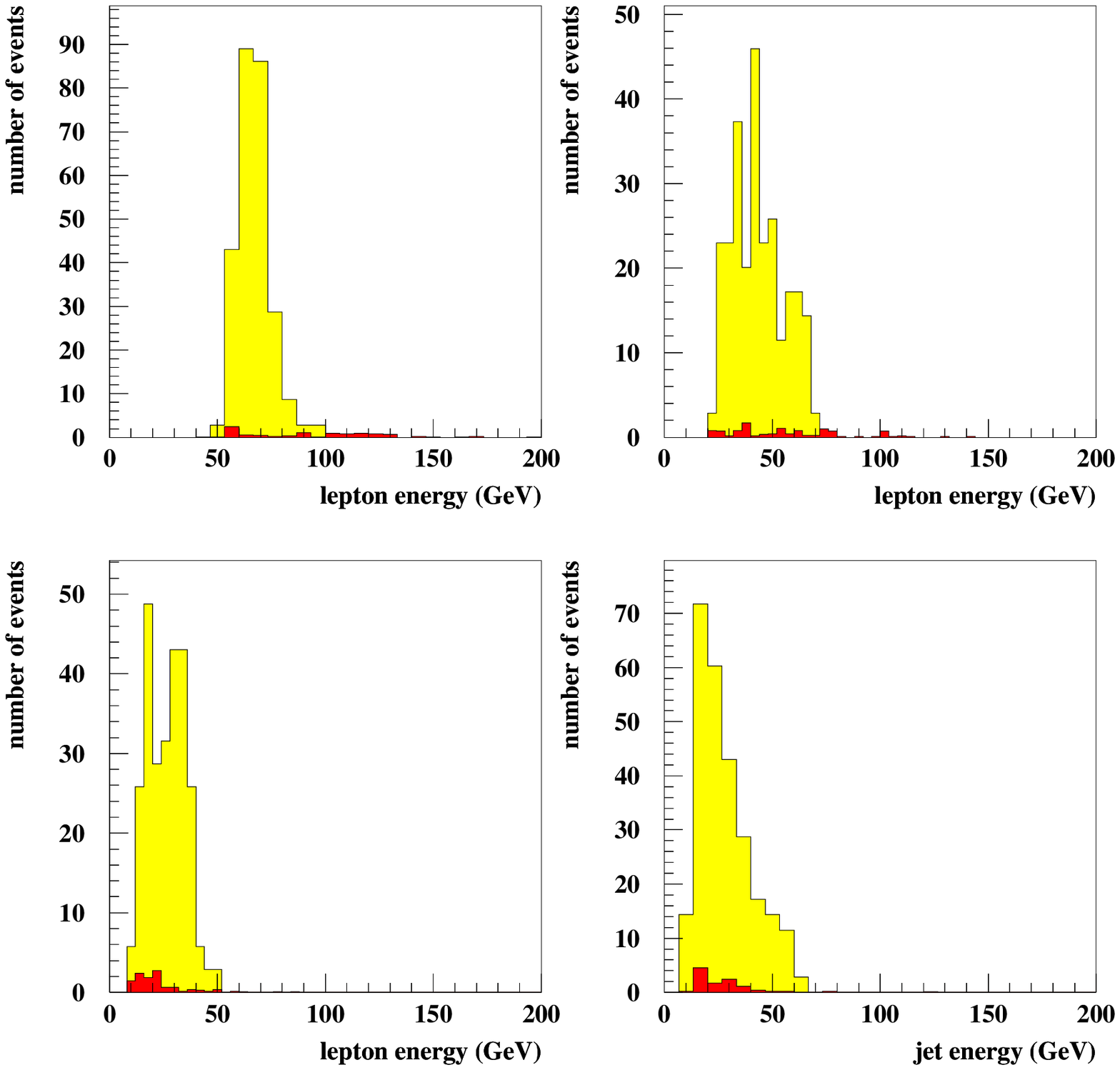,height=4.in}}
\end{center}
\caption{\footnotesize  \it
Energy distributions (in $GeV$) of the 3 leading charged leptons
and the second leading jet in the events containing at least 3 charged
leptons selected by applying cut $3$ and
produced by the \sm background (curve in black), namely the
$W^{\pm} Z^0$, $Z^0 Z^0$ and $t \bar t$ productions,
and the SUSY signal (curve in grey), for $\l'_{211}=0.09$,
$M_2=150GeV$, $m_0=200GeV$, $\tan \beta=1.5$ and $sign(\mu)<0$.
The upper left plot represents the leading lepton distribution,
the upper right plot the second leading lepton distribution and
the lower left plot the third leading lepton distribution.
The numbers of events correspond to an integrated luminosity of
${\cal L}=10fb^{-1}$.
\rm \normalsize }
\label{apres}
\end{figure}

The selected events require high energy charged leptons and jets
and can thus easily be triggered at Tevatron. This is illustrated in
Fig.\ref{apres} where we show the energy distributions of the 3 leptons
and the second leading jet in the 3 leptons events selected by
applying cut $3$ and generated by the SUSY signal and \sm background.

\begin{table}[t]
\begin{center}
\begin{tabular}{|c|c|c|c|c|}
\hline
& $W^{\pm} Z^0$   & $Z^0 Z^0$       & $t \bar t$
& Total   \\
\hline
cut $1$
& $1.39 \pm 0.11$ & $1.37 \pm 0.11$ & $39.80 \pm 1.00$
& $42.56 \pm 1.01$ \\
\hline
cut $2$
& $0.26 \pm 0.05$ & $0.21 \pm 0.04$ & $4.23  \pm 0.39$
& $4.70 \pm 0.40$ \\
\hline
cut $3$
& $0.24 \pm 0.04$ & $0.17 \pm 0.04$ & $1.14 \pm 0.17$
& $1.55 \pm 0.18$  \\
\hline
cut $1^{\star}$
& $0.51 \pm 0.06$ & $0.73 \pm 0.08$ & $27.80 \pm 0.80$
& $29.04 \pm 0.80$ \\
\hline
cut $2^{\star}$
& $0.26 \pm 0.05$ & $0.21 \pm 0.04$ & $2.92 \pm 0.27$
& $3.39 \pm 0.28$ \\
\hline
cut $3^{\star}$
& $0.23 \pm 0.04$ & $0.17 \pm 0.04$ & $0.64 \pm 0.13$
& $1.04 \pm 0.14$  \\
\hline
\end{tabular}
\caption{Numbers of three lepton events generated by the
\sm background ($W^{\pm} Z^0$, $Z^0 Z^0$ and $t \bar t$ productions)
at Tevatron Run II for the cuts described in the text, assuming an
integrated
luminosity of ${\cal L}=1 fb^{-1}$ and a center of mass energy of $\sqrt s=2
TeV$.
The cuts marked by a $\star$ do not include the reconstruction
of the tau-jets.
These results have been obtained by generating and simulating
$3 \ 10^5$ events for the $W^{\pm} Z^0$ production,
$    10^4$ events for the $Z^0 Z^0$ and
$3 \ 10^5$ events for the $t \bar t$.}
\label{cuteff}
\end{center}
\end{table}

In Table \ref{cuteff}, we give the numbers of three lepton events expected
from the
\sm background at Tevatron Run II with the various cuts described above.
We see in Table \ref{cuteff} that the main source
of \sm background to the three lepton signature at Tevatron
is the $t \bar t$ production.
This is due to the important cross section of the
$t \bar t$ production compared to the other \sm
backgrounds (see Section \ref{back1}).
Table \ref{cuteff} also shows that the $t \bar t$ background
is relatively more suppressed than the other sources of
\sm background by the lepton isolation cuts. The reason is that
in the $t \bar t$ background, one of the 3 charged leptons
of the final state is generated in a $b$-jet and is thus
not well isolated.

\begin{table}[t]
\begin{center}
\begin{tabular}{|c|c|c|c|c|c|}
\hline
$m_{1/2} \ \backslash \ m_0$
&  $100GeV$ & $200GeV$   & $300GeV$  & $400GeV$
&  $500GeV$   \\
\hline
$100GeV$
& $93.94$       & $125.59$       & $80.53$       & $66.62$
& $63.90$ \\
\hline
$200GeV$
& $5.11$        & $4.14$         & $3.86$        & $4.02$
&  $4.26$ \\
\hline
$300GeV$
& $2.26$        & $0.66$         & $0.52$        & $0.55$
&  $0.55$ \\
\hline
\end{tabular}
\caption{Number of 3 lepton events generated by the
SUSY background (all superpartners pair productions)
at Tevatron Run II
as a function of the $m_0$ and $m_{1/2}$ parameters
for $\tan \beta=1.5$, $sign(\mu)<0$ and $\l'_{211}=0.05$.
Cut 3 (see text) has been applied.
These results have been obtained by generating 7500 events
and correspond to an integrated luminosity
of ${\cal L}=1 fb^{-1}$
and a center of mass energy of $\sqrt s=2 TeV$.}
\label{cutSUSY}
\end{center}
\end{table}

In Table \ref{cutSUSY}, we give the number of three lepton
events generated by the
SUSY background (all superpartners pair productions)
at Tevatron Run II as a function of
the $m_0$ and $m_{1/2}$ parameters for the cut 3.
This number of events decreases as $m_0$ and $m_{1/2}$
increase due to the behaviour of the summed superpartners
pair productions cross section in the SUSY parameter space
(see Section \ref{susyback1}).

\subsection{Results}
\label{res}

\subsubsection{Discovery potential for the
$\l'_{2jk}$ coupling constant}
\label{lp211}

We first present the reach in the
mSUGRA parameter space obtained from the analysis of
the trilepton signature at Tevatron Run II generated
by the single chargino production
through the $\l'_{211}$ coupling, namely
$p \bar p \to \tilde \chi^{\pm}_1 \mu^{\mp}$.
The sensitivity that can be obtained on the $\l'_{2jk}$
($j$ and $k$ being not equal to $1$ simultaneously) couplings
based on the $\tilde \chi^{\pm}_1 \mu^{\mp}$ production analysis
will be discussed at the end of this section for a given mSUGRA point.
We give more detailed results for the case of a single dominant
$\l'_{211}$ coupling since this \rpv coupling gives the highest partonic
luminosity to the $\tilde \chi^{\pm}_1 \mu^{\mp}$ production
(see Section \ref{cross1}) and leads thus to
the highest sensitivities.

\begin{figure}[t]
\begin{center}
\leavevmode
\centerline{\psfig{figure=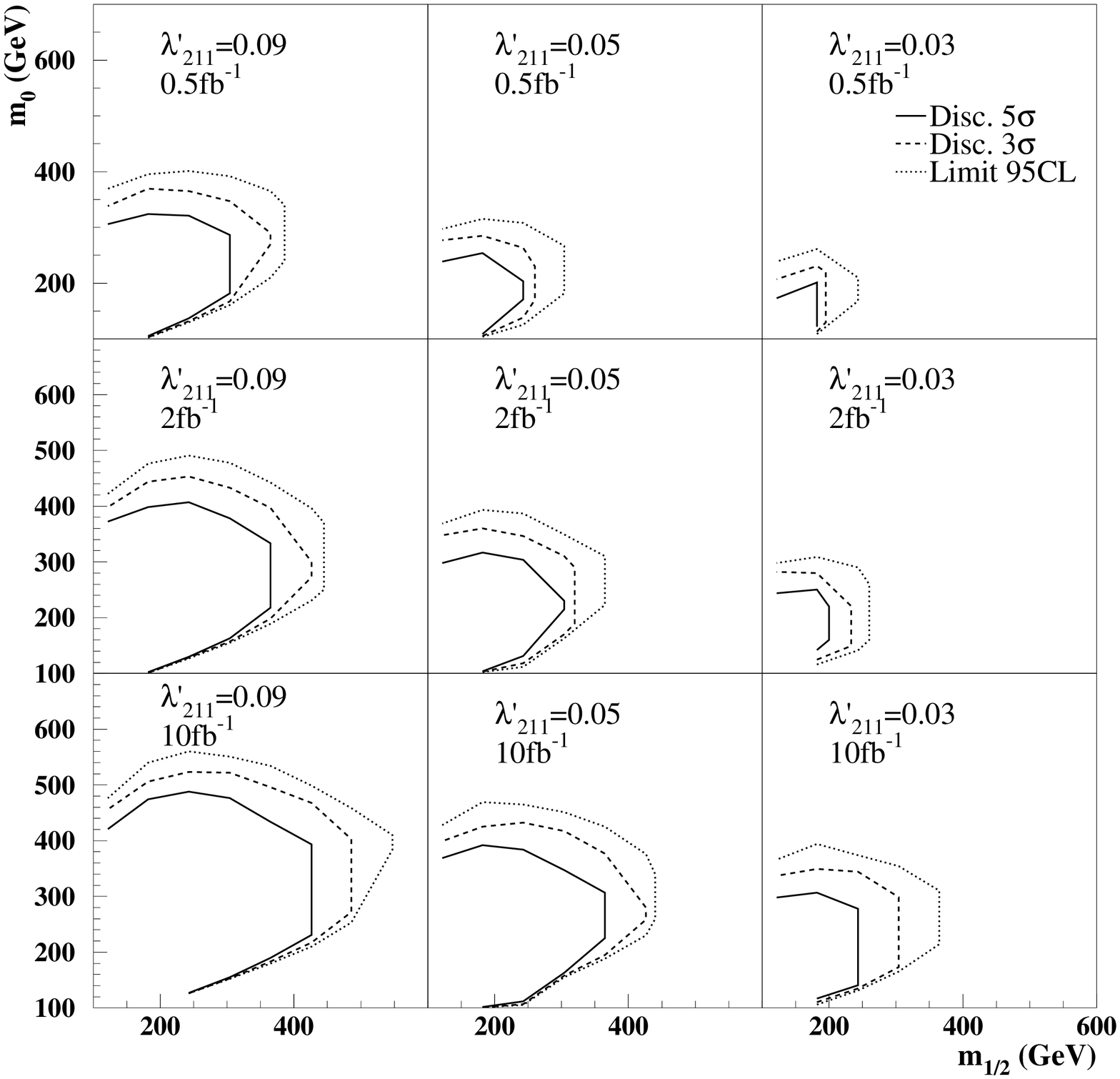,height=4.5in}}
\end{center}
\caption{Discovery contours at $5 \sigma$ (full line),
$ 3 \sigma$ (dashed line)
and limit at $95 \% \ C.L.$ (dotted line)
obtained from the trilepton signature analysis
at Tevatron Run II assuming
a center of mass energy of $\sqrt s=2 TeV$.
This discovery reach is
presented in the plane $m_0$ versus $m_{1/2}$,
for $sign(\mu)<0$, $\tan \beta=1.5$
and different values of $\l'_{211}$ and of luminosity.}
\label{fig2}
\end{figure}

\begin{figure}[t]
\begin{center}
\leavevmode
\centerline{\psfig{figure=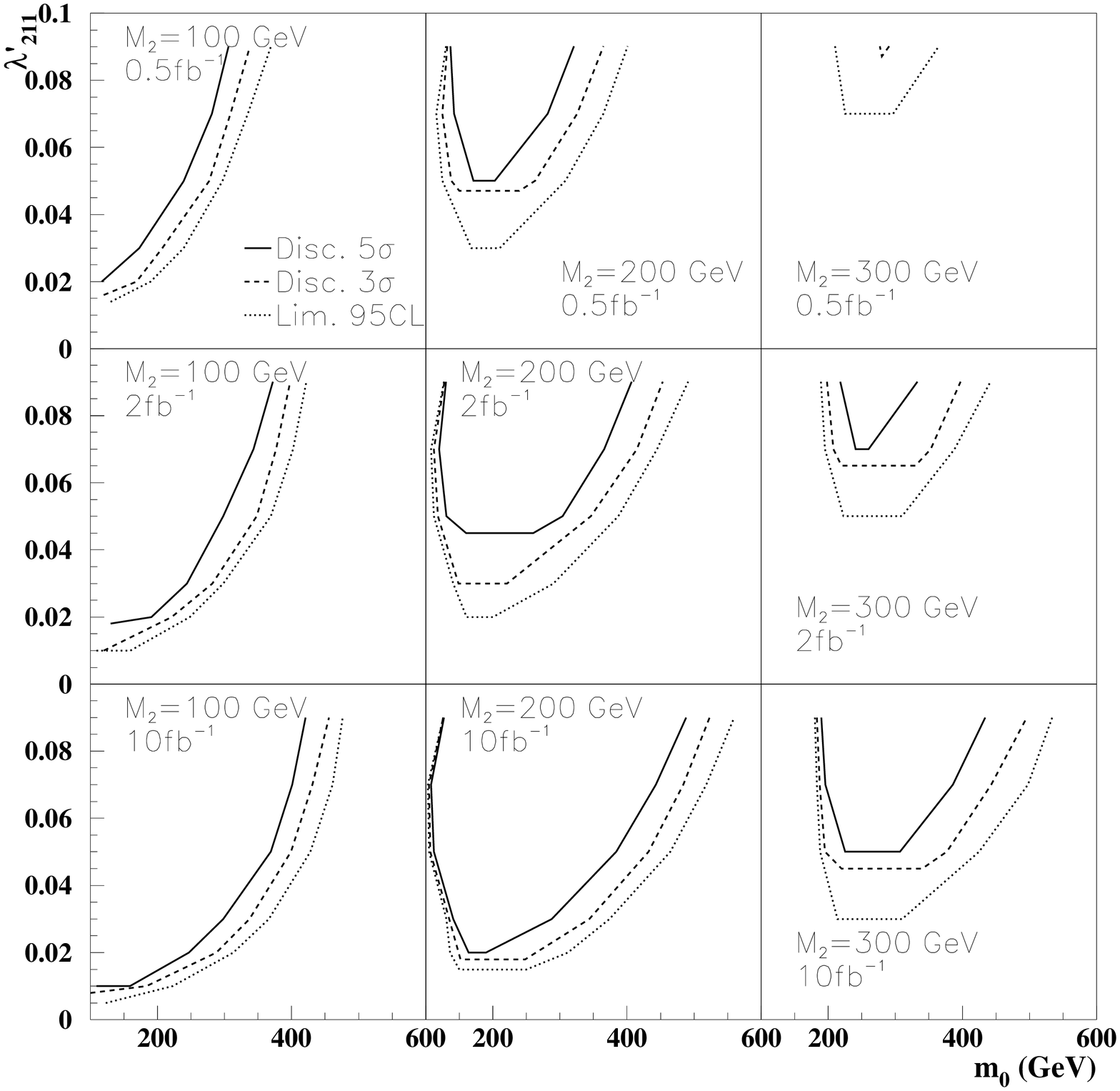,height=4.5in}}
\end{center}
\caption{Discovery contours at $5 \sigma$ (full line), $ 3 \sigma$
(dashed line) and limit at $95 \% \ C.L.$ (dotted line) presented
in the plane $\l'_{211}$ versus the $m_0$ parameter,
for $sign(\mu)<0$, $\tan \beta=1.5$
and different values of $M_2$ and of luminosity.}
\label{fig1}
\end{figure}

In Fig.\ref{fig2},
we present the $3 \sigma$ and $ 5 \sigma$ discovery
contours and the limits at $95 \%$
confidence level in the plane $m_0$ versus $m_{1/2}$,
for $sign(\mu)<0$, $\tan \beta=1.5$ and
using a set of values for $\l'_{211}$ and the luminosity.
This discovery potential was obtained by considering
the $\tilde \chi^{\pm}_1 \mu^{\mp}$ production and
the background originating from the Standard Model.
The signal and background were selected by
using cut $3$ described in Section \ref{cut1}.
The results presented for a luminosity of
${\cal L}=0.5 fb^{-1}$
in Fig.\ref{fig2} and Fig.\ref{fig1}
were obtained with cut 2 only in order to optimize the
sensitivity on the SUSY parameters.
The reduction of the sensitivity on $\l'_{211}$
observed in Fig.\ref{fig2}
when either $m_0$ or $m_{1/2}$ increases is due to the
decrease of the $\tilde \chi^{\pm}_1 \mu^{\mp}$ production cross section
with $m_0$ or $m_{1/2}$ (or equivalently $M_2$), 
which can be observed in Fig.\ref{XS02}.
In Fig.\ref{fig2}, we also see that
for all the considered values of $\l'_{211}$ and the luminosity,
the sensitivity on $m_{1/2}$ is reduced to low masses
in the domain $m_0 \stackrel{<}{\sim} 200GeV$. This important reduction
of the sensitivity as $m_0$ decreases is due to
the decrease of the phase space
factor associated to the decay
$\tilde \nu_{\mu} \to \tilde \chi^{\pm} \mu^{\mp}$
(see Section \ref{cross1}).
%
%
Finally, we note from Fig.\ref{XSmu} that for $sign(\mu)>0$
the $\tilde \chi^{\pm}_1 \mu^{\mp}$ production cross section,
and thus the sensitivities presented in Fig.\ref{fig2},
would incur a little increase compared to the case $sign(\mu)<0$.

In Fig.\ref{fig1}, the discovery potential is shown
in the $\l'_{211}$-$m_0$ plane for different values of $M_2$
and the luminosity.
For a given value of $M_2$, we note that the sensitivity on the $\l'_{211}$
coupling decreases at high and low values of $m_0$.
The main explanation is the decrease of the
$p \bar p \to \tilde \chi^{\pm}_1 \mu^{\mp}$
rate at high and low values of $m_0$
which appears clearly in Fig.\ref{XScl}.
We also observe, as in Fig.\ref{fig2},
a decrease of the sensitivity on the $\l'_{211}$
coupling when $M_2$ (or equivalently $m_{1/2}$) increases
for a fixed value of $m_0$.

The strongest bounds on the \susyq masses obtained at LEP 
in an \rpv model with a non-vanishing $\l'$ Yukawa coupling are
$m_{\tilde \chi^{0}_1}>26 GeV$ 
(for $m_0=200 GeV$ and $\tan \beta =\sqrt 2$ \cite{neut}), 
$m_{\tilde \chi^{\pm}_1}>100 GeV$, 
$m_{\tilde l}>93 GeV$, $m_{\tilde \nu}> 86 GeV$ \cite{Mass}.
For the minimum values of $m_0$ and $m_{1/2}$
spanned by the parameter space described in Figures \ref{fig2}
and \ref{fig1}, namely
$m_0=100GeV$ and $M_2=100GeV$, the mass spectrum is
$m_{\tilde \chi^{\pm}_1}= 113GeV$, $m_{\tilde \chi^{0}_1}= 54GeV$,
$m_{\tilde \nu_L}= 127 GeV$, $m_{\tilde l_L}= 137 GeV$,
$m_{\tilde l_R}= 114 GeV$, so that we are well above these limits.
Since both the scalar and gaugino
masses increase with $m_0$ and $m_{1/2}$,
the parameter space described in Figures \ref{fig2}
and \ref{fig1} lies outside the SUSY parameters ranges
excluded by LEP data \cite{Mass,neut}.
Therefore, the discovery potential of Figures \ref{fig2}
and \ref{fig1} represents an important improvement with respect to
the \susyq masses limits derived from LEP data \cite{Mass,neut}.
Figures \ref{fig2} and \ref{fig1} show also that the low-energy bound
on the considered \rpv coupling,
$\l'_{211}<0.09 (m_{\tilde d_R}/100GeV)$ at $1 \sigma$ (from $\pi$ decay)
\cite{Bhatt}, can be greatly improved.

\begin{figure}[t]
\begin{center}
\leavevmode
\centerline{\psfig{figure=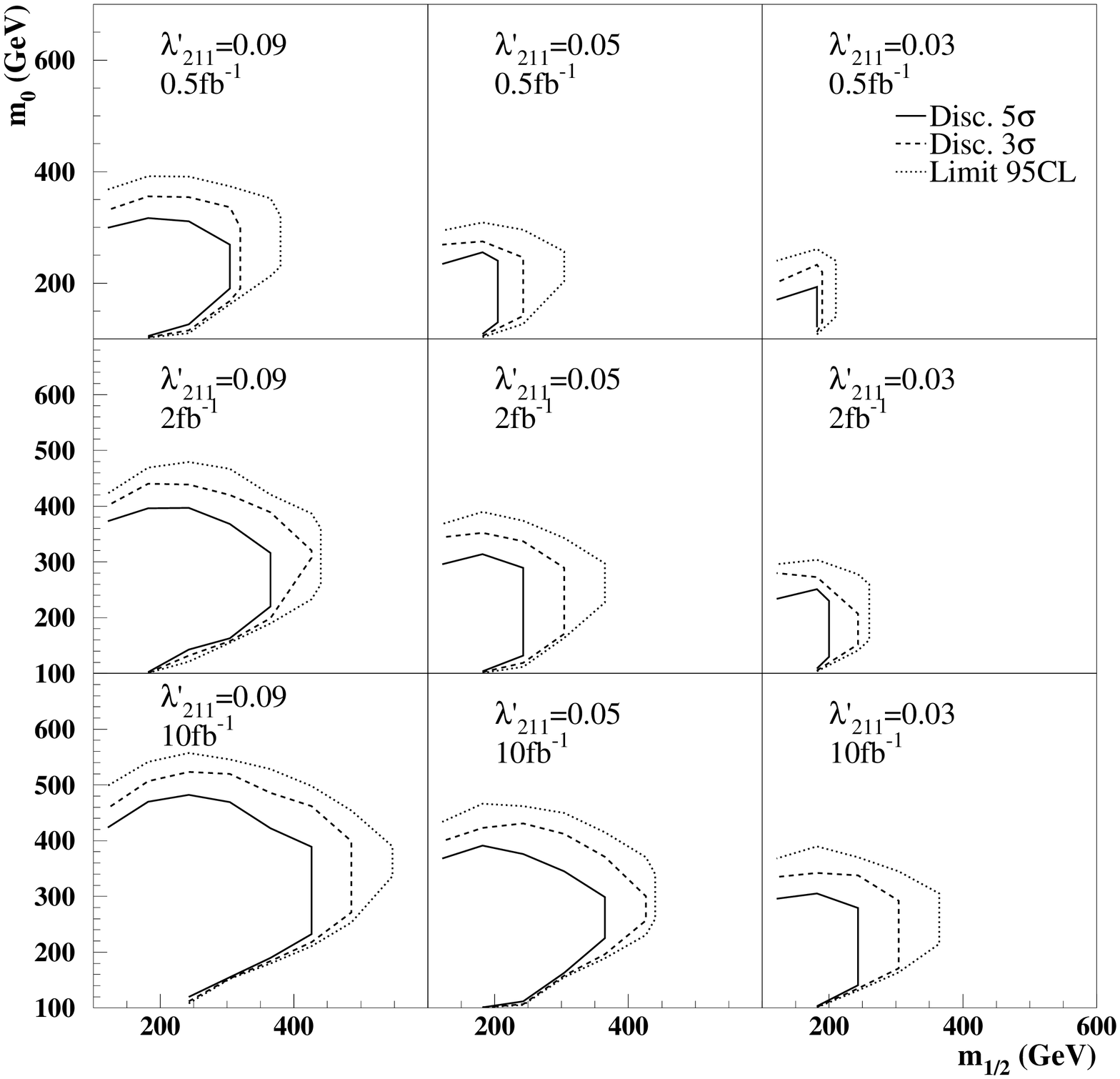,height=4.5in}}
\end{center}
\caption{Discovery contours at $5 \sigma$ (full line),
$ 3 \sigma$ (dashed line)
and limit at $95 \% \ C.L.$ (dotted line) presented
in the plane $m_0$ versus $m_{1/2}$
and obtained without reconstruction of the tau leptons decaying into jets
for $sign(\mu)<0$, $\tan \beta=1.5$
and different values of $\l'_{211}$ and of luminosity.}
\label{fig2p}
\end{figure}

\begin{figure}[t]
\begin{center}
\leavevmode
\centerline{\psfig{figure=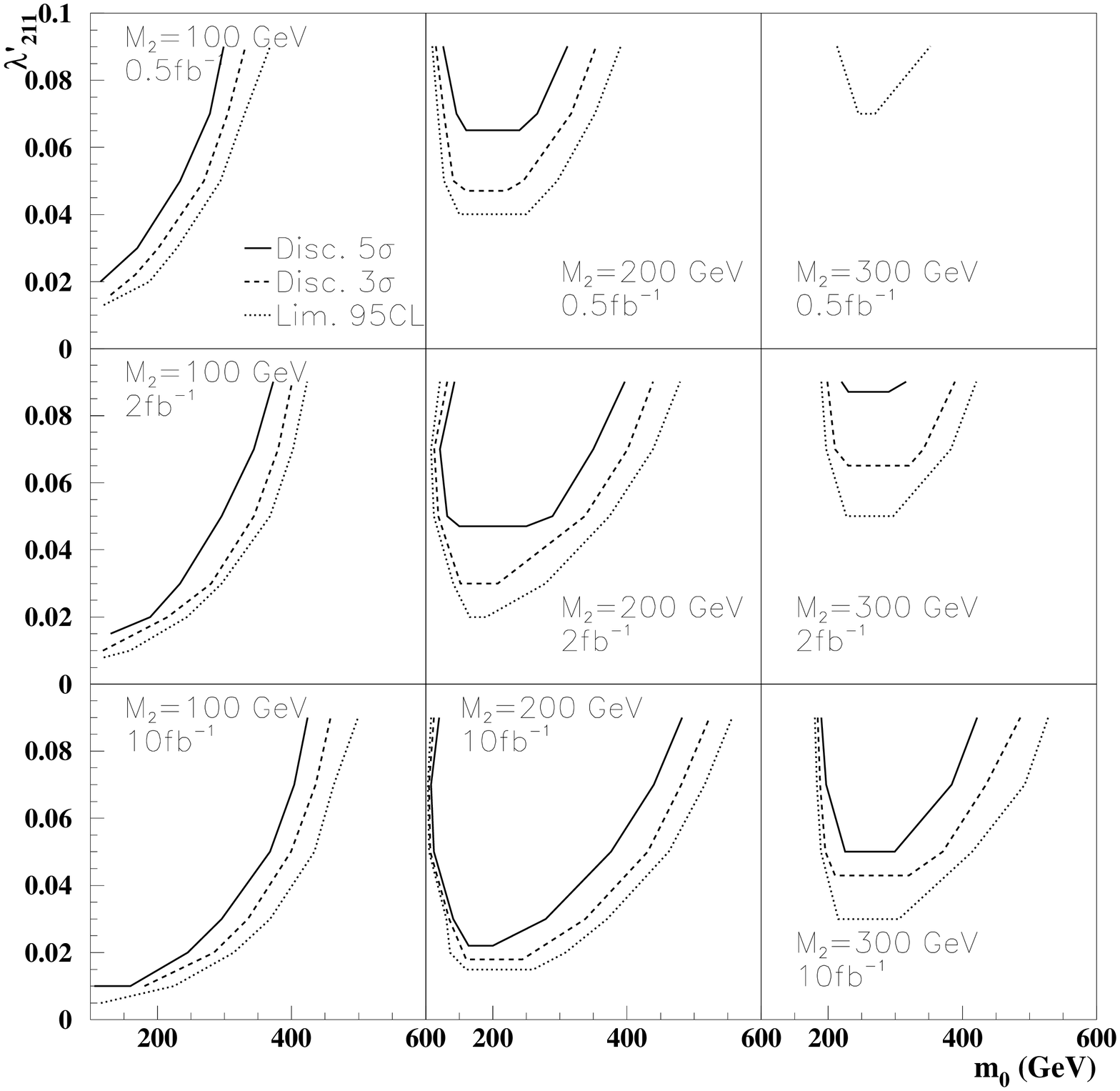,height=4.5in}}
\end{center}
\caption{Discovery contours at $5 \sigma$ (full line), $ 3 \sigma$
(dashed line) and limit at $95 \% \ C.L.$ (dotted line) presented
in the plane $\l'_{211}$ versus the $m_0$ parameter
and obtained without reconstruction of the tau leptons decaying into jets
for $sign(\mu)<0$, $\tan \beta=1.5$
and different values of $M_2$ and of luminosity.}
\label{fig1p}
\end{figure}

Interesting sensitivities on the SUSY parameters can
already be obtained within the first year of Run II at Tevatron
with a low luminosity (${\cal L}=0.5 fb^{-1}$)
and no reconstruction of the tau-jets. To illustrate this point,
we present in Fig.\ref{fig2p}
and Fig.\ref{fig1p} the same discovery potentials as
in Fig.\ref{fig2} and Fig.\ref{fig1}, respectively,
obtained without reconstruction of the tau leptons decaying into jets.
By comparing Fig.\ref{fig2}, Fig.\ref{fig1} and
Fig.\ref{fig2p}, Fig.\ref{fig1p},
we observe that the sensitivity on the SUSY parameters is
weakly affected by the reconstruction of the tau-jets
\footnote{This is actually an artefact of the method:
cut 3 is our most efficient cut to reduce the \sm background with 
electrons and muons but is not applied with taus. Thus, the
relative ratio signal over background is not so good with taus.
Finding another efficient cut could improve our discovery potential
and limits using taus.}.

Using the ratios of the cross sections for the
$\tilde \chi^+_1 \mu^-$ productions via different $\l'_{2jk}$ couplings,
one can deduce
from the sensitivity obtained on $\l'_{211}$ via the 3 lepton
final state analysis an estimation of
the sensitivity on any $\l'_{2jk}$ coupling.
For instance, let us consider the SUSY point
$m_0=180GeV$, $M_2=200GeV$, $\tan \beta=1.5$ and $\mu=-200GeV$
($m_{\tilde u_L}=601GeV$, $m_{\tilde d_L}=603GeV$,
$m_{\tilde u_R}=582GeV$, $m_{\tilde d_R}=580GeV$,
$m_{\tilde l_L}=253GeV$, $m_{\tilde l_R}=205GeV$
$m_{\tilde \nu_L}=248GeV$,
$m_{\tilde \chi^{\pm}_1}=199GeV$, $m_{\tilde \chi^0_1}=105GeV$)
which corresponds, as can be seen in Fig.\ref{fig1},
to the point where the sensitivity
on $\l'_{211}$ is maximized for $M_2=200GeV$.
We can see on Fig.\ref{XScl} that for this SUSY point,
the ratio between the rates of the $\tilde \chi^+_1 \mu^-$
productions via $\l'_{211}$ and $\l'_{221}$ is
$\sigma(\l'_{211}) / \sigma(\l'_{221}) \approx 7.9$
for same values of the \rpv couplings. Therefore,
since the single chargino production rate scales as $\l'^2$ (see
Appendix \ref{formulas}),
the sensitivity on $\l'_{221}$ at this SUSY point
is equal to the sensitivity obtained on $\l'_{211}$
($\sim 0.02$ at $95 \% CL$ with ${\cal L}=2fb^{-1}$
as shows Fig.\ref{fig1})
multiplied by the factor $\sqrt {7.9}$, namely
$\sim 0.05$. This result represents a significant
improvement with respect to the low-energy
indirect limit $\l'_{221}<0.18 (m_{\tilde d_R}/100GeV)$ \cite{Bhatt}.
Using the same method, we find at the same SUSY point the sensitivities
on the $\l'_{2jk}$ coupling constants given in Table \ref{coup}.
\begin{table}[t]
\begin{center}
\begin{tabular}{|c|c|c|c|c|c|c|c|}
\hline
$\l'_{212}$ & $\l'_{213}$ & $\l'_{221}$ & $\l'_{222}$
& $\l'_{223}$ & $\l'_{231}$ & $\l'_{232}$ & $\l'_{233}$  \\
\hline
0.04 & 0.07 & 0.05 & 0.12 & 0.21 & 0.10 & 0.36 & 0.63 \\
\hline
\end{tabular}
\caption{\em
Sensitivities at $95 \% CL$ on the $\l'_{2jk}$
coupling constants derived from the sensitivity
on $\l'_{211}$ for a luminosity of ${\cal L}=2fb^{-1}$
and the following set of SUSY parameters,
$\tan\beta=1.5$,  $M_2=200GeV$, $\mu=-200GeV$ and
$m_0=180GeV$.}
\label{coup}
\end{center}
\end{table}
All the sensitivities on the $\l'_{2jk}$ coupling constants
given in Table \ref{coup} are stronger than
the low-energy bounds of \cite{Bhatt} which we rewrite here:
$\l'_{21k}<0.09 (m_{\tilde d_{kR}}/100GeV)$ at $1 \sigma$ ($\pi$ decay),
$\l'_{22k}<0.18 (m_{\tilde d_{kR}}/100GeV)$ at $1 \sigma$ ($D$ decay),
$\l'_{231}<0.22 (m_{\tilde b_L}/100GeV)$ at $2 \sigma$
($\nu_{\mu}$ deep inelastic scattering),
$\l'_{232}<0.36 (m_{\tilde q}/100GeV)$ at $1 \sigma$ ($R_{\mu}$),
$\l'_{233}<0.36 (m_{\tilde q}/100GeV)$ at $1 \sigma$ ($R_{\mu}$).
\\ In the case of a single dominant $\l'_{2j3}$
coupling,
the neutralino decays as $\tilde \chi^0_1 \to \mu u_j b$
and the semileptonic decay of the b-quark could affect
the analysis efficiency. Therefore in this case, the precise sensitivity
cannot be simply calculated by scaling the value obtained for
$\lambda^\prime_{211}$.
Nevertheless, the order of magnitude of the sensitivity
which can be inferred from our analysis should be correct.
\\ The range of SUSY parameters in which the constraint on a given
$\l'_{2jk}$ coupling constant obtained via the three leptons analysis
is stronger than the relevant low-energy bound
depends on the low-energy bound itself as well as
on the values of the cross section for the
single chargino production via the considered $\l'_{2jk}$ coupling.
\\ Finally, we remark that
while the low-energy constraints on the $\l'_{2jk}$ couplings
become weaker as the squark masses increase, the sensitivities
on those couplings obtained from the three leptons analysis
are essentially independent of the squark masses as long as
$m_{\tilde q}>m_{\tilde \chi^{\pm}_1}$ (recall that the
branching ratio of the decay
$\tilde \chi^{\pm}_1 \to q \bar q \tilde \chi^0_1$ is
greatly enhanced when $m_{\tilde q}<m_{\tilde \chi^{\pm}_1}$).

We end this section by some comments on the effect of
the \susyq $R_p$ conserving background to the 3 lepton
signature. In order to illustrate this discussion, we consider
the results on the $\l'_{211}$ coupling constant.\\
We see from Table \ref{cutSUSY} that
the SUSY background to the 3 lepton final
state can affect the sensitivity on the
$\l'_{211}$ coupling constant
obtained by considering only the \sm background,
which is shown in Fig.\ref{fig2}, only in the
region of small superpartner masses, namely in the domain
$m_{1/2} \stackrel{<}{\sim} 300GeV$ for $\tan \beta=1.5$,
$sign(\mu)<0$ and assuming a luminosity of ${\cal L}=1fb^{-1}$.\\
In contrast with the SUSY signal amplitude which is increased
if $\l'_{211}$ is enhanced,
the SUSY background amplitude is typically
independent on the value of the $\l'_{211}$ coupling constant
since the superpartner pair production
does not involve \rpv couplings.
Therefore, even if we consider the SUSY background
in addition to the \sm one, it is still true that
large values of the $\l'_{211}$ coupling can
be probed over a wider domain of the SUSY parameter space than
low values, as can be observed in Fig.\ref{fig2} for
$m_{1/2} \stackrel{>}{\sim} 300GeV$. Note that in Fig.\ref{fig2}
larger values of $\l'_{211}$ could have been considered as the
low-energy bound on this \rpv coupling, namely $\l'_{211}<0.09
(m_{\tilde d_R}/100GeV)$ \cite{Bhatt}, is proportional to
the squark mass.\\
Finally, we mention that further cuts,
as for instance some cuts based on the
superpartner mass reconstructions (see Section \ref{recons}),
could allow to reduce the SUSY background to the 3 lepton
signature.

\subsubsection{High $\tan \beta$ scenario}
\label{htanb}

\begin{figure}[t]
\begin{center}
\leavevmode
\centerline{\psfig{figure=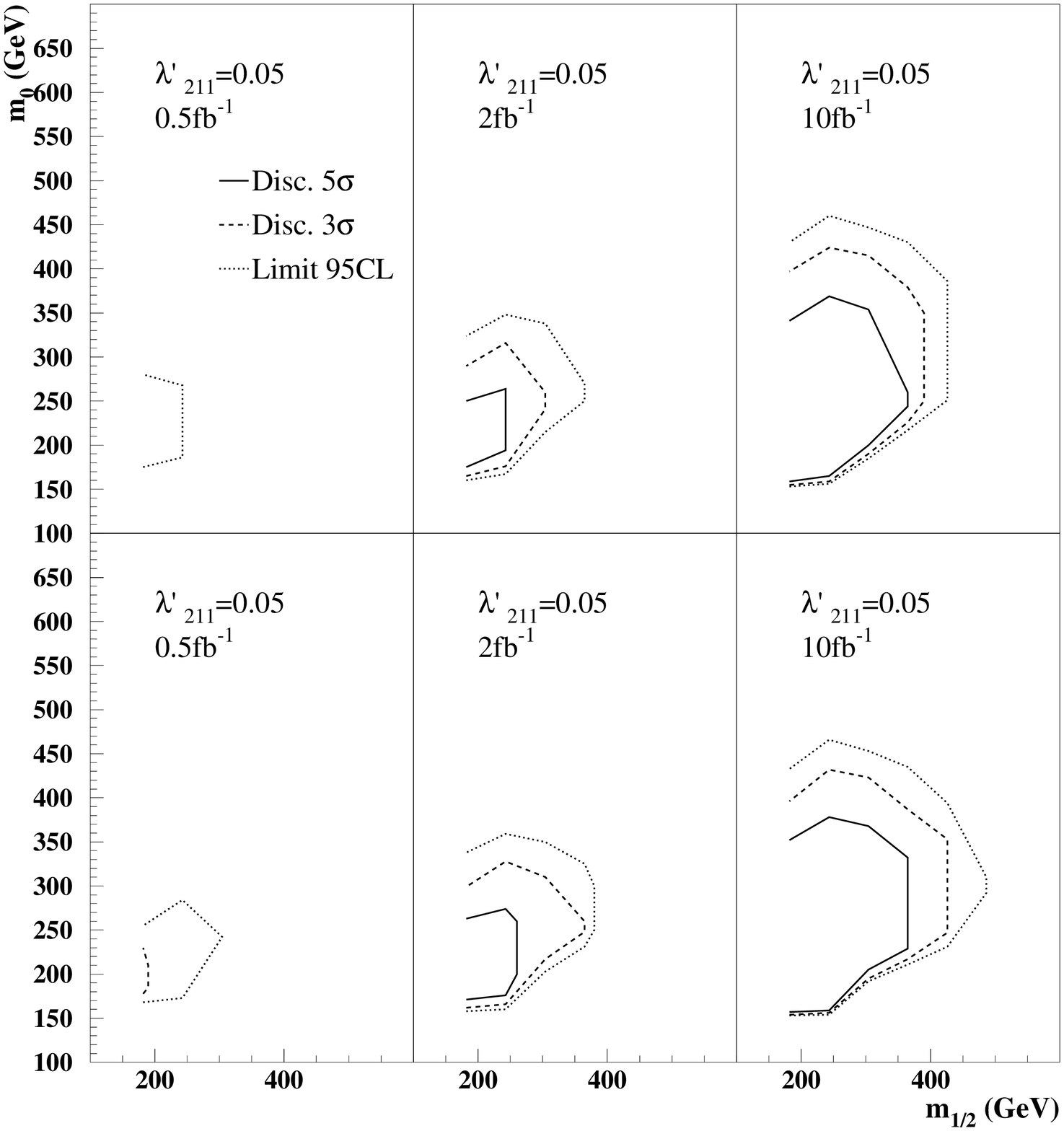,height=4.in}}
\end{center}
\caption{Discovery contours at $5 \sigma$ (full line),
$ 3 \sigma$ (dashed line)
and limit at $95 \% \ C.L.$ (dotted line) presented
in the plane $m_0$ versus $m_{1/2}$,
for $sign(\mu)<0$, $\tan \beta=50$, $\l'_{211}=0.09$
and different values of luminosity. The upper (lower) curves
are obtained without (with) the reconstruction of the tau-jets.}
\label{fig2tanb}
\end{figure}

\begin{figure}[t]
\begin{center}
\leavevmode
\centerline{\psfig{figure=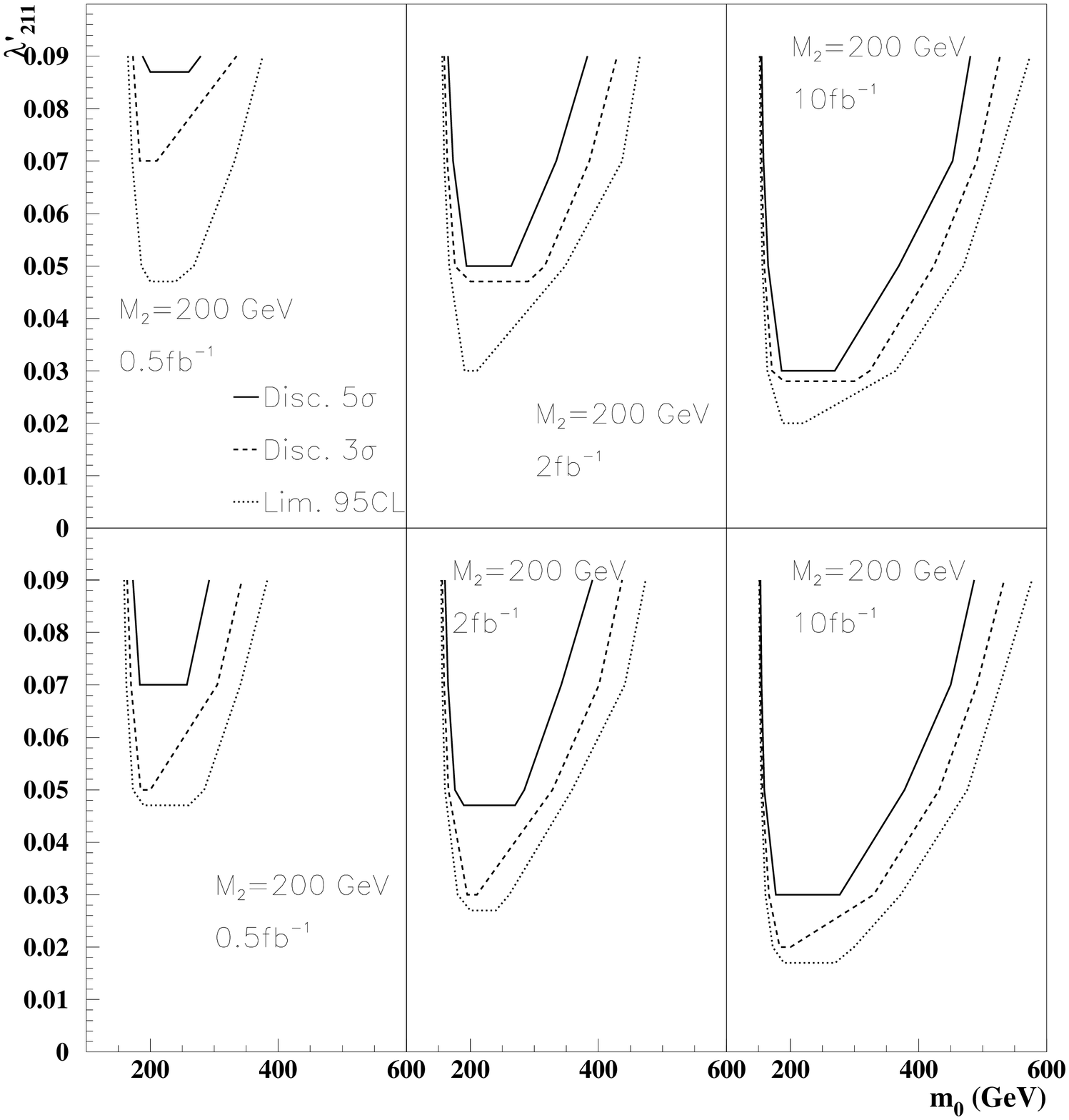,height=4.in}}
\end{center}
\caption{Discovery contours at $5 \sigma$ (full line), $ 3 \sigma$
(dashed line) and limit at $95 \% \ C.L.$ (dotted line) presented
in the plane $\l'_{211}$ versus the $m_0$ parameter,
for $sign(\mu)<0$, $\tan \beta=50$, $M_2=200GeV$
and different values of luminosity. The upper (lower) curves
are obtained without (with) the reconstruction of the tau-jets.}
\label{fig1tanb}
\end{figure}

In mSUGRA, for large values of $\tan \beta$ and small values
of $m_0$ compared to $m_{1/2}$,
due to the large mixing in the third generation sfermions,
the mass of the lighter $\tilde \tau_1$ slepton can become
smaller than $m_{\tilde \chi^{\pm}_1}$,
with the sneutrino remaining heavier than the $\tilde \chi^{\pm}_1$
so that the $\tilde \chi^{\pm}_1 l^{\mp}$ production rate can still
be significant. In this situation, the efficiency for the
3 lepton signature arising mainly through,
$\tilde \chi^{\pm}_1 \to \tilde \tau_1^{\pm} \nu_{\tau},
\ \tilde \tau_1^{\pm} \to \tilde \chi^0_1 \tau^{\pm},
\ \tilde \chi^0_1 \to l^{\pm}_i u_j d_k$, can be enhanced
compared to the case where the 3 lepton signal comes from,
$\tilde \chi^{\pm}_1 \to \tilde \chi^0_1 l^{\pm} \nu,
\ \tilde \chi^0_1 \to l^{\pm}_i u_j d_k$.
Indeed, the branching ratio
$B(\tilde \chi^{\pm}_1 \to \tilde \tau_1^{\pm} \nu_{\tau})$
can reach $\sim 100 \%$,
$B(\tilde \tau_1^{\pm} \to \tilde \chi^0_1 \tau^{\pm}) \approx 100 \%$,
since the $\tilde \chi^0_1$ is the LSP,
$B(\tau \to l \nu_l \nu_{\tau})=35 \%$ ($l=e,\mu$) and
the $\tau$-jets can be reconstructed
at Tevatron Run II. However, in such a scenario
the increased number of tau leptons in the final state
leads to a softer charged lepton spectrum
which tends to reduce the efficiency after cuts.
Therefore, for relatively small values
of $m_0$ compared to $m_{1/2}$, the sensitivity obtained
in the high $\tan \beta$ scenario
is essentially unaffected with respect to the low $\tan \beta$
situation, unless $m_0$ is small enough to render
$m_{\tilde \tau_1}$ and $m_{\tilde \chi^0_1}$
almost degenerate. As a matter of fact,
in such a situation, the energy of the tau produced in the decay
$\tilde \tau_1^{\pm} \to \tilde \chi^0_1 \tau^{\pm}$ often
falls below the analysis cuts. Therefore,
this degeneracy results in a loss of signal efficiency after cuts,
at small values of $m_0$ compared to $m_{1/2}$,
and thus in a loss of sensitivity, with respect to the
low $\tan \beta$ situation.
This can be seen by comparing Fig.\ref{fig2}, Fig.\ref{fig1}
and Fig.\ref{fig2tanb}, Fig.\ref{fig1tanb}.
Indeed, the decrease of the sensitivity on
$m_{1/2}$ at low $m_0$ is stronger for high $\tan \beta$ (see
Fig.\ref{fig2tanb})
than for low $\tan \beta$ (see Fig.\ref{fig2}).
Similarly, the decrease of the sensitivity on
$\l'_{211}$ at low $m_0$ is stronger for high $\tan \beta$ (see
Fig.\ref{fig1tanb})
than for low $\tan \beta$ (see Fig.\ref{fig1}).

The effect on the discovery potential
of the single chargino production
rate increase at large $\tan \beta$ values
shown in Fig.\ref{XStan} is hidden by the large $\tan \beta$
scenario influences on the cascade decays described above.


In contrast with the low $\tan \beta$ scenario (see Section \ref{lp211}),
the sensitivity on the SUSY parameters depends in a significant way
on the reconstruction of the tau-jets in case where
$\tan \beta$ is large,
as can be seen in Fig.\ref{fig2tanb} and Fig.\ref{fig1tanb}.
The reason is the increased number of tau leptons among the final state
particles in a large $\tan \beta$ model. This is due to the
decrease of the lighter stau mass which tends to increase the
$B(\tilde \chi^{\pm}_1 \to \tilde \chi^0_1 \tau^{\pm} \nu_{\tau})$
branching ratio.

\subsubsection{Discovery potential for the
$\l'_{1jk}$ and $\l'_{3jk}$ coupling constants}
\label{lp311}

\begin{figure}[t]
\begin{center}
\leavevmode
\centerline{\psfig{figure=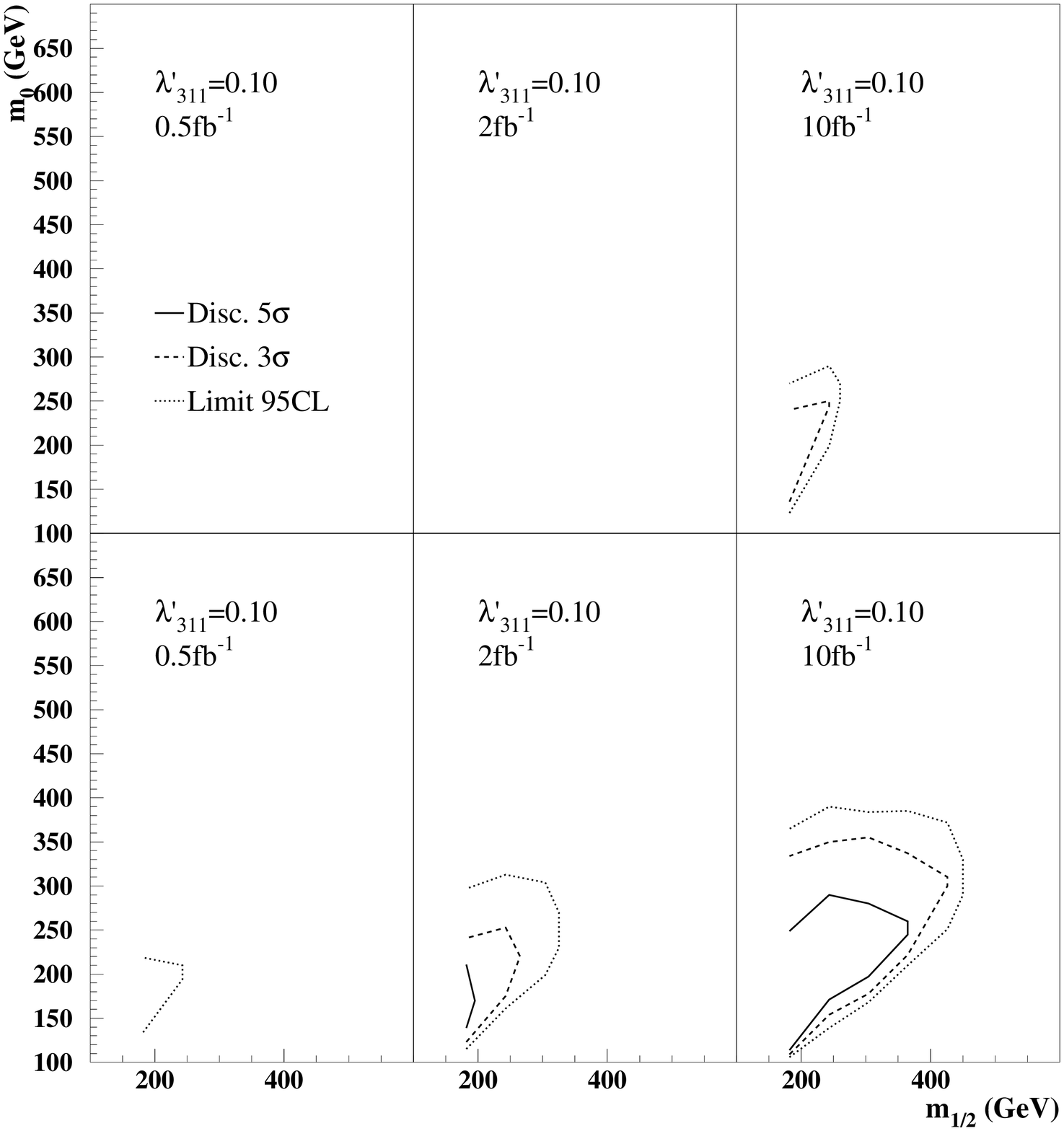,height=4.in}}
\end{center}
\caption{Discovery contours at $5 \sigma$ (full line),
$ 3 \sigma$ (dashed line)
and limit at $95 \% \ C.L.$ (dotted line) presented
in the plane $m_0$ versus $m_{1/2}$,
for $sign(\mu)<0$, $\tan \beta=1.5$, $\l'_{311}=0.10$
and different values of luminosity. The upper (lower) curves
are obtained without (with) the reconstruction of the tau-jets.}
\label{fig2tau}
\end{figure}

\begin{figure}[t]
\begin{center}
\leavevmode
 \centerline{\psfig{figure=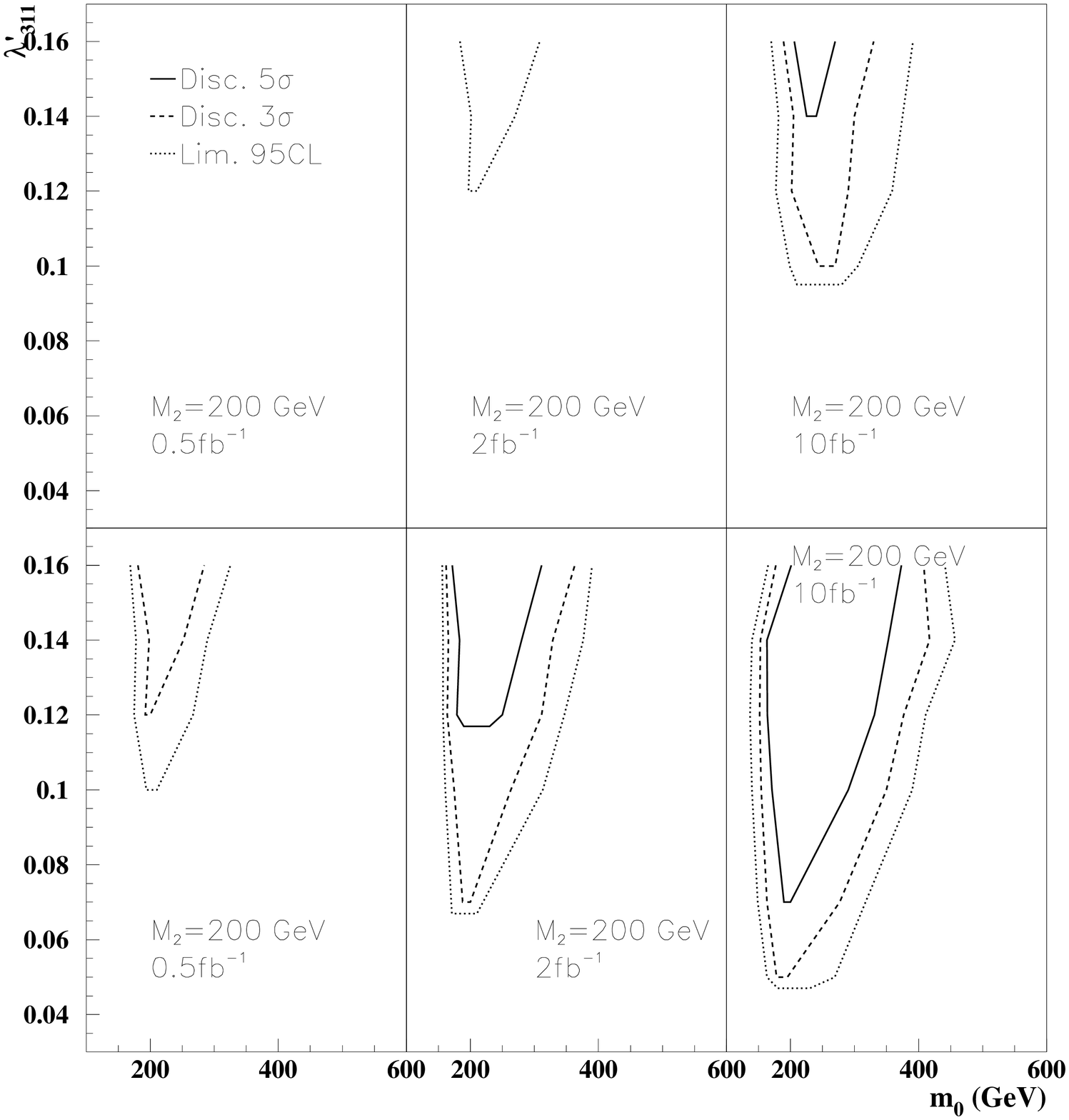,height=4.in}}
\end{center}
\caption{Discovery contours at $5 \sigma$ (full line), $ 3 \sigma$
(dashed line) and limit at $95 \% \ C.L.$ (dotted line) presented
in the plane $\l'_{311}$ versus the $m_0$ parameter,
for $sign(\mu)<0$, $\tan \beta=1.5$, $M_2=200GeV$
and different values of luminosity. The upper (lower) curves
are obtained without (with) the reconstruction of the tau-jets.}
\label{fig1tau}
\end{figure}

In Fig.\ref{fig2tau},
we present the $3 \sigma$ and $ 5 \sigma$ discovery
contours and the limits at $95 \%$
confidence level in the plane $m_0$ versus $m_{1/2}$,
for $sign(\mu)<0$, $\tan \beta=1.5$, $\l'_{311}=0.10$
and various values of the luminosity.
In Fig.\ref{fig1tau}, the discovery potential is shown
in the $\l'_{311}$-$m_0$ plane for $M_2=200GeV$.
Comparing Fig.\ref{fig2tau}, Fig.\ref{fig1tau} and
Fig.\ref{fig2}, Fig.\ref{fig1}, we see that
the sensitivity on the SUSY parameters is weaker
in the case of a single dominant $\l'_{311}$ coupling
than in the case of a single
dominant $\l'_{211}$ coupling. The reason is that in the case of
a single dominant $\l'_{3jk}$ coupling constant,
tau leptons are systematically produced
at the chargino production level
$p \bar p \to \tilde \chi^{\pm}_1 \tau^{\mp}$
(see Fig.\ref{graphes}(a)) as well as in the LSP decay
$\tilde \chi^0_1 \to \tau u_j d_k$ (see Section \ref{signal1}),
so that the number of tau leptons among the 3 charged leptons
of the final state is increased
compared to the dominant $\l'_{2jk}$ case.
The decrease in sensitivity is due to the fact that
a lepton (electron or muon) generated in a tau decay
has an higher probability not to pass the analysis requirements
concerning the particle energy
and that the reconstruction efficiency for a tau decaying into
a jet is limited.\\
Nevertheless, the discovery potentials of Fig.\ref{fig2tau}
and Fig.\ref{fig1tau}
represent also an important improvement with respect to
the experimental mass limits from LEP measurements \cite{Mass,neut}
and to the low-energy indirect constraint
$\l'_{311}<0.10 (m_{\tilde d_R}/100GeV)$ at $1 \sigma$
(from $\tau^- \to \pi^- \nu_{\tau}$) \cite{Bhatt}.\\
We also observe in Fig.\ref{fig2tau} and Fig.\ref{fig1tau} that
the results obtained from the $\tilde \chi^{\pm}_1 \tau^{\mp}$
production analysis in the case of a single dominant $\l'_{3jk}$
coupling depend strongly on the reconstruction of
the tau-jets. This is due to the large number of tau leptons
among the 3 charged leptons of the considered final state.

Using the same method and same SUSY point
as in Section \ref{lp211},
we have estimated the sensitivity on all the $\l'_{3jk}$ coupling
constants from the sensitivity obtained on $\l'_{311}$ at
$95 \% CL$ for a luminosity of ${\cal L}=2fb^{-1}$. The
results are given in Table \ref{couptau}.
\begin{table}[t]
\begin{center}
\begin{tabular}{|c|c|c|c|c|c|c|c|}
\hline
$\l'_{312}$ & $\l'_{313}$ & $\l'_{321}$ & $\l'_{322}$
& $\l'_{323}$ & $\l'_{331}$ & $\l'_{332}$ & $\l'_{333}$  \\
\hline
0.13 & 0.23 & 0.18 & 0.41 & 0.70 & 0.33 & 1.17 & 2.05 \\
\hline
\end{tabular}
\caption{\em
Sensitivities at $95 \% CL$ on the $\l'_{3jk}$
coupling constants derived from the sensitivity
on $\l'_{311}$ for a luminosity of ${\cal L}=2fb^{-1}$
and the following set of SUSY parameters,
$\tan\beta=1.5$,  $M_2=200GeV$, $\mu=-200GeV$ and
$m_0=180GeV$.}
\label{couptau}
\end{center}
\end{table}
All the sensitivities on the \rpv couplings
presented in Table \ref{couptau}, except those on $\l'_{32k}$,
are stronger than the present
indirect limits on the same \rpv couplings:
$\l'_{31k}<0.10 (m_{\tilde d_{kR}}/100GeV)$
at $1 \sigma$ ($\tau^- \to \pi^- \nu_{\tau}$),
$\l'_{32k}<0.20$ (for $m_{\tilde l}=m_{\tilde q}=100GeV$)
at $1 \sigma$ ($D^0-\bar D^0$ mix),
$\l'_{33k}<0.48 (m_{\tilde q}/100GeV)$ at $1 \sigma$ ($R_{\tau}$)
\cite{Bhatt}.\\
We mention that in the case of a single dominant $\l'_{3j3}$
coupling,
the neutralino decays as $\tilde \chi^0_1 \to \tau u_j b$
so that the b semileptonic decay could affect
a little the analysis efficiency.

We discuss now the sensitivities that could be obtained on a
single dominant
$\l'_{1jk}$ coupling constant via the analysis of the reaction
$p \bar p \to \tilde \chi^{\pm}_1 e^{\mp}$ (see
Fig.\ref{graphes}(a)). Since the cross section of the
$\tilde \chi^{\pm}_1 e^{\mp}$ production through $\l'_{1jk}$
is equal to the rate of
the $\tilde \chi^{\pm}_1 \mu^{\mp}$ production via $\l'_{2jk}$,
for same $j$ and $k$ indices (see Section \ref{cross1}),
the sensitivity obtained on a $\l'_{1jk}$ coupling constant
is expected to be identical to the sensitivity on $\l'_{2jk}$.
If we assume that the sensitivities obtained on the $\l'_{1jk}$
couplings are equal to those presented in Table \ref{coup},
we remark that for the SUSY point chosen in this table
only the sensitivities expected for the $\l'_{112}$,
$\l'_{113}$, $\l'_{121}$, $\l'_{131}$ and $\l'_{132}$ couplings
are stronger than the corresponding low-energy bounds:
$\l'_{11k}<0.02 (m_{\tilde d_{kR}}/100GeV)$ at $2 \sigma$
(Charged current universality),
$\l'_{1j1}<0.035 (m_{\tilde q_{jL}}/100GeV)$ at $2 \sigma$
(Atomic parity violation),
$\l'_{132}<0.34$ at $1 \sigma$ for $m_{\tilde q}=100GeV$
($R_e$) \cite{Bhatt}.
The reason is that the low-energy constraints on the $\l'_{1jk}$
couplings are typically more stringent than the limits on the
$\l'_{2jk}$ couplings \cite{Bhatt}.

\subsubsection{Mass reconstructions}
\label{recons}

\begin{figure}[t]
\begin{center}
\leavevmode
\centerline{\psfig{figure=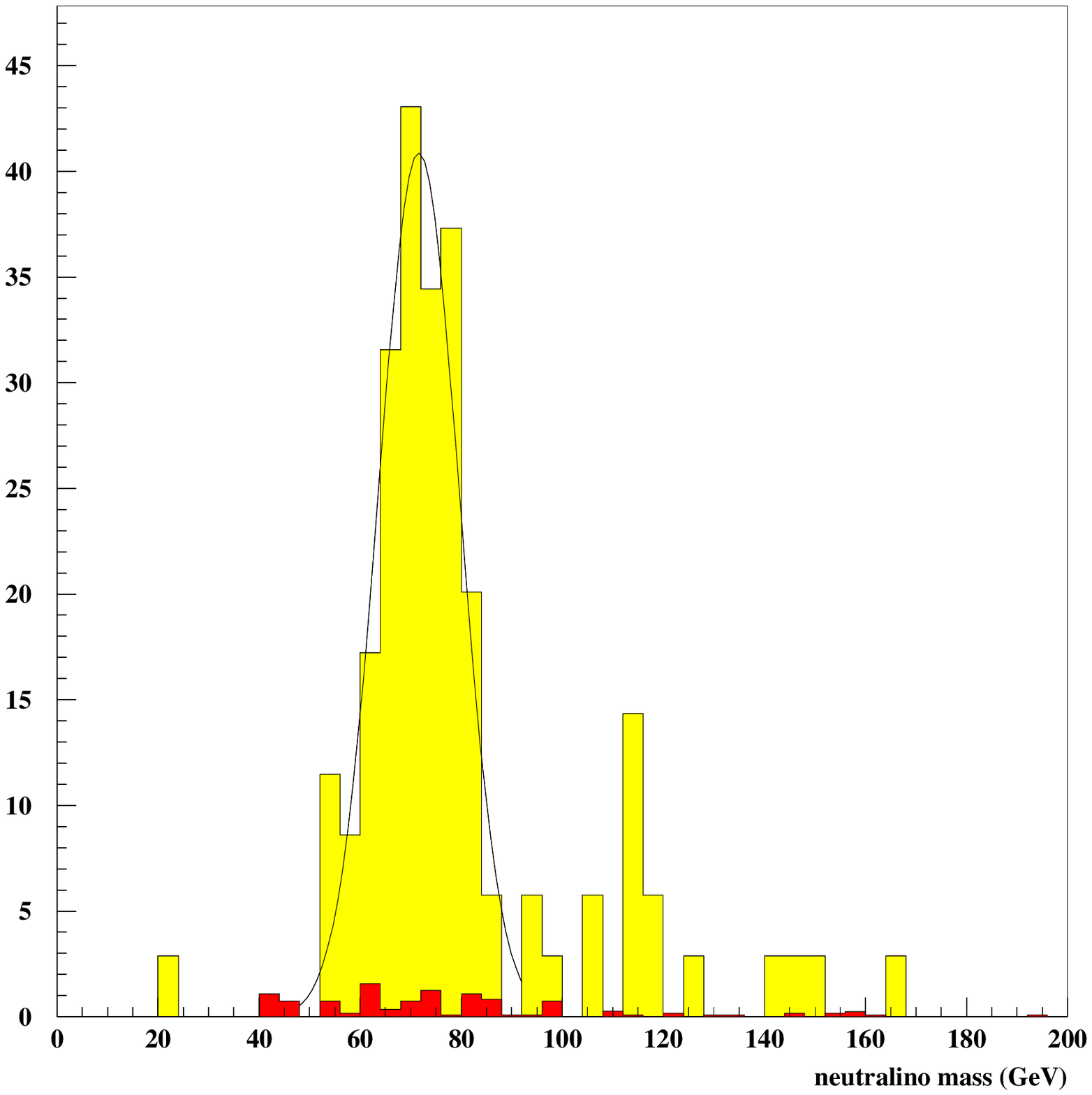,height=4.in}}
\end{center}
\caption{Distribution of the $softer \ \mu + 2j$
invariant mass in the $e + \mu + \mu + 2j + \nu$ events,
for a luminosity of ${\cal L}=10fb^{-1}$. The sum of the
$WZ$, $ZZ$ and $t \bar t$ backgrounds is in black and the
SUSY signal is in grey. The mSUGRA point taken for this figure is
$m_0=200GeV$, $M_2=150GeV$, $\tan \beta=1.5$ and $sign(\mu)<0$
($m_{\tilde \chi^0_1}=77.7GeV$)
and the considered \rpv coupling is $\l'_{211}=0.09$.}
\label{rec3n}
\end{figure}

The $\tilde \chi^0_1$ neutralino decays in our
framework as $\tilde \chi^0_1 \to l_i u_j d_k$ through the
$\l'_{ijk}$ coupling constant.
The invariant mass distribution of the lepton and the 2 jets
coming from this decay channel is peaked at the
$\tilde \chi^0_1$ mass. The experimental analysis of this
invariant mass distribution would thus be particularly interesting
since it would allow a model independent determination of the
lightest neutralino mass.

We have performed the $\tilde \chi^0_1$ mass reconstruction
based on the 3 lepton signature analysis.
The difficulty of this mass reconstruction
lies in the selection of the lepton and the 2 jets coming from the
$\tilde \chi^0_1$ decay.
In the signal we are considering, the only jets come from the
$\tilde \chi^0_1$ decay, and of course
from the initial and final QCD
radiations. Therefore, if there are more than 2 jets
in the final state we have selected the 2 hardest ones.
It is more subtle for the selection of the lepton since
our signal contains 3 leptons.
We have considered the case of a single dominant
coupling of type $\l'_{2jk}$ and focused on the
$e \mu \mu$ final state.
In these events, one of the $\mu^{\pm}$ is generated
in the decay of the produced sneutrino as
$\tilde \nu_{\mu} \to \tilde \chi^{\pm}_1 \mu^{\mp}$
and the other one in the decay of the
$\tilde \chi^0_1$ as $\tilde \chi^0_1 \to \mu^{\pm} u_j d_k$,
while the electron comes from the chargino decay
$\tilde \chi^{\pm}_1 \to \tilde  \chi^0_1 e^{\pm} \nu_e$.
Indeed, the dominant contribution to the single
chargino production is the resonant sneutrino production
(see Fig.\ref{graphes}).
In order to select the muon from the $\tilde \chi^0_1$ decay
we have chosen the
softer muon, since for relatively important values of the
$m_{\tilde \nu_{\mu}}-m_{\tilde \chi^{\pm}_1}$ mass
difference the muon
generated in the sneutrino decay is the most energetic.
Notice that for too degenerate $\tilde \nu_{\mu}$ and
$\tilde \chi^{\pm}_1$ masses, the sensitivity
on the SUSY parameters
suffers a strong decrease as shown in Section \ref{lp211}.

We present in Fig.\ref{rec3n}
the invariant mass distribution of the muon and the 2 jets
produced in the $\tilde \chi^0_1$ decay.
This distribution has been obtained by using the selection
criteria described above and by considering the mSUGRA point:
$m_0=200GeV$, $M_2=150GeV$, $\tan \beta=1.5$, $sign(\mu)<0$
and $\l'_{211}=0.09$ ($m_{\tilde \chi^0_1} = 77.7 GeV$,
$m_{\tilde \chi^{\pm}_1} = 158.3 GeV$,
$m_{\tilde \nu_L} = 236 GeV$).
We also show on the plot of Fig.\ref{rec3n}
the fit of the invariant mass distribution.
As can be seen from this fit, the distribution is well
peaked around the $\tilde \chi^0_1$ generated mass.
The average reconstructed $\tilde \chi^0_1$ mass is of
$71 \pm 9GeV$.

\begin{figure}[t]
\begin{center}
\leavevmode
\centerline{\psfig{figure=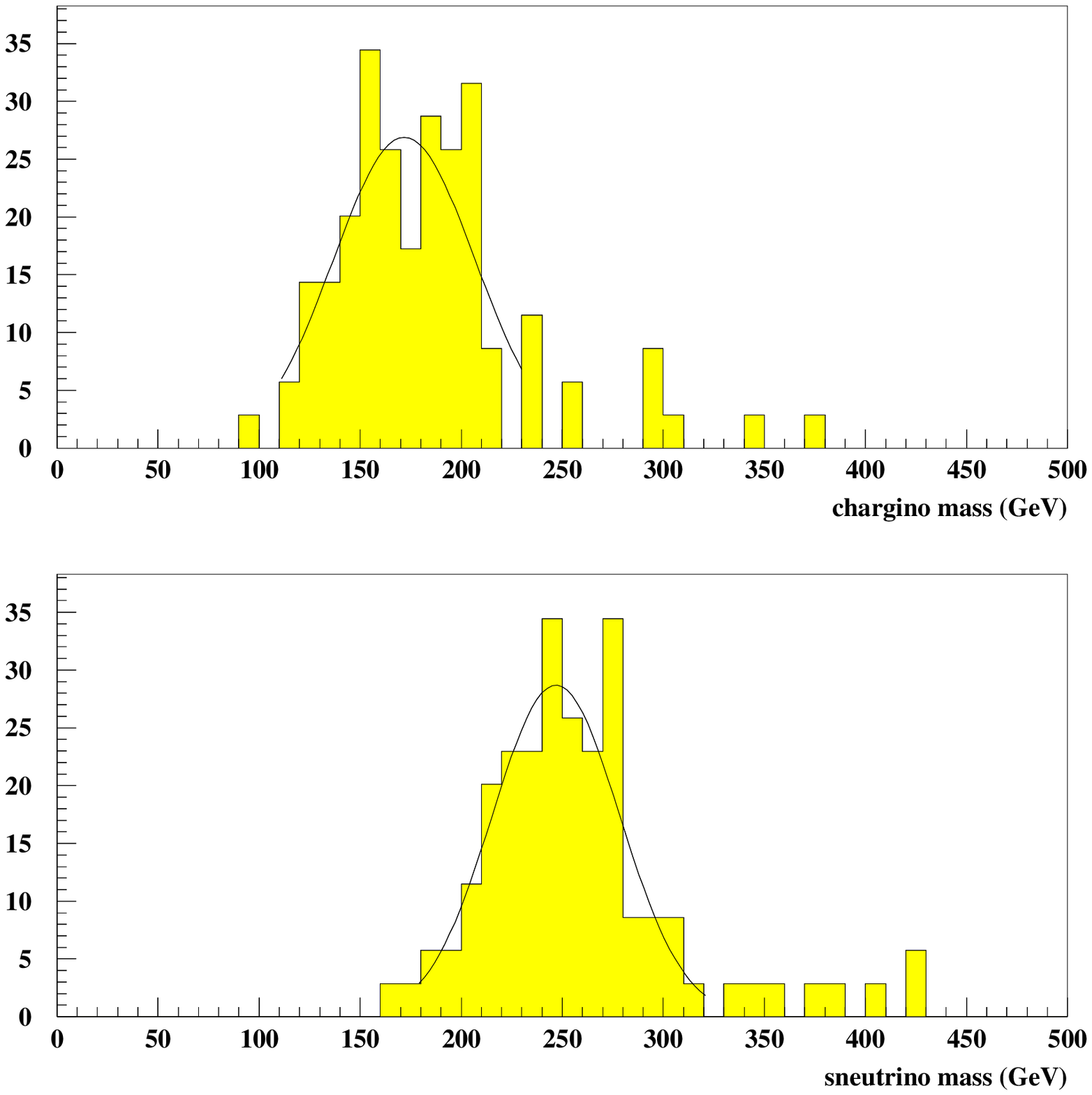,height=4.in}}
\end{center}
\caption{Distributions of the $e + softer \ \mu + 2j +\nu$
(upper plot) and $e + \mu + \mu + 2j + \nu$ (lower plot)
invariant masses in the $e + \mu + \mu + 2j + \nu$ events,
for a luminosity of ${\cal L}=10fb^{-1}$.
The mSUGRA point taken for these figures is
$m_0=200GeV$, $M_2=150GeV$, $\tan \beta=1.5$ and $sign(\mu)<0$
($m_{\tilde \chi^{\pm}_1}= 158.3 GeV$, $m_{\tilde \nu_{\mu L}}= 236 GeV$)
and the considered \rpv coupling is $\l'_{211}=0.09$.}
\label{rec3c}
\end{figure}

We have also performed the $\tilde \chi^{\pm}_1$ and
$\tilde \nu_{\mu}$ mass reconstructions
based on the 3 lepton signature analysis in the scenario
of a single dominant coupling of type $\l'_{2jk}$.
The $\tilde \chi^{\pm}_1$ and $\tilde \nu_{\mu}$ masses
reconstructions are based on the 4-momentum of the
neutrino present in the $3l+2j+\nu$ final state (see Section
\ref{signal1}). The transverse component
of this momentum can be deduced from the momentum of the
charged leptons and jets present in the final state. However, the
longitudinal component of the neutrino momentum remains unknown
due to the poor detection at small polar angle values.
Therefore, in this study we have assumed a vanishing
longitudinal component of the neutrino momentum.
Besides, we have focused on the $e \mu \mu$ events as in the
$\tilde \chi^0_1$ mass reconstruction study. In this context,
the cascade decay initiated by the produced
lightest chargino is
$\tilde \chi^{\pm}_1 \to \tilde \chi^0_1 e^{\pm} \nu_e$,
$\tilde \chi^0_1 \to \mu^{\pm} u_j d_k$. Therefore, the
$\tilde \chi^{\pm}_1$ has been reconstructed
from the softer muon, the 2 jets, the electron and
the neutrino present in the final state,
since the softer muon is mainly generated
in the $\tilde \chi^0_1$ decay as explained above.
The $\tilde \nu_{\mu}$ has then been reconstructed
from the $\tilde \chi^{\pm}_1$
and the leading muon of the final state.
This was motivated by the fact
that the dominant contribution to the single
chargino production is the reaction
$p \bar p \to \tilde \nu_{\mu} \to \tilde \chi^{\pm}_1 \mu^{\mp}$
(see Fig.\ref{graphes}).

In Fig.\ref{rec3c}, we present the $\tilde \chi^{\pm}_1$ and
$\tilde \nu_{\mu}$ mass reconstructions performed through the
method presented above. We also show on the plots of Fig.\ref{rec3c}
the fits of the invariant mass distributions.
As can be seen from those fits, the distributions are well
peaked around the $\tilde \chi^{\pm}_1$ and $\tilde \nu_{\mu L}$
generated masses. The average reconstructed masses are
$m_{\tilde \chi^{\pm}_1} = 171 \pm 35 GeV$ and
$m_{\tilde \nu_{\mu L}} = 246 \pm 32 GeV$.
This study on the $\tilde \chi^{\pm}_1$ and $\tilde \nu_{\mu L}$
masses shows that based on a simplified mass
reconstruction analysis promising results are obtained from
the 3 lepton signature generated by the single
$\tilde \chi^{\pm}_1$ production. The $\tilde \chi^{\pm}_1$ and
$\tilde \nu_{\mu L}$ mass reconstructions can be
improved using constrained fits.

In the hypothesis of a single dominant coupling constant
of type $\l'_{1jk}$, exactly the same kind of
$\tilde \chi^0_1$, $\tilde \chi^{\pm}_1$
and $\tilde \nu_{\mu}$ mass reconstructions can
be performed by selecting the $e + e + \mu + 2j + \nu$ events.
In contrast, the case of a single dominant
$\l'_{3jk}$ coupling requires more sophisticated methods.

As a conclusion, in the case of a single dominant
coupling constant of type $\l'_{1jk}$ or $\l'_{2jk}$,
the $\tilde \chi^0_1$, $\tilde \chi^{\pm}_1$
and $\tilde \nu_{\mu}$ mass reconstructions based on
the 3 lepton signature generated by the single
$\tilde \chi^{\pm}_1$ production at Tevatron can easily give
precise results, in contrast with the mass reconstructions
performed in the superpartner pair production analysis
at hadronic colliders which suffer a high
combinatorial background \cite{Atlas}.

\subsubsection{Model dependence of the results}

In this Section, we discuss qualitatively
the impact on our results of the choice of our theoretical model, namely
mSUGRA with the infrared fixed point hypothesis for the top quark
Yukawa coupling.
We focus on the discovery potentials obtained in Sections \ref{lp211},
\ref{htanb} and \ref{lp311}, since the choice of the theoretical framework
does not influence the study of the neutralino mass reconstruction made
in Section \ref{recons} which is model independent.

The main effect of the infrared fixed point approach is to fix
the value of the $\tan \beta$ parameter, up to the
ambiguity on the low or high solution.
Therefore, the infrared fixed point
hypothesis has no important effects on the results
since the dependences of the single gaugino productions rates
on $\tan \beta$ are smooth,
in the high $\tan \beta$ scenario (see Section \ref{cross1}).

As we have mentioned in Section \ref{theoretical},
in the mSUGRA scenario,
the $\vert \mu \vert$ parameter is fixed. This point does
not influence much our results since the single gaugino
production cross sections vary weakly with $\vert \mu \vert$
as shown in Section \ref{cross1}.

Another particularity of the mSUGRA model is that
the LSP is the $\tilde \chi^0_1$ in most of the parameter space.
For instance, in a model where the LSP would be
the lightest chargino or a squark,
the contribution to the three lepton signature from the
$\tilde \chi^{\pm}_1 l^{\mp}$ production would vanish.

Finally in mSUGRA,
the squark masses are typically larger than the lightest chargino mass
so that the decay $\tilde \chi^{\pm}_1 \to \tilde \chi^0_1 l^{\pm} \nu$
has a branching ratio of at least
$\sim 30\%$ (see Section \ref{signal1}).
In a scenario where $m_{\tilde \chi^{\pm}_1}>m_{\tilde q}$,
the two-body decay $\tilde \chi^{\pm}_1 \to \tilde q q$
would be dominant so that the contribution to the
three lepton signature from the
$\tilde \chi^{\pm}_1 l^{\mp}$ production would be small.
Besides, in mSUGRA, the
$\tilde \chi^{\pm}_1-\tilde \chi^0_1$ mass difference
is typically large enough to avoid
significant branching ratio for the
\rpv decay of the lightest chargino which would result in a decrease
of the sensitivities on the SUSY parameters
presented in Sections \ref{lp211}, \ref{htanb} and \ref{lp311}.

In a model where the contribution to the three lepton signature from the
$\tilde \chi^{\pm} l^{\mp}$ production would be suppressed,
the three lepton final state could be generated in a significant way
by other single gaugino productions, namely the $\tilde \chi^{\pm} \nu$,
$\tilde \chi^0 l^{\mp}$ or $\tilde \chi^0 \nu$ productions.

\section{Like sign dilepton signature analysis}
\label{analysis2}

\subsection{Signal}
\label{signal2}

Within the context of the mSUGRA model,
three of the single gaugino
productions via $\l'_{ijk}$
presented in Section \ref{resonant} can generate
a final state containing a pair of same sign leptons.
As a matter of fact,
the like sign dilepton signature can be produced
through the reactions
$p \bar p \to \tilde \chi^0_1 l^{\pm}_i$;
$p \bar p \to \tilde \chi^0_2 l^{\pm}_i$,
$\tilde \chi^0_2 \to \tilde \chi^0_1 + X$ ($X \neq l^{\pm}$);
$p \bar p \to \tilde \chi^{\pm}_1 l^{\mp}_i$,
$\tilde \chi^{\pm}_1 \to \tilde \chi^0_1 q \bar q$ and
$p \bar p \to \tilde \chi^{\pm}_1 \nu_i$,
$\tilde \chi^{\pm}_1 \to \tilde \chi^0_1 l^{\pm} \nu$,
$i$ corresponding to the flavour index of the
$\l'_{ijk}$ coupling.
Indeed, since the $\tilde \chi^0_1$ is a Majorana particle,
it decays via
$\l'_{ijk}$ into a lepton, as $\tilde \chi^0_1 \to l_i u_j \bar d_k$,
and into an anti-lepton, as $\tilde \chi^0_1 \to \bar l_i \bar u_j d_k$,
with the same probability.
The $\tilde \chi^0_{3,4} l^{\pm}_i$, $\tilde \chi^{\pm}_2 l^{\mp}_i$ and
$\tilde \chi^{\pm}_2 \nu_i$ productions
do not bring significant contributions to the
like sign dilepton signature due to their relatively small cross sections
(see Section \ref{cross1}).

In mSUGRA, the most important contribution to the
like sign dilepton signature originates from the $\tilde \chi^0_1 l^{\pm}_i$
production since this reaction has a dominant cross section
in most of the mSUGRA parameter space, as shown in
Section \ref{cross1}.
The other reason is that if $\tilde \chi^0_1$ is the LSP,
the $\tilde \chi^0_1 l^{\pm}_i$ production rate
is not affected by branching ratios of any cascade decay since the
$\tilde \chi^0_1$ only decays through \rpv coupling.

\subsection{Standard Model
background of the like sign dilepton signature at Tevatron}
\label{back2}

The $b \bar b$ production can lead to the like sign dilepton signature
if both of the b quarks decay semi-leptonically. The leading order
cross section
of the $\bar b b$ production at Tevatron for an energy of $\sqrt s=2TeV$ is
$\sigma (p \bar p \to b \bar b) \approx 4.654 \ 10^{10}fb$.
This rate has been calculated with PYTHIA \cite{PYTHIA}
using the CTEQ2L structure function.

The $t \bar t$ production, followed by the decays
$t \to W^+ b \to l^+ \nu b$,
$\bar t \to W^- \bar b \to \bar q q \bar b \to \bar q q l^+ \nu \bar c$,
or $t \to W^+ b \to \bar q q b \to \bar q q l^- \bar \nu c$,
$\bar t \to W^- \bar b \to l^- \bar \nu \bar b$,
also generates a final state with two same sign leptons.
The leading order cross section of the $t \bar t$ production at
$\sqrt s=2TeV$, including the relevant branching ratios,
is $\sigma (p \bar p \to t \bar t) \times 2
\times B(W \to l_p \nu_p) \times B(W \to q_p \bar q_{p'})
\approx 3181fb$ ($2800fb$) for $m_{top}=170GeV$ ($175GeV$) with
$p,p'=1,2,3$.

The third important source of \sm background is the
$t \bar b / \bar t b$ production
since the (anti-)$b$ quark can undergo a semi-leptonic decay as
$b \to l^- \bar \nu c$ ($\bar b \to l^+ \nu \bar c$)
and the (anti-) top quark can decay simultaneously as $t \to b W^+ \to b l^+
\nu$
($\bar t \to \bar b W^- \to \bar b l^- \bar \nu$).
The leading order cross section at
$\sqrt s=2TeV$ including the branching fraction is
$\sigma(p \bar p \to t q, \bar t q) \times B(W \to l_p \nu_p) \approx 802fb$
($687fb$) for $m_{top}=170GeV$ ($175GeV$) with $p=1,2,3$.

Other small sources of \sm background are
the $W^{\pm} W^{\mp}$ production,
followed by the decays: $W \to l \nu$ and $W \to b u_p$ ($p=1,2$)
or $W \to b u_p$ and $W \to b u_p$ ($p=1,2$),
the $W^{\pm} Z^0$ production,
followed by the decays: $W \to l \nu$ and $Z \to b \bar b$
or $W \to q_p \bar q_{p'}$ and $Z \to b \bar b$,
and the $Z^0 Z^0$ production,
followed by the decays: $Z \to l \bar l$ and $Z \to b \bar b$
or $Z \to q_p \bar q_p$ and $Z \to b \bar b$.

Finally, the 3 lepton final states generated by the $Z^0 Z^0$
and $W^{\pm} Z^0$ productions
(see Section \ref{back1}) can be mistaken
for like sign dilepton events
in case where one of the leptons is lost in the detection.
Non-physics sources of background can also be caused by some fake leptons
or by the misidentification of the charge of a lepton.

Therefore for the study of the \sm background associated to
the like sign dilepton signal at Tevatron Run II,
we consider the $b \bar b$, the $t \bar t$, the
$W^{\pm} W^{\mp}$ and the single top
production and both the physics and non-physics
contributions generated by the $W^{\pm} Z^0$
and $Z^0 Z^0$ productions.

\subsection{Supersymmetric background
of the like sign dilepton signature at Tevatron}
\label{susyback2}

All the pair productions of superpartners are a source of SUSY background
for the like sign dilepton signature originating from the
single gaugino productions. Indeed, both of the produced superpartners
initiate a cascade decay ended by the \rpv decay of the LSP
through $\l'_{ijk}$, and if the two LSP's
undergo the same decay $\tilde \chi^0_1 \to l_i u_j \bar d_k$ or
$\tilde \chi^0_1 \to \bar l_i \bar u_j d_k$,
two same sign charged leptons are generated. Another possible way
for the SUSY pair production to generate the like sign dilepton signature
is that only one of the LSP's decays into a charged lepton of a given sign,
the other decaying as $\tilde \chi^0_1 \to \nu_i d_j d_k$, and
a second charged lepton of the same sign is produced in the cascade decays.

The cross sections of the superpartners pair productions
have been studied in Section \ref{susyback1}.

\subsection{Cuts}
\label{cut2}

In order to simulate the single chargino productions
$p \bar p \to \tilde \chi^{\pm}_1 l^{\mp}$,
$p \bar p \to \tilde \chi^{\pm}_1 \nu$
and the single neutralino production
$p \bar p \to \tilde \chi^0_1 l^{\mp}$ at Tevatron,
the matrix elements (see Appendix \ref{formulas}) of these
processes have been implemented
in a version of the SUSYGEN event generator \cite{SUSYGEN3}
allowing the generation of $p \bar p$ reactions \cite{priv}.
The \sm background
($W^{\pm} W^{\mp}$, $W^{\pm} Z^0$, $Z^0 Z^0$,
$t \bar b / \bar t b$, $t \bar t$ and $b \bar b$ productions)
has been simulated
using the PYTHIA event generator \cite{PYTHIA}
and the SUSY background (all SUSY particles pair productions) using the
HERWIG event generator \cite{HERWIG}.
SUSYGEN, PYTHIA and HERWIG have been interfaced with
the SHW detector simulation package \cite{SHW}
(see Section \ref{cut1}).

Several selection criteria have been applied
in order to reduce the background.\\
First, we have selected the events containing two
same sign muons. The reason is that in the like sign dilepton
signature analysis we have focused on the
case of a single dominant \rpv coupling constant
of the type $\l'_{2jk}$. In such a scenario, the two same charge
leptons generated in the $\tilde \chi^0_1 l^{\mp}$ production,
which represents the main contribution to the like sign
dilepton final state (see Section \ref{signal2}), are muons
(see Fig.\ref{graphes} and Section \ref{signal2}).
This requirement that the 2 like sign leptons
have the same flavour
allows to reduce the \sm background with respect to the
signal.

We require a number of jets greater or equal to two with a
transverse momentum
higher than $10 GeV$, namely $N_j \geq 2$ with $P_t(j) > 10 GeV$.
This jet veto reduces the non-physics backgrounds
generated by the $W^{\pm} Z^0$ and $Z^0 Z^0$
productions (see Section \ref{back2}) which produce
at most one hard jet (see Section \ref{cut1}).

\begin{figure}[t]
\begin{center}
\leavevmode
\centerline{\psfig{figure=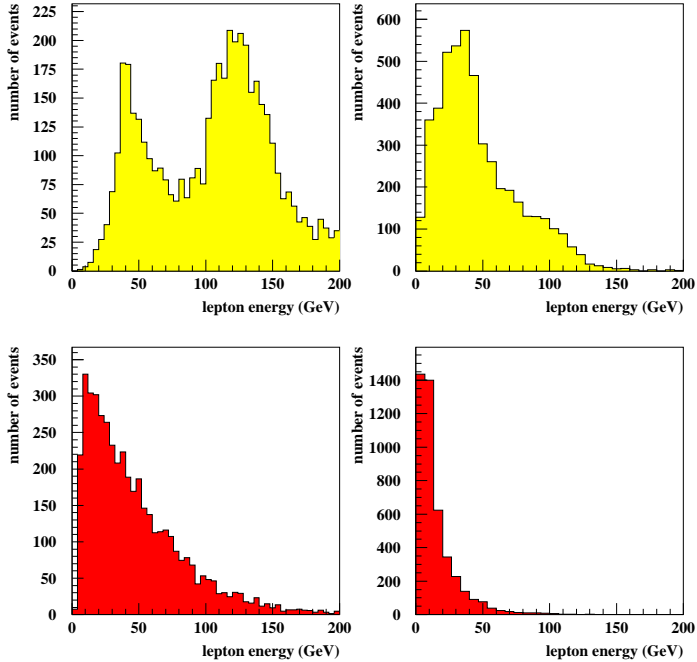,height=4.in}}
\end{center}
\caption{\footnotesize  \it
Distributions of the 2 muon energies (in $GeV$)
in the events containing 2 same sign muons and at least 2 jets
generated by the \sm background (lower curve),
namely the $W^{\pm} W^{\mp}$,
$W^{\pm} Z^0$, $Z^0 Z^0$, $t \bar t$, $t \bar b / \bar t b$
and $b \bar b$ productions,
and the SUSY signal (upper curve), for $\l'_{211}=0.05$,
$M_2=250GeV$, $m_0=200GeV$, $\tan \beta=1.5$
and $sign(\mu)<0$.
The left plots represent the leading muon distributions and
the right plots the second leading muon distributions.
The numbers of events correspond to an integrated luminosity of
${\cal L}=10fb^{-1}$.
\rm \normalsize }
\label{dienmu}
\end{figure}

Besides, some effective cuts concerning the energies of the 2 selected muons
have been applied. In Fig.\ref{dienmu},
we present the distributions of the 2 muon energies
in the like sign dimuon events generated by the \sm background
($W^{\pm} W^{\mp}$, $W^{\pm} Z^0$, $Z^0 Z^0$, $t \bar t$, $t \bar b / \bar t
b$
and $b \bar b$) and the SUSY signal.
Based on these distributions, we have chosen the following cuts
on the muon energies: $E(\mu_2)>20GeV$ and $E(\mu_1)>20GeV$.

We will refer to all the selection criteria described above, namely
2 same sign muons with $E(\mu_2)>20GeV$ and $E(\mu_1)>20GeV$,
and $N_j \geq 2$ with $P_t(j) > 10 GeV$, as cut $1$.

Let us explain the origin of the two peaks in the upper left
plot of Fig.\ref{dienmu}. This will be helpful for the mass
reconstruction study of Section \ref{reconsp}.\\
The main contribution to the like sign dimuon signature
from the SUSY signal is
the $\tilde \chi^0_1 \mu^{\pm}$ production (see Section
\ref{signal2}) in the case of a single dominant $\l'_{2jk}$
coupling. Furthermore, the dominant contribution to this
production is the reaction $p \bar p \to \tilde \mu^{\pm}_L \to
\tilde \chi^0_1 \mu^{\pm}$. In this reaction, the $\mu^{\pm}$
produced together with the $\tilde \chi^0_1$ has an
energy around $E(\mu^{\pm}) \approx (m_{\tilde \mu^{\pm}_L}^2
+m_{\mu^{\pm}}^2-m_{\tilde \chi^0_1}^2)/2 m_{\tilde \mu^{\pm}_L}
=121.9GeV$ for the SUSY point considered in Fig.\ref{dienmu},
namely $M_2=250GeV$, $m_0=200GeV$, $\tan \beta=1.5$
and $sign(\mu)<0$, which gives rise to the mass spectrum:
$m_{\tilde \chi^0_1}=127.1GeV$, $m_{\tilde \chi^0_2}=255.3GeV$,
$m_{\tilde \chi^{\pm}_1}=255.3GeV$, $m_{\tilde l^{\pm}_L}=298GeV$
and $m_{\tilde \nu^{\pm}_L}=294GeV$.
This energy value corresponds approximatively
to the mean value of the right peak of
the leading muon energy distribution presented in
the upper left plot of Fig.\ref{dienmu}. This is due to the
fact that the leading muon in the dimuon events generated by
the reaction $p \bar p \to \tilde \chi^0_1 \mu^{\pm}$ is the
$\mu^{\pm}$ produced together with the $\tilde \chi^0_1$
for relatively important values of the $m_{\tilde \mu^{\pm}_L}-
m_{\tilde \chi^0_1}$ mass difference. The right peak
in the upper left plot of Fig.\ref{dienmu} is thus associated
to the $\tilde \chi^0_1 \mu^{\pm}$ production. \\
Similarly,
the left peak in the upper left plot of Fig.\ref{dienmu}
corresponds to the reactions $p \bar p \to \tilde \mu^{\pm}_L
\to \tilde \chi^0_2 \mu^{\pm}$ and $p \bar p
\to \tilde \nu_{\mu L} \to \tilde \chi^{\pm}_1 \mu^{\mp}$
which produce $\mu^{\pm}$ of energies around $E(\mu^{\pm})
\approx (m_{\tilde \mu^{\pm}_L}^2+m_{\mu^{\pm}}^2
-m_{\tilde \chi^0_2}^2)/2 m_{\tilde \mu^{\pm}_L}=39.6GeV$ and
$E(\mu^{\pm}) \approx (m_{\tilde \nu_{\mu L}}^2
+m_{\mu^{\pm}}^2-m_{\tilde \chi^{\pm}_1}^2)/2
m_{\tilde \nu_{\mu L}}=36.2GeV$, respectively.
The $\tilde \chi^{\pm}_1 \nu_{\mu}$ production represents a
less important contribution to the like sign dimuon events
compared to the 3 above single gaugino productions
since the 2 same sign leptons generated
in this production are not systematically muons and the
involved branching ratios have smaller values
(see Section \ref{signal2}).

\begin{figure}[t]
\begin{center}
\leavevmode
\centerline{\psfig{figure=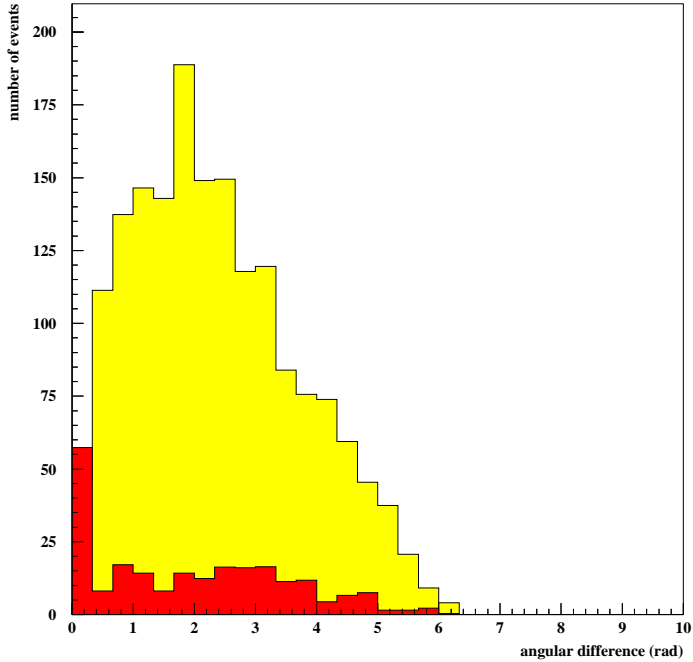,height=4.in}}
\end{center}
\caption{\footnotesize  \it
Distributions of the $\Delta R$ angular difference (in $rad$)
between the second leading muon and the second leading
jet in the like sign dimuons events selected by applying cut $1$ and
generated by the \sm background (curve in black),
namely the
$W^{\pm} W^{\mp}$, $W^{\pm} Z^0$, $Z^0 Z^0$, $t \bar t$, $t \bar b / \bar t
b$
and $b \bar b$ productions, and the SUSY signal
(curve in grey), for $\l'_{211}=0.05$,
$M_2=250GeV$, $m_0=200GeV$, $\tan \beta=1.5$ and $sign(\mu)<0$.
The numbers of events correspond to an integrated luminosity of
${\cal L}=10fb^{-1}$.
\rm \normalsize }
\label{dianmu}
\end{figure}

Finally, since the leptons produced in the quark $b$
decays are not well isolated (as in the $W^{\pm} W^{\mp}$,
$W^{\pm} Z^0$, $Z^0 Z^0$, $t \bar t$, $t \bar b / \bar t b$
and $b \bar b$ productions),
we have applied some cuts on the lepton isolation.
We have imposed the isolation cut
$\Delta R=\sqrt{\delta \phi^2+\delta \theta^2}>0.4$ where
$\phi$ is the azimuthal angle and $\theta$ the polar angle
between the 2 same sign muons and the 2 hardest jets.
This cut is for example motivated
by the distributions shown in Fig.\ref{dianmu}
of the $\Delta R$ angular difference
between the second leading muon and the second leading
jet, in the like sign dimuons events generated by the SUSY signal
and \sm background.
We call cut $\Delta R>0.4$ together with 
cut $1$, cut $2$.

\begin{figure}[t]
\begin{center}
\leavevmode
\centerline{\psfig{figure=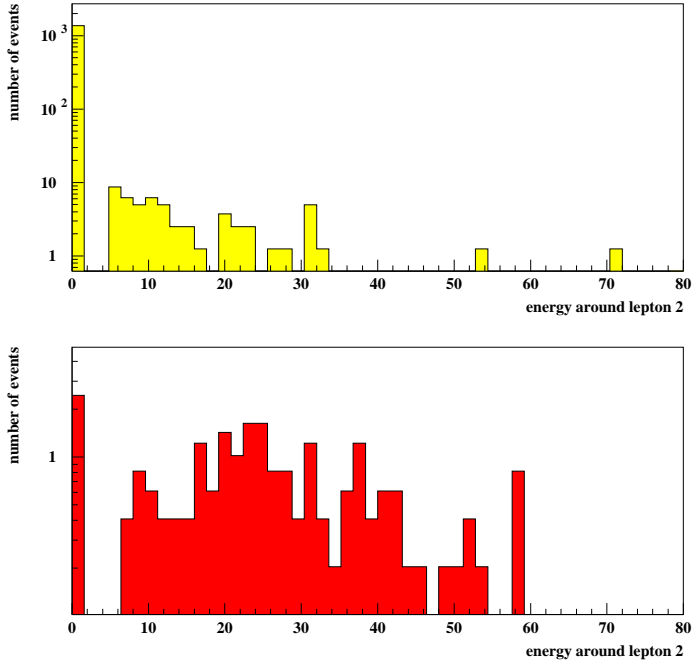,height=4.in}}
\end{center}
\caption{\footnotesize  \it
Distributions of the summed energies ($E$, in $GeV$) of the jets being
close to the second leading muon, namely the jets contained
in the cone centered on the second leading muon
and defined by $\Delta R<0.25$, in the like sign dimuons events
selected by applying cut $2$ and
generated by the \sm background (lower curve), namely the
$W^{\pm} W^{\mp}$, $W^{\pm} Z^0$, $Z^0 Z^0$, $t \bar t$, $t \bar b / \bar t
b$
and $b \bar b$ productions,
and the SUSY signal (upper curve), for $\l'_{211}=0.05$,
$M_2=250GeV$, $m_0=200GeV$, $\tan \beta=1.5$ and $sign(\mu)<0$.
These distributions were obtained after cut $E<2GeV$, where $E$
represents the
summed energies of the jets being close to the leading muon, has been
applied
in these like sign dimuons events.
The numbers of events correspond to an integrated luminosity of
${\cal L}=10fb^{-1}$.
\rm \normalsize }
\label{ismu}
\end{figure}

In order to eliminate poorly isolated muons,
we have also imposed that $E<2GeV$, where $E$ represents the
summed energies of the jets being close to a muon,
namely the jets contained in the cone centered on a muon
and defined by $\Delta R<0.25$. This cut is for instance motivated by
the distributions shown in Fig.\ref{ismu} which represent
the summed energies $E$ of the jets being
close to the second leading muon in the like sign dimuons events
generated by the SUSY signal and \sm background.
We denote cut $E<2GeV$ plus cut $2$ as cut $3$.


\begin{figure}[t]
\begin{center}
\leavevmode
\centerline{\psfig{figure=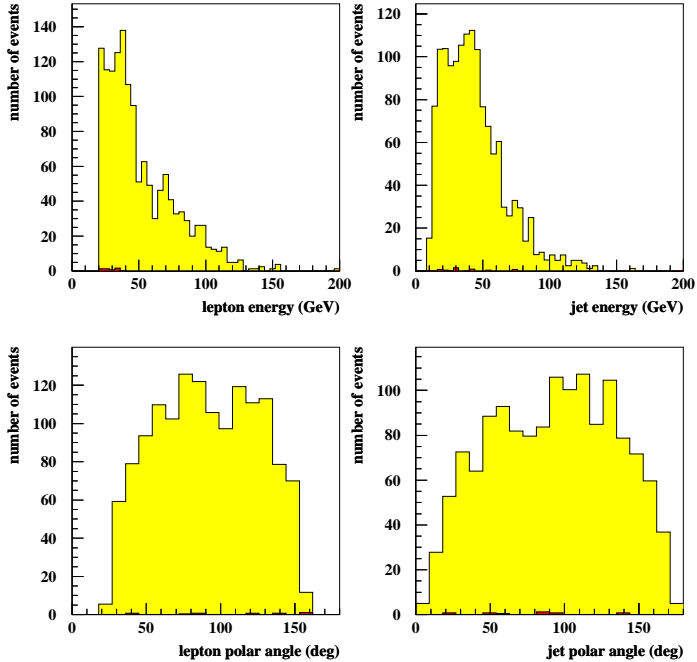,height=4.in}}
\end{center}
\caption{\footnotesize  \it
Energy (in $GeV$) and polar angle ($\theta$, in $deg$) distributions
of the leading muon and the leading jet in the like sign dimuon events
selected by applying cut $3$ and
generated by the \sm background (curve in black), namely the
$W^{\pm} W^{\mp}$, $W^{\pm} Z^0$, $Z^0 Z^0$, $t \bar t$, $t \bar b / \bar t
b$
and $b \bar b$ productions,
and the SUSY signal (curve in grey), for $\l'_{211}=0.05$,
$M_2=250GeV$, $m_0=200GeV$, $\tan \beta=1.5$ and $sign(\mu)<0$.
The numbers of events correspond to an integrated luminosity of
${\cal L}=10fb^{-1}$.
\rm \normalsize }
\label{aprmu}
\end{figure}

The selected events require high energy charged leptons and jets
and can thus be easily triggered at Tevatron. Moreover, the
considered charged leptons and jets are typically emitted at
intermediate polar angles and would thus be often detected at Tevatron.
These points are illustrated in
Fig.\ref{aprmu} where are shown the energy and polar angle distributions
of the leading muon and the leading jet in the like sign dimuons events
selected
by applying cut $3$ and generated by the SUSY signal and \sm background.

\begin{table}[t]
\begin{center}
\begin{tabular}{|c|c|c|c|c|c|}
\hline
        & $W^{\pm} Z^0$  & $Z^0 Z^0$      & $t \bar t$       & $t \bar b / \bar t b$   & Total       \\
\hline
cut $1$ & $0.21\pm 0.06$ & $0.11\pm 0.04$ & $21.80 \pm 0.70$ & $0.69 \pm 0.13 $        & $22.81 \pm 0.71$  \\
\hline
cut $2$ & $0.05\pm 0.03$ & $0.03\pm 0.03$ & $8.80 \pm 0.50 $ & $0.28 \pm 0.08 $        & $9.16  \pm 0.51$  \\
\hline
cut $3$ & $0.03\pm 0.03$ & $0.01\pm 0.02$ & $0.64 \pm 0.13 $ & $0.10 \pm 0.05 $        & $0.78  \pm 0.14$  \\
\hline
\end{tabular}
\caption{Numbers of like sign dilepton events generated by the
\sm background
($W^{\pm} W^{\mp}$,
$W^{\pm} Z^0$, $Z^0 Z^0$, $t \bar t$, $t \bar b / \bar t b$
and $b \bar b$ productions)
at Tevatron Run II for the cuts described in the text, assuming an
integrated
luminosity of ${\cal L}=1 fb^{-1}$ and a center of mass energy of $\sqrt s=2
TeV$. The numbers of events coming from the
$W^{\pm} W^{\mp}$ and $b \bar b$ backgrounds have been found 
to be negligible after cut 3 is applied.
These results have been obtained by generating
$2 \ 10^4$ events for the $W^{\pm} Z^0$ production,
$    10^4$ events for the $W^{\pm} Z^0$ (non-physics contribution),
$3 \ 10^4$ events for the $Z^0 Z^0$,
$    10^4$ events for the $Z^0 Z^0$ (non-physics contribution),
$3 \ 10^5$ events for the $t \bar t$ and
$    10^5$ events for the $t \bar b / \bar t b$.}
\label{cutefp}
\end{center}
\end{table}

In Table \ref{cutefp}, we give the numbers of like sign dilepton events
expected from the
\sm background at Tevatron Run II with the various cuts described above.
We see in Table \ref{cutefp} that the main source
of \sm background to the like sign dilepton signature at Tevatron
is the $t \bar t$ production. This is due to its important
cross section compared to the other \sm backgrounds
(see Section \ref{back2}) and to the fact that in the $t \bar t$
background, in contrast with the $b \bar b$ background,
only one charged lepton of the final state
is produced in a $b$-jet and is thus not isolated.

\begin{table}[t]
\begin{center}
\begin{tabular}{|c|c|c|c|c|c|}
\hline
$m_{1/2} \ \backslash \ m_0$
&  $100GeV$ & $200GeV$   & $300GeV$  & $400GeV$
&  $500GeV$   \\
\hline
$100GeV$
& $101.64$      & $54.92$        & $44.82$       & $39.26$
& $38.77$ \\
\hline
$200GeV$
& $3.74$        & $4.08$         & $4.33$        & $4.56$
& $4.99$ \\
\hline
$300GeV$
& $1.04$        & $0.63$         & $0.61$        & $0.70$
& $0.66$ \\
\hline
\end{tabular}
\caption{Number of like sign dilepton events generated by the
SUSY background (all superpartner pair productions)
at Tevatron Run II
as a function of the $m_0$ and $m_{1/2}$ parameters
for $\tan \beta=1.5$, $sign(\mu)<0$ and $\l'_{211}=0.05$.
Cut 3 (see text) has been applied.
These results have been obtained by generating 7500 events
and correspond to an integrated luminosity
of ${\cal L}=1 fb^{-1}$
and a center of mass energy of $\sqrt s=2 TeV$.}
\label{cutSUSp}
\end{center}
\end{table}

In Table \ref{cutSUSp}, we give the number of like sign dilepton
events generated by the
SUSY background (all superpartners pair productions)
at Tevatron Run II as a function of
the $m_0$ and $m_{1/2}$ parameters for cut 3.
This number of events decreases as $m_0$ and $m_{1/2}$
increase due to the behaviour of the summed superpartners
pair production cross section in the SUSY parameter space
(see Section \ref{susyback1}).

\subsection{Results}
\label{resp}

\subsubsection{Discovery potential}
\label{lp211p}

We first present the reach in the
mSUGRA parameter space obtained from the analysis of
the like sign dilepton final state at Tevatron Run II produced
by the single neutralino and chargino productions
via $\l'_{211}$:
$p \bar p \to \tilde \chi^0_{1,2} \mu^{\pm}$,
$p \bar p \to \tilde \chi^{\pm}_1 \mu^{\mp}$ and
$p \bar p \to \tilde \chi^{\pm}_1 \nu_{\mu}$.
The sensitivities that can be obtained on the $\l'_{2jk}$
($j$ and $k$ being not equal to $1$ simultaneously), $\l'_{1jk}$
and $\l'_{3jk}$ coupling constants
will be discussed at the end of this section.

\begin{figure}[t]
\begin{center}
\leavevmode
\centerline{
\psfig{figure=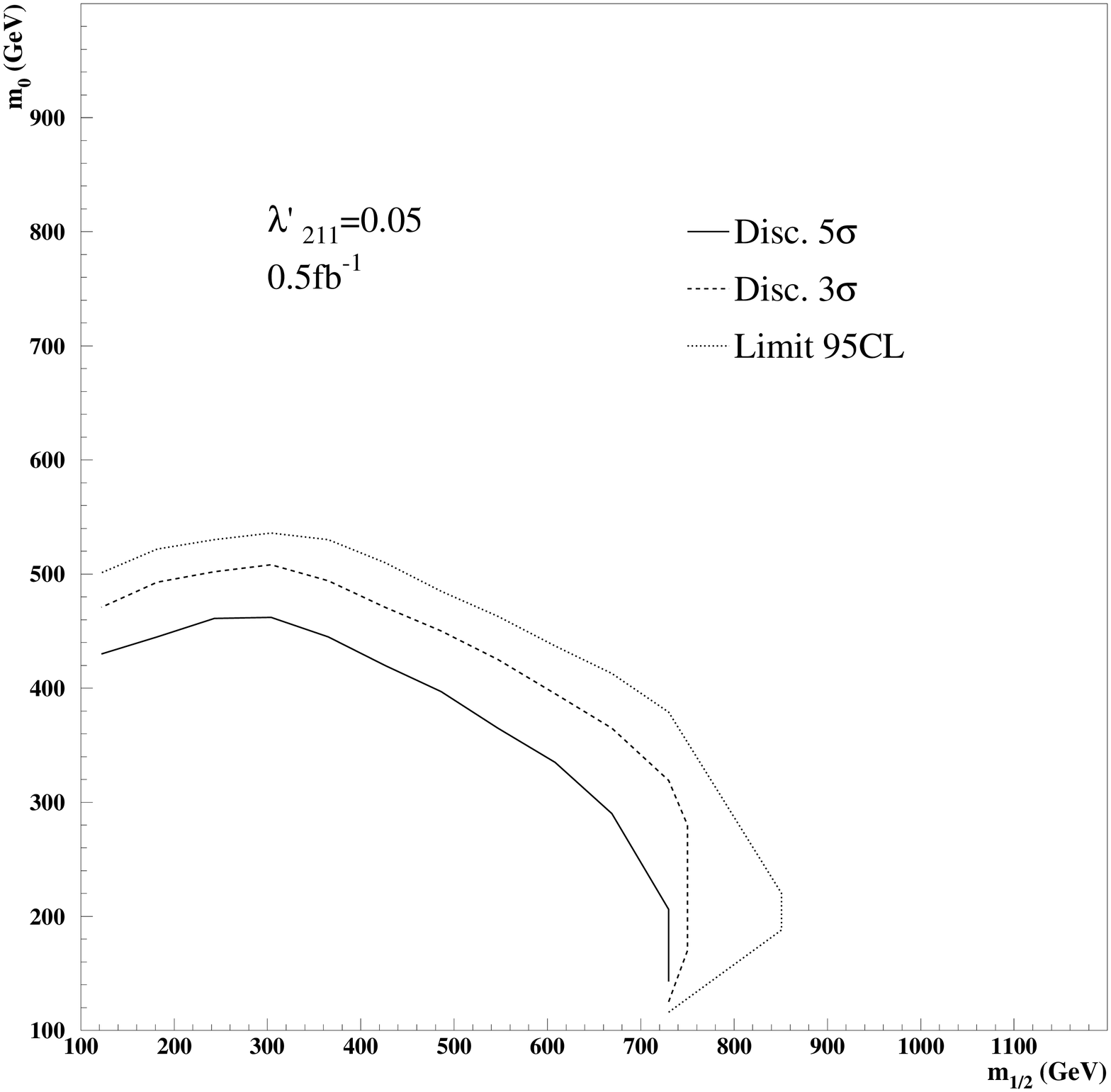,height=2.5in}
\psfig{figure=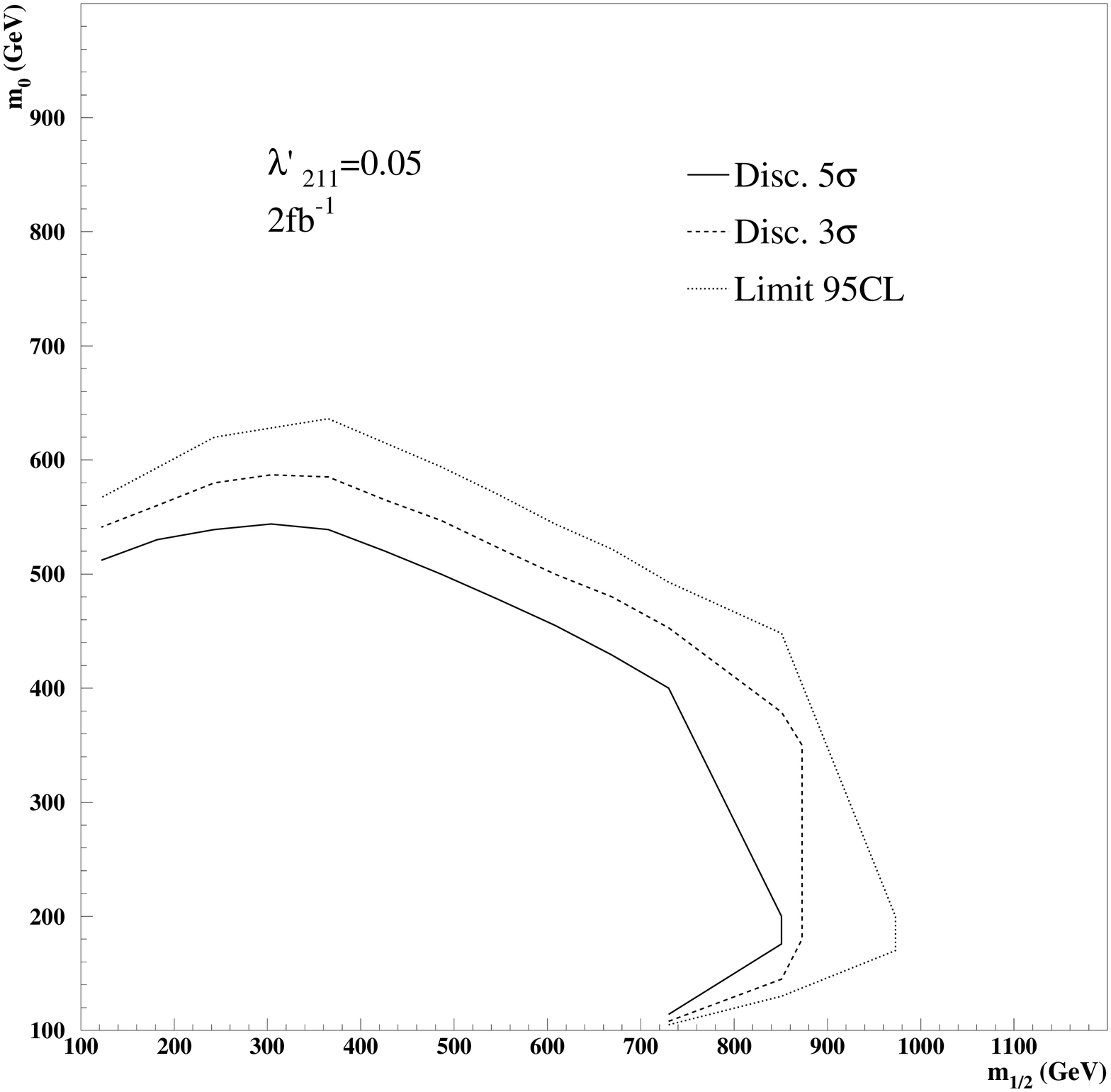,height=2.5in}}
\centerline{\psfig{figure=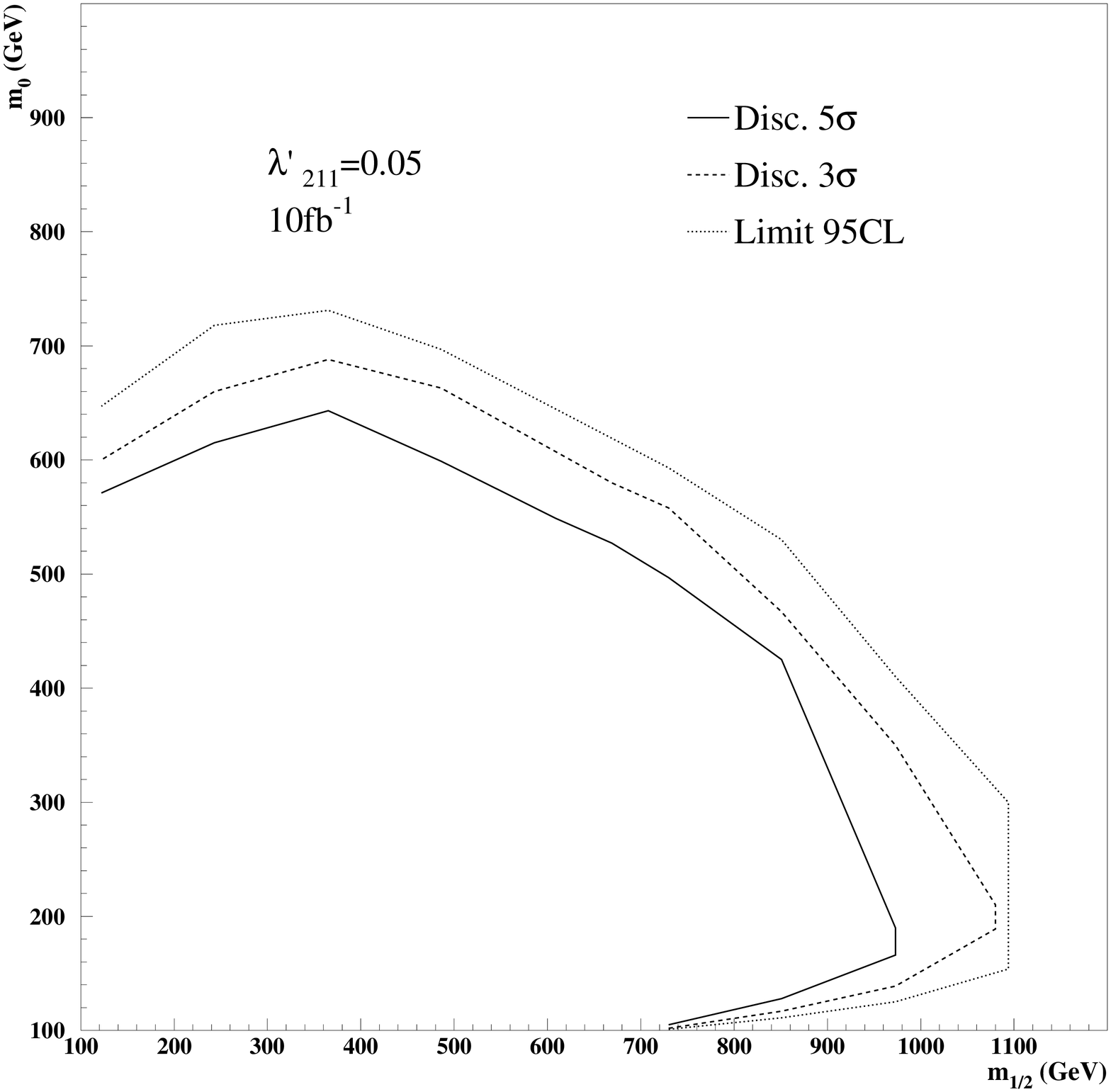,height=2.5in}}
\end{center}
\caption{Discovery contours at $5 \sigma$ (full line),
$ 3 \sigma$ (dashed line)
and limit at $95 \% \ C.L.$ (dotted line)
obtained from the like sign dilepton signature analysis
at Tevatron Run II assuming
a center of mass energy of $\sqrt s=2 TeV$.
These discovery potentials are
presented in the plane $m_0$ versus $m_{1/2}$,
for $sign(\mu)<0$, $\tan \beta=1.5$, $\l'_{211}=0.05$
and different values of luminosity.}
\label{fig2d}
\end{figure}

In Fig.\ref{fig2d},
we present the $3 \sigma$ and $ 5 \sigma$ discovery
contours and the limits at $95 \%$
confidence level in the plane $m_0$ versus $m_{1/2}$,
for $sign(\mu)<0$, $\tan \beta=1.5$, $\l'_{211}=0.05$ and
using a set of values for the luminosity.
Those discovery potentials were obtained by considering
the $\tilde \chi^0_{1,2} \mu^{\pm}$,
$\tilde \chi^{\pm}_1 \mu^{\mp}$ and
$\tilde \chi^{\pm}_1 \nu_{\mu}$ productions and
the background originating from the Standard Model.
The signal and background were selected by
using cut $3$ described in Section \ref{cut2}.
The reduction of the sensitivity on $m_{1/2}$
observed in Fig.\ref{fig2d} as $m_0$ increases is due to the
decrease of the $\tilde \chi^0_{1,2} \mu^{\pm}$,
$\tilde \chi^{\pm}_1 \mu^{\mp}$ and
$\tilde \chi^{\pm}_1 \nu_{\mu}$ productions cross sections
with the $m_0$ increase observed in Fig.\ref{XScl}
and Fig.\ref{allXS}. In Fig.\ref{fig2d}, we also see that
the sensitivity on $m_{1/2}$ is reduced
in the domain $m_0 \stackrel{<}{\sim} 200GeV$.
This reduction of the sensitivity is due to the fact that
in mSUGRA at low $\tan \beta$ and
for large values of $m_{1/2}$ and small values
of $m_0$, the LSP is the Right slepton $\tilde l^{\pm}_{iR}$
($i=1,2,3$). Therefore, in this mSUGRA region
the dominant decay channel
of the lightest neutralino is $\tilde \chi^0_1
\to \tilde l^{\pm}_{iR} l^{\mp}_i$ ($i=1,2,3$) so that the
$\tilde \chi^0_1 \mu^{\pm}$ production, which is the main
contribution to the like sign dilepton signature, leads to
the $2 \mu^{\pm}+2 \ jets$ final state only in a few cases.
There are two reasons. First, in this mSUGRA scenario
the charged lepton produced in the main
$\tilde \chi^0_1$ decay is not systematically a muon.
Secondly, if the LSP is the Right slepton $\tilde l^{\pm}_{iR}$
it cannot decay in the case of a single dominant
$\l'_{ijk}$ coupling constant and it is thus a stable particle.

The sensitivities presented in the discovery reach of
Fig.\ref{fig2d} which are obtained from the like sign dilepton
signature analysis are higher than the sensitivities shown
in Fig.\ref{fig2} which correspond to the trilepton final state
analysis. This is due to the 3 following points.
First, the rate of the $\tilde \chi^0_1 \mu^{\pm}$ production
(recall that it represents the main contribution to the like
sign dilepton final state)
is larger than the $\sigma(p \bar p \to \tilde \chi^{\pm}_1
\mu^{\mp})$ cross section in most of the mSUGRA parameter
space (see Section \ref{cross1}).
Secondly, the $\tilde \chi^0_1$ decay leading to the like sign dilepton
final state
in the case of the $\tilde \chi^0_1 \mu^{\pm}$ production
has a larger branching ratio than the cascade decay initiated by the
$\tilde \chi^{\pm}_1$ which generates the
trilepton final state (see Sections \ref{signal1} and \ref{signal2}).
Finally, at Tevatron Run II
the \sm background of the like sign dilepton
signature is weaker than the trilepton \sm background
(see Tables \ref{cutSUSY} and \ref{cutSUSp}).

It is clear from Fig.\ref{fig2d} that at low values of the $m_0$ and
$m_{1/2}$ parameters,
high sensitivities can be obtained on the $\l'_{211}$ coupling constant. We
have found that for
instance at the mSUGRA point defined as $m_0=200GeV$, $m_{1/2}=200GeV$,
$sign(\mu)<0$ and
$\tan \beta=1.5$, $\l'_{211}$ values of $\sim 0.03$ can be probed through the
like sign dilepton analysis at
Tevatron Run II assuming a luminosity of ${\cal L}= 1 fb^{-1}$. This result
was obtained by applying
cut $3$ described in Section \ref{cut2} on the SUSY signal ($\tilde
\chi^0_{1,2} \mu^{\pm}$,
$\tilde \chi^{\pm}_1 \mu^{\mp}$ and $\tilde \chi^{\pm}_1 \nu_{\mu}$
productions) and the Standard
Model background.

We expect that, as in the three lepton signature analysis,
interesting sensitivities could be obtained on other
$\l'_{2jk}$ coupling constants.\\
The sensitivities obtained on the $\l'_{3jk}$ couplings
from the like sign dilepton signature analysis should be
weaker than the sensitivities on the $\l'_{2jk}$ couplings
deduced from the same study. Indeed,
in the case of a single dominant $\l'_{3jk}$ coupling
the same sign leptons generated by the
$\tilde \chi^0_1 \tau^{\pm}$ production
would be 2 tau leptons (see Fig.\ref{graphes}(d)
and Section \ref{signal2}). Therefore,
the like sign dileptons ($e^{\pm} e^{\pm}$ or $\mu^{\pm}
\mu^{\pm}$) produced by the \rpv signal
would be mainly generated in tau decays and
would thus have higher probabilities to not pass the analysis cuts
on the particle energy. Moreover, the requirement of
$e^{\pm} e^{\pm}$ or $\mu^{\pm}
\mu^{\pm}$ events would decrease the efficiency after
cuts of the \rpv signal
due to the hadronic decay of the tau.
Finally, the selection of two same flavour like sign dileptons
($e^{\pm} e^{\pm}$ or $\mu^{\pm}
\mu^{\pm}$) would reduce the \rpv signal,
since each of the 2 produced taus could
decay either into an electron or a muon, and hence would not
be an effective cut anymore.\\
The sensitivities obtained on the $\l'_{1jk}$ couplings
from the like sign dilepton signature study
are expected to be identical to the sensitivities on the
$\l'_{2jk}$ couplings obtained from the same study.
Indeed, in the case of a single dominant $\l'_{1jk}$
coupling constant, the only difference in the
like sign dilepton signature analysis
would be that $e^{\pm} e^{\pm}$ events should be
selected instead of $\mu^{\pm} \mu^{\pm}$ events
(see Fig.\ref{graphes}(d) and Section \ref{signal2}).
Nevertheless, a smaller number of $\l'_{1jk}$ couplings
is expected to be probed since the low-energy constraints
on the $\l'_{1jk}$ couplings are generally
stronger than the limits on the
$\l'_{2jk}$ couplings \cite{Bhatt}.


In the high $\tan \beta$ case, the
lightest stau $\tilde \tau_1$ can become the LSP
instead of the lightest neutralino,
due to a large mixing in the third generation of charged
sleptons. In such a situation,
the dominant decay channel of the lightest neutralino
is $\tilde \chi^0_1 \to \tilde \tau^{\pm}_1 \tau^{\mp}$.
Two scenarios must then be discussed:
if the single dominant \rpv coupling is not of the type
$\l'_{3jk}$, the $\tilde \tau^{\pm}_1$-LSP is a stable particle
so that the reaction $p \bar p \to \tilde \chi^0_1 l^{\pm}_i$,
representing the main contribution to the like sign dilepton
final state, does not often lead to the
$2 \mu^{\pm}+2 \ jets$ signature.
If the single dominant \rpv coupling is of the type
$\l'_{3jk}$, the $\tilde \chi^0_1 \tau^{\pm}$ production
can receive a contribution from the resonant
$\tilde \tau^{\pm}_2$ production (see Fig.\ref{graphes}(d))
and the $\tilde \tau^{\pm}_1$-LSP decays via $\l'_{3jk}$ as
$\tilde \tau^{\pm}_1 \to u_j d_k$ so that the
$2 \mu^{\pm}+2 \ jets$ signature can still be generated
in a significant way by the $p \bar p \to \tilde \chi^0_1
\tau^{\pm}$ reaction.

We end this Section by some comments on the effect of
the \susyq $R_p$ conserving background to the like sign dilepton
signature. In order to illustrate this discussion, we consider
the results on the $\l'_{211}$ coupling constant.\\
We see from Table \ref{cutSUSp} that
the SUSY background to the like sign dilepton final
state can affect the sensitivity on the
$\l'_{211}$ coupling constant
obtained by considering only the \sm background,
which is shown in Fig.\ref{fig2d}, only in the
region of small superpartners masses, namely in the domain
$m_{1/2} \stackrel{<}{\sim} 300GeV$ for $\tan \beta=1.5$,
$sign(\mu)<0$ and assuming a luminosity of ${\cal L}=1fb^{-1}$.\\
In contrast with the SUSY signal amplitude which is increased
if $\l'_{211}$ is enhanced,
the SUSY background amplitude is typically
independent on the value of the $\l'_{211}$ coupling constant
since the superpartner pair production
does not involve \rpv couplings.
Therefore, even if we consider the SUSY background
in addition to the \sm one, it is still true that
large values of the $\l'_{211}$ coupling can
be probed over a wider domain of the SUSY parameter space than
low values, as can be observed in Fig.\ref{fig2d} for
$m_{1/2} \stackrel{>}{\sim} 300GeV$. Note that in Fig.\ref{fig2d}
larger values of $\l'_{211}$ still respecting the indirect limit
could have been considered.\\
Finally, we mention that further cuts,
as for instance some cuts based on the
superpartners mass reconstructions (see Section \ref{reconsp}),
could allow to reduce the SUSY background to the like sign
dilepton signature.

\subsubsection{Mass reconstructions}
\label{reconsp}

\begin{figure}[t]
\begin{center}
\leavevmode
\centerline{\psfig{figure=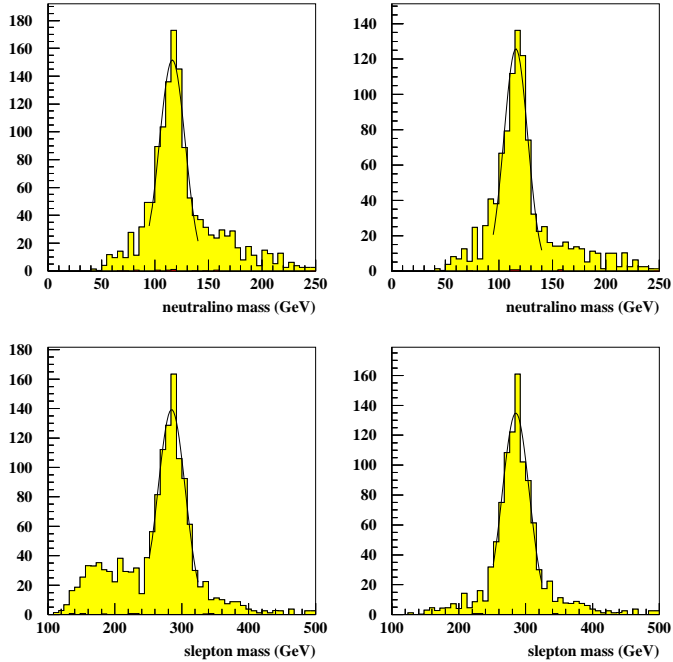,height=4.in}}
\end{center}
\caption{Distributions of the $softer \ \mu^{\pm}+2 \
leading \ jets$ (upper plots) and $\mu^{\pm}+\mu^{\pm}+2 \
leading \ jets$ (lower plots) invariant masses in the
$\mu^{\pm}+\mu^{\pm}+jets+\Eslash$ events generated
by the SUSY signal ($\tilde \chi^0_{1,2} \mu^{\pm}$,
$\tilde \chi^{\pm}_1 \mu^{\mp}$ and $\tilde \chi^{\pm}_1
\nu_{\mu}$ productions),
for a luminosity of ${\cal L}=10fb^{-1}$.
The 2 right plots are obtained by applying a cut in the
upper left plot of Fig.\ref{dienmu} selecting only
the peak associated to the $\tilde \chi^0_1 \mu^{\pm}$
production.
The mSUGRA point taken for this figure is,
$m_0=200GeV$, $M_2=250GeV$, $\tan \beta=1.5$ and $sign(\mu)<0$
($m_{\tilde \chi^0_1}=127.1GeV$, $m_{\tilde \mu^{\pm}_L}=298.0GeV$)
and the considered \rpv coupling is $\l'_{211}=0.05$.}
\label{rec2s}
\end{figure}

The $\tilde \chi^0_1$ and $\tilde l^{\pm}_L$
mass reconstructions can be performed in a model independent
way via the like sign dilepton analysis.
We have simulated these mass reconstructions based on the
like sign dimuon events generated in the scenario
of a single dominant $\l'_{2jk}$ coupling constant. In this
scenario, the main SUSY contribution to the
like sign dilepton signature, namely the $\tilde \chi^0_1
\mu^{\pm}$ production, has the final state
$\mu^{\pm}+\mu^{\pm}+2jets$ (see Section \ref{signal2}).
Indeed, the produced $\tilde \chi^0_1$ decays into
$\mu^{\pm} u_j d_k$ through $\l'_{2jk}$. The muon generated
together with the $\tilde \chi^0_1$ can be identified
as the leading muon for relatively large $m_{\tilde \mu^{\pm}_L}-
m_{\tilde \chi^0_1}$ mass differences (see Section \ref{cut2}).
Note that for nearly degenerate values of $m_{\tilde \mu^{\pm}_L}$
and $m_{\tilde \chi^0_1}$ the $\tilde \chi^0_1
\mu^{\pm}$ production rate and thus the sensitivity on
the SUSY parameters would be reduced (see Section \ref{cross1}).
The muon created in the $\tilde \chi^0_1$
decay can thus be identified as the softer muon so that
the $\tilde \chi^0_1$ can be reconstructed from the
the softer muon and the 2 jets present in the $\tilde \chi^0_1
\mu^{\pm}$ production final state.
The other contributions to the like sign dimuons events
can lead to some missing energy and
at most 4 jets in the final state
(see Section \ref{signal2}). Hence,
we have chosen to reconstruct the
$\tilde \chi^0_1$ from the 2 leading jets when the
final state contains more than 2 jets.
Once the $\tilde \chi^0_1$ has been reconstructed, the
$\tilde \mu^{\pm}_L$ has been reconstructed from the
$\tilde \chi^0_1$ and the leading muon since the dominant
contribution to the $\tilde \chi^0_1 \mu^{\pm}$ production is the
reaction $p \bar p \to \tilde \mu^{\pm}_L \to \tilde \chi^0_1
\mu^{\pm}$. These mass
reconstructions are represented in Fig.\ref{rec2s}.
In this figure, we also represent the same mass
reconstructions obtained by applying a cut in the
upper left plot of Fig.\ref{dienmu} excluding
the peak associated to the
$\tilde \chi^0_2 \mu^{\pm}$ and $\tilde \chi^{\pm}_1
\mu^{\mp}$ productions (see Section \ref{cut2}).
The interest of this cut, as can be seen in Fig.\ref{rec2s},
is to select the $\tilde \chi^0_1
\mu^{\pm}$ production and thus to improve the accuracy on the
$\tilde \chi^0_1$ and $\tilde \mu^{\pm}_L$ reconstructions
which are based on this production.
We observe in Fig.\ref{rec2s} that the $\tilde \chi^0_1$
reconstruction has less combinatorial background than the
$\tilde \mu^{\pm}_L$ reconstruction.
This comes from the fact that the selection of
the softer muon and the 2 leading jets
allows to reconstruct the $\tilde \chi^0_1$ even in the dimuon
events generated by the
$\tilde \chi^0_2 \mu^{\pm}$ and $\tilde \chi^{\pm}_1
\mu^{\mp}$ productions, while the selection of
the 2 muons and the 2 leading jets does not
allow to reconstruct the $\tilde \mu^{\pm}_L$
in the dimuon events generated by the
$\tilde \chi^0_2 \mu^{\pm}$ and $\tilde \chi^{\pm}_1
\mu^{\mp}$ productions (see Section \ref{signal2}).
We have represented on the plots of Fig.\ref{rec2s}
the fits of the invariant mass distributions.
We see from these fits that the distributions are well
peaked around the $\tilde \chi^0_1$ and $\tilde \mu^{\pm}_L$
generated masses. The average reconstructed masses are
$m_{\tilde \chi^0_1} = 116 \pm 11 GeV$ and
$m_{\tilde \mu^{\pm}_L} = 285 \pm 20 GeV$.

We note that the accuracy on the $\tilde \chi^0_1$
(and thus on the $\tilde \mu^{\pm}_L$)
mass reconstruction could be improved if the distributions
in the upper plots of
Fig.\ref{rec2s} were recalculated by selecting the muon giving
the $\tilde \chi^0_1$ mass the closer to the mean value of the
peak obtained in the relevant upper plot of Fig.\ref{rec2s}.

In the hypothesis of a single dominant coupling constant
of type $\l'_{1jk}$ or $\l'_{3jk}$, exactly the same kind of
$\tilde \chi^0_1$
and $\tilde \mu^{\pm}_L$ mass reconstructions can
be performed by selecting the $e^{\pm}+e^{\pm}+jets+\Eslash$
or $l_i^{\pm}+l_j^{\pm}+jets+\Eslash$ events, respectively.

As a conclusion, the $\tilde \chi^0_1$
and $\tilde \mu^{\pm}_L$ mass reconstructions based on
the like sign dilepton signature generated by the
$\tilde \chi^0_{1,2} \mu^{\pm}$,
$\tilde \chi^{\pm}_1 \mu^{\mp}$ and $\tilde \chi^{\pm}_1
\nu_{\mu}$ productions at Tevatron can easily give
precise results, in contrast with the mass reconstructions
performed in the superpartner pair production analysis
at hadronic colliders which suffer an high
combinatorial background \cite{Atlas}.

\subsubsection{Model dependence of the results}

In our theoretical framework (see Section \ref{theoretical}),
the values of the $\vert \mu \vert$ and $\tan \beta$
(up to the ambiguity of low/high solution) parameters
are predicted. This has no important effects on the results
presented in Sections \ref{lp211p}
as the single gaugino production cross sections vary weakly
with these parameters (see Section \ref{cross1}).

However, since we have worked within the mSUGRA model,
the $\tilde l_L^{\pm}$ mass was typically larger than the
$\tilde \chi^0_1$ mass. In a situation where
$m_{\tilde l_L^{\pm}}$ would approach $m_{\tilde \chi^0_1}$,
the rate of the $\tilde \chi^0_1 l^{\pm}_i$ production, representing
in mSUGRA
the main contribution to the like sign dilepton signature
(see Section \ref{signal2}),
would decrease. Therefore,
within a model allowing degenerate $\tilde l_L^{\pm}$ and $\tilde \chi^0_1$
masses or even a $\tilde l_L^{\pm}$ lighter than the $\tilde \chi^0_1$,
other single gaugino productions than
the $p \bar p \to \tilde \chi^0_1 l^{\pm}_i$ reaction
could represent the major contribution to the
like sign dilepton signature in some parts of the SUSY
parameter space.

Besides, in a situation
where the LSP would not be the $\tilde \chi^0_1$,
the branching ratios of the $\tilde \chi^0_1$ decays
violating $R_p$ would be reduced with respect to the
case where the LSP is the $\tilde \chi^0_1$,
as often occurs in mSUGRA. However, in such a situation,
the like sign dilepton signature
could receive a significant contribution from
a decay of the $\tilde \chi^0_1$ different from the
\rpv channel. In those kinds of scenarios
where the LSP is not the $\tilde \chi^0_1$,
the $\tilde \chi^0_1 l^{\pm}_i$ production
would not represent systematically the main contribution
to the like sign dilepton signature.

In the several scenarios described above
where the $\tilde \chi^0_1 l^{\pm}_i$ production
is not the major contribution
to the like sign dilepton signature, this signature
could receive quite important contribution from the other
single gaugino productions described in Section \ref{resonant}.

\section{Conclusion}

The single gaugino productions at Tevatron reach important cross sections
thanks to the contributions of the resonant slepton productions. Hence,
the analysis of the 3 charged leptons and like sign dilepton signatures
generated by the single gaugino productions at Tevatron Run II would allow
to obtain
high sensitivities on many \rpv coupling constants, compared to the
low-energy limits,
in wide domains of the SUSY parameter space. This is also due to the fact
that
the Standard Model backgrounds associated to the 3 charged leptons and like
sign dilepton
final states at Tevatron can be greatly suppressed.

From the supersymmetry discovery point of view, superpartner masses well
beyond the
present experimental limits could be tested through the analysis of the
the 3 charged leptons and like sign dilepton signatures
generated by the single gaugino productions at Tevatron Run II.
If some of the \rpv coupling constants values were close to their low-energy
bounds,
the single gaugino productions study based on the 3 charged leptons and like
sign dilepton
signatures would even allow to
extend the region in the $m_0$-$m_{1/2}$ plane probed by the superpartner
pair
production analyses in the 3 charged leptons and like sign dilepton channels
at Tevatron Run II.
The reason is that the single superpartner production has a larger phase
space factor than
the superpartner pair production.\\
Besides, the 3 charged leptons and like sign dilepton signatures
generated by the single gaugino productions at Tevatron Run II would allow
to reconstruct
in a model independent way the $\tilde \chi^0_1$, $\tilde \chi^{\pm}_1$,
$\tilde \nu_L$ and
$\tilde l^{\pm}_L$ masses with a smaller combinatorial background than in
the superpartner pair
production analysis.

We end this summary by a comparison between the results obtained from
the studies of the 3 charged lepton and like sign dilepton signatures
generated by the single gaugino productions at Tevatron Run II.
In the mSUGRA model, the like sign dilepton signature analysis would give
rise to higher sensitivities on
the SUSY parameters than the study of the 3 charged lepton final state.
This comes notably from the
fact that in mSUGRA, the $\tilde \chi^0_1$ is lighter than the $\tilde
\chi^{\pm}_1$ so that
the cross section of the $\tilde \chi^0_1 l^{\pm}$ production, which is the
main contribution to the like sign
dilepton signature, reaches larger values than the cross section of the
$\tilde \chi^{\pm}_1 l^{\mp}$
production, representing the main contribution to the 3 charged lepton
final state.

Other interesting prospective studies concerning hadronic colliders are 
the analyses of the single gaugino productions
occuring through resonant squark productions via $\lambda''$ 
coupling constants which we will perform in the next future.

\section{Acknowledgments}
We would like to thank Emmanuelle Perez, Robi Peschanski and Auguste Besson for 
fruitful discussions and reading the manuscript.

\newpage

\appendix

\renewcommand{\thesubsection}{A.\arabic{subsection}}
\renewcommand{\theequation}{A.\arabic{equation}}
\setcounter{subsection}{0}
\setcounter{equation}{0}

\section{Formulas for spin summed amplitudes}
\label{formulas}

In this Appendix, we give the amplitudes for all the single productions
of \susyq particle at hadronic colliders, which can receive a contribution
from a slepton resonant production. These single productions occur
via the \rpv coupling $\l'_{ijk}$
and correspond to the four reactions,
$q \bar q \to \tilde \chi_a^+ \bar \nu_i$, $q \bar q \to \tilde \chi_a^0
\bar \nu_i$,
$ q \bar q \to \tilde \chi_a^0 \bar l_i$, $ q \bar q \to \tilde \chi_a^-
\bar l_i$.
Each of those four processes receives contributions from both the t and u
channel
(see Fig.\ref{graphes}) and have charge conjugated diagrams. Note also that
the contributions coming from the exchange of a right squark in the
u channel involve the higgsino components of the gauginos. These
contributions,
in the case of the single chargino production, do not interfere with the s
channel
slepton exchange since the initial or final states are different
(see Fig.\ref{graphes}).
In the following, we give the formulas for the probability amplitudes,
squared and summed over the polarizations. Our notations closely follow
the notations of \cite{Gunion}. In particular, the matrix elements $N'_{ij}$
are defined in the basis of the photino and the zino, as in \cite{Gunion}.

{
\hoffset-1in
\voffset-1in
\if@twoside\oddsidemargin25mm
\evensidemargin25mm\marginparwidth25mm
\else\oddsidemargin25mm\evensidemargin25mm\marginparwidth25mm\fi
\footheight12pt\footskip30pt
\textwidth 16cm
\baselineskip15pt
\textheight 45\baselineskip

\normalsize

\begin{eqnarray}
\vert M_s (u^j \bar d^k \to \tilde \chi_a^+ \bar \nu_i) \vert ^2 & = &
{ {\l'_{ijk}}^2  g^2 \vert U_{a1} \vert ^2
\over  12  (s-m_{\tilde l^i_L}^2)^2 }
 (m_{u^j}^2+m_{d^k}^2-s)(m_{\tilde \chi_a^+}^2-s)   \cr
\vert M_t (u^j \bar d^k \to \tilde \chi_a^+ \bar \nu_i) \vert ^2 & = & {
{\l'_{ijk}}^2  g^2
 \over 12  (t-m_{\tilde d^j_L}^2)^2 }
(m_{d^k}^2-t)
\bigg [ (\vert U_{a1} \vert ^2 +{m_{u^j}^2 \vert V_{a2} \vert ^2 \over
2  m_W^2 \sin^2 \beta} ) (m_{u^j}^2+m_{\tilde \chi_a^+}^2-t) \cr
& - & {4  m_{u^j}^2 m_{\tilde \chi_a^+} Re(U_{a1}V_{a2})
\over \sqrt{2} m_W \sin \beta} \bigg ] \cr
\vert M_u (u^k \bar d^j \to \tilde \chi_a^+ \nu_i) \vert ^2 & = &
{ {\l'_{ijk}}^2  g^2 m_{d^k}^2 \vert U_{a2} \vert^2
 \over 24 m_W^2 \cos \beta ^2  (u-m_{\tilde d^k_R}^2)^2 }
  (m_{\tilde \chi_a^+}^2+m_{u^k}^2-u) (m_{d^j}^2-u) \cr
2 Re [M_s M_t^* (\tilde \chi_a^+ \bar \nu_i)] & = & { {\l'_{ijk}}^2  g^2
 \over  6  (s-m_{\tilde l^i_L}^2) (t-m_{\tilde d^j_L}^2) }
\bigg [ { \vert U_{a1} \vert ^2 \over 2}  [(m_{u^j}^2+m_{\tilde
\chi_a^+}^2-t)
 (m_{d^k}^2-t) \cr
 &  +  & (m_{u^j}^2+m_{d^k}^2-s) (m_{\tilde \chi_a^+}^2-s)
-(m_{u^j}^2-u) (m_{\tilde \chi_a^+}^2+m_{d^k}^2-u)] \cr
 &  - &  (m_{d^k}^2-t)
{Re(U_{a1}V_{a2}) m_{\tilde \chi_a^+} m_{u^j}^2  \over \sqrt{2} m_W \sin
\beta} \bigg ],
\label{fchanu}
\end{eqnarray}

where, $s=(p(u^j)-p(\bar d_k))^2$, $t=(p(u^j)-p(\tilde \chi_a^+))^2$
and $u=(p(\bar d^j)-p(\nu_i))^2$.

\begin{eqnarray}
  \vert M_s (d_j \bar d_k \to  \tilde \chi_a^0 \bar \nu_i) \vert ^2 & = &
{ {\l'_{ijk}}^2  g^2 \vert N'_{a2} \vert ^2
  \over 24  \cos^2 \theta_W (s-m_{\tilde \nu^i_L}^2)^2 }
   (s-m_{d^k}^2-m_{d^j}^2) (s-m_{\tilde \chi_a^0}^2)  \cr
  \vert M_t (d_j \bar d_k \to \tilde \chi_a^0 \bar \nu_i) \vert ^2 & = & {
{\l'_{ijk}}^2  g^2
  \over 6  (t-m_{\tilde d^j_L}^2)^2 } (m^2_{d^k}-t)
 \bigg [ (m_{d^j}^2+m_{\tilde \chi_a^0}^2-t)
\bigg ( { g^2 m_{d^j}^2 \vert N'_{a3} \vert ^2 \over 4  m_W^2 \cos^2 \beta }
 +{e^2 \over 9} \vert N'_{a1} \vert ^2 \cr
& + & {g^2 \vert N'_{a2} \vert ^2 (\sin^2 \theta_W/3 -1/2)^2 \over \cos^2
\theta_W }
 - { 2  e g Re(N'_{a1}N'_{a2}) (\sin^2 \theta_W/3 -1/2) \over 3  \cos
\theta_W } \bigg ) \cr
 & + & {2 m_{\tilde \chi_a^0} m_{d^j}^2 g \over m_W \cos \beta}
\bigg ( -{e Re(N'_{a1}N'_{a3})\over 3} +
 {g Re(N'_{a2}N'_{a3}) \over \cos \theta_W}
({\sin^2 \theta_W \over 3} -{1 \over 2}) \bigg ) \bigg ] \cr
  \vert M_u (d_j \bar d_k \to \tilde \chi_a^0 \bar \nu_i) \vert ^2 & = & {
{\l'_{ijk}}^2
  \over 6  (u-m_{\tilde d^k_R}^2)^2 }
   (m_{d^j}^2-u)
 \bigg [ (m_{\tilde \chi_a^0}^2+m_{d^k}^2-u) \bigg
( {g^2 m_{d^k}^2 \vert N'_{a3} \vert ^2 \over 4  m_W^2 \cos^2 \beta}
 +{e^2 \vert N'_{a1} \vert ^2 \over 9} \cr
& + & {g^2 \sin^4 \theta_W \vert N'_{a2} \vert ^2 \over 9  \cos^2 \theta_W}
 - { 2  e g Re(N'_{a1}N'_{a2}) \sin^2 \theta_W \over 9  \cos \theta_W} \bigg
) \cr
& - & { 2  m_{\tilde \chi_a^0} m_{d^k}^2 g \over m_W \cos \beta}
\bigg ( -{e Re(N'_{a1}N'_{a3}) \over 3} +{g \sin^2 \theta_W
Re(N'_{a2}N'_{a3})
\over3  \cos \theta_W} \bigg ) \bigg ] \cr
 2 Re [M_s M_t^* (\tilde \chi_a^0 \bar \nu_i)] & = & - { {\l'_{ijk}}^2 g
 \over 12  \cos \theta_W (s-m_{\tilde \nu^i_L}^2) (t-m_{\tilde d^j_L}^2) }
 \bigg [ (m_{d^k}^2-t) {m_{\tilde \chi_a^0} m_{d^j}^2 g Re(N'_{a2}N'_{a3})
\over m_W \cos \beta} \cr
& + & \bigg (-{e Re(N'_{a1}N^*_{a2}) \over 3} +{g \vert N'_{a2}
\vert ^2 \over \cos \theta_W} ({\sin^2 \theta_W \over 3} -{1 \over 2}) \bigg
)
 [ (m_{d^j}^2+m_{\tilde \chi_a^0}^2-t) (m_{d^k}^2-t) \cr
& + & (m_{d^j}^2+m_{d^k}^2-s) (m_{\tilde \chi_a^0}^2-s)
 -(m_{\tilde \chi_a^0}^2+m_{d^k}^2-u) (m_{d^j}^2-u) ] \bigg ] \cr
 2 Re [M_t M_u^* (\tilde \chi_a^0 \bar \nu_i)] & = &  { {\l'_{ijk}}^2
 \over 6  (u-m_{\tilde d^k_R}^2) (t-m_{\tilde d^j_L}^2) }
 \bigg [ (m_{d^k}^2-t) {  g m_{\tilde \chi_a^0} m_{d^j}^2
\over m_W \cos \beta} \bigg ( {g \sin^2 \theta_W Re(N'_{a2}N'_{a3})
\over 3  \cos \theta_W}-{e Re(N'_{a1}N'_{a3}) \over 3 } \bigg ) \cr
& + & [(m_{d^j}^2-u) (m_{\tilde \chi_a^0}^2+m_{d^k}^2-u)
  +  (m_{d^k}^2-t) (m_{d^j}^2+m_{\tilde \chi_a^0}^2-t)
 -(m_{\tilde \chi_a^0}^2-s) (m_{d^j}^2+m_{d^k}^2-s)] \cr
& & \bigg (  -{egRe(N'_{a1}N'_{a2}) \over 3 \cos \theta_W}
( {2 \sin^2 \theta_W \over 3} -{1 \over 2} )
+{e^2 \vert N'_{a1} \vert ^2 \over 9 }+{g^2 \sin^2 \theta_W \vert N'_{a2}
\vert ^2
\over 3 \cos^2 \theta_W} ({\sin^2 \theta_W \over 3}-{1 \over 2}) \bigg ) \cr
& - & {m_{\tilde \chi_a^0} m_{d^k}^2 g \over m_W \cos \beta}
 \bigg ( -{e Re(N'_{a1}N'_{a3}) \over 3} +{g Re(N'_{a2}N'_{a3})
 \over \cos \theta_W} ({\sin^2 \theta_W \over 3}-{1 \over 2}) \bigg )
 (m_{d^j}^2-u)  \cr
& + & {m_{d^j}^2 m_{d^k}^2  g^2 \vert N'_{a3} \vert ^2
 \over 2  m_W^2 \cos^2 \beta}(m_{\tilde \chi_a^0}^2-s) \bigg ] \cr
 2 Re [M_s M_u^* (\tilde \chi_a^0 \bar \nu_i)] & = & { {\l'_{ijk}}^2 g
\over 12  \cos \theta_W (s-m_{\tilde \nu^i_L}^2) (u-m_{\tilde d^k_R}^2) }
\bigg [-  {m_{\tilde \chi_a^0} m_{d^k}^2 g Re(N'_{a2}N'_{a3})
\over m_W \cos \beta} (m_{d^j}^2-u) \cr
&  + & \bigg ( -{e Re(N^*_{a1}N'_{a2}) \over 3} +{\vert N'_{a2} \vert ^2
g \sin^2 \theta_W \over 3  \cos \theta_W} \bigg )
[(m_{d^j}^2+m_{d^k}^2-s) (m_{\tilde \chi_a^0}^2-s) \cr
& + & (m_{\tilde \chi_a^0}^2+m_{d^k}^2-u) (m_{d^j}^2-u)
-(m_{d^j}^2+m_{\tilde \chi_a^0}^2-t) (m_{d^k}^2-t)]  \bigg ],
\label{fnenu}
\end{eqnarray}

where, $s=(p(d^j)-p(\bar d_k))^2$, $t=(p(d^j)-p(\tilde \chi_a^0))^2$
and $u=(p(d^j)-p(\bar \nu_i))^2$.

\begin{eqnarray}
 \vert M_s (u_j \bar d_k \to  \tilde \chi_a^0 \bar l_i) \vert ^2 & = & {
{\l'_{ijk}}^2
 \over 6  (s-m_{\tilde l^i_L}^2)^2 }
 (s-m_{u^j}^2-m_{d^k}^2)
 \bigg [ \bigg (  {g^2 m_{l^i}^2 \vert N'_{a3} \vert ^2 \over 4  m_W^2
\cos^2 \beta}
+e^2 \vert N'_{a1} \vert ^2
 + {g^2 \vert N'_{a2} \vert ^2  \over \cos^2 \theta_W}
(\sin^2 \theta_W- {1 \over 2})^2 \cr
& - &{2  e g Re(N'_{a1}N'_{a2}) \over \cos \theta_W}(\sin^2 \theta_W-{1
\over 2}) \bigg )
 (s-m_{l^i}^2-m_{\tilde \chi_a^0}^2)
 - {2  g m_{\tilde \chi_a^0} m_{l^i}^2 \over m_W \cos \beta}
\bigg ( -e Re(N'_{a1}N'_{a3}) \cr
 & + & {g Re(N'_{a2}N'_{a3}) \over \cos \theta_W}
(\sin^2 \theta_W-{1 \over 2}) \bigg ) \bigg ] \cr
 \vert M_t (u_j \bar d_k \to \tilde \chi_a^0 \bar l_i) \vert ^2 & = & {
{\l'_{ijk}}^2
 \over 6  (t-m_{\tilde u^j_L}^2)^2 }
 (-t+m_{l^i}^2+m_{d^k}^2)
 \bigg [ \bigg (  {g^2 m_{u^j}^2 \vert N'_{a4} \vert ^2 \over 4  m_W^2
\sin^2 \beta}
 +{4  e^2 \vert N'_{a1} \vert ^2 \over 9} \cr
 & + & {g^2 \vert N'_{a2} \vert ^2  \over \cos^2 \theta_W}
({1 \over 2}-{2  \sin^2 \theta_W \over 3} )^2
 +{4  e g Re(N'_{a1}N'_{a2})  \over 3  \cos \theta_W}
 ({1 \over 2}-{2  \sin^2 \theta_W \over 3}) \bigg )
 (-t+m_{u^j}^2+m_{\tilde \chi_a^0}^2) \cr
 &  +  & {2  g m_{u^j}^2 m_{\tilde \chi_a^0} \over m_W \sin \beta}
\bigg ( {2  e Re(N'_{a1}N'_{a4}) \over 3} +{g Re(N'_{a2}N'_{a4})
  \over \cos \theta_W} ({1 \over 2}-{2  \sin^2 \theta_W \over 3} ) \bigg )
\bigg ] \cr
 \vert M_u (u_j \bar d_k \to  \tilde \chi_a^0 \bar l_i) \vert ^2 & = & {
{\l'_{ijk}}^2
  \over 6  (u-m_{\tilde d^k_R}^2)^2 }
  (m_{u^j}^2+m_{l^i}^2-u)
 \bigg [ \bigg ( {e^2 \vert N'_{a1} \vert ^2 \over 9}
 + {g^2 \sin^4 \theta_W \vert N'_{a2} \vert ^2 \over 9  \cos^2 \theta_W}
  -{2  e g Re(N'_{a1}N'_{a2}) \sin^2 \theta_W \over 9  \cos \theta_W} \cr
 & + & {g^2 m_{d^k}^2 \vert N'_{a3} \vert ^2 \over 4  m_W^2 \cos^2 \beta}
\bigg )
 (m_{\tilde \chi_a^0}^2+m_{d^k}^2-u)
 - {2  g m_{d^k}^2 m_{\tilde \chi_a^0}  \over m_W \cos \beta}
\bigg ( -{e Re(N'_{a1}N'_{a3}) \over 3} \cr
& + & {g \sin^2 \theta_W Re(N'_{a2}N'_{a3}) \over 3  \cos \theta_W} \bigg )
\bigg ] \cr
2 Re [M_s M_t^* (\tilde \chi_a^0 \bar l_i)] & = & - {   {\l'_{ijk}}^2
  \over 6  (s-m_{\tilde l^i_L}^2) (t-m_{\tilde u^j_L}^2) }
  \bigg [  -{m_{l^i}^2 m_{u^j}^2   g^2 Re(N'_{a3}N^*_{a4})
  \over 2  m_W^2 \sin \beta \cos \beta} (m_{\tilde \chi_a^0}^2+m_{d^k}^2-u)
\cr
 & + & \bigg ( {-2  e^2 \vert N'_{a1} \vert ^2 \over 3} +{  e g
Re(N^*_{a1}N'_{a2})
  \over 3  \cos \theta_W}  (4 \sin^2 \theta_W-{5 \over 2})  \cr
 & + &  {g^2 \vert N'_{a2} \vert ^2
 \over \cos^2 \theta_W} ({1 \over 2}-{2  \sin^2 \theta_W \over 3} )
 (\sin^2 \theta_W-{1 \over 2}) \bigg ) \cr
 & & [ (m_{u^j}^2+m_{d^k}^2-s) (m_{\tilde \chi_a^0}^2+m_{l^i}^2-s)
 +(m_{u^j}^2+m_{\tilde \chi_a^0}^2-t) (m_{l^i}^2+m_{d^k}^2-t) \cr
 & - & (m_{u^j}^2+m_{l^i}^2-u) (m_{\tilde \chi_a^0}^2+m_{d^k}^2-u) ]
 + {g m_{u^j}^2 m_{\tilde \chi_a^0} \over m_W \sin \beta}
 \bigg ( -e Re(N'_{a1}N'_{a4})+{g Re(N'_{a2}N'_{a4})  \over \cos \theta_W}
\cr
& & ( \sin^2 \theta_W-{1 \over 2}) \bigg )
 (m_{l^i}^2+m_{d^k}^2-t)
 -(s-m_{u^j}^2-m_{d^k}^2) {g m_{l^i}^2 m_{\tilde \chi_a^0} \over m_W \cos
\beta}
\bigg ( {2  e Re(N'_{a1}N'_{a3}) \over 3} \cr
& + & {g Re(N'_{a2}N'_{a3}) \over \cos \theta_W}
({1 \over 2}-{2 \sin^2 \theta_W \over 3}) \bigg ) \bigg ] \cr
 2 Re [M_t M_u^* (\tilde \chi_a^0 \bar l_i)] & = &  {   {\l'_{ijk}}^2
 \over 6  (u-m_{\tilde d^k_R}^2) (t-m_{\tilde u^j_L}^2) }
\bigg  [ {g m_{u^j}^2 m_{\tilde \chi_a^0}
     \over m_W \sin \beta} ( m_{l^i}^2+m_{d^k}^2-t )
   \bigg ( -{e Re(N'_{a1}N'_{a4}) \over 3} \cr
    & + & {g \sin^2 \theta_W Re(N'_{a2}N'_{a4}) \over 3  \cos \theta_W}
\bigg )
 - {m_{\tilde \chi_a^0} g m_{d^k}^2  \over m_W \cos \beta}
  \bigg ( {2  e Re(N'_{a1}N'_{a3}) \over 3} +{g Re(N'_{a2}N'_{a3})
  \over \cos \theta_W} ({1 \over 2}-{2  \sin^2 \theta_W \over 3}) \bigg )
\cr
      & &  ( m_{l^i}^2+m_{u^j}^2-u )
 - {g^2 Re(N'_{a3}N^*_{a4}) m_{u^j}^2 m_{d^k}^2
   \over 2  m_W^2 \cos \beta \sin \beta} (s-m_{l^i}^2-m_{\tilde \chi_a^0}^2)
+ \bigg ( -{2  e^2 \vert N'_{a1} \vert ^2 \over 9} \cr
 & + & {e g Re(N^*_{a1}N'_{a2}) \over 3  \cos \theta_W}
 (-{1 \over 2}+{ 4 \sin^2 \theta_W \over 3})
  + {g^2 \sin^2 \theta_W \vert N'_{a2} \vert ^2 \over 3  \cos^2 \theta_W}
\cr
& &  ( {1 \over 2}-{2  \sin^2 \theta_W \over 3} )  \bigg )
 [  (m_{l^i}^2+m_{u^j}^2-u) (m_{\tilde \chi_a^0}^2+m_{d^k}^2-u)
 +(m_{l^i}^2+m_{d^k}^2-t) (m_{\tilde \chi_a^0}^2+m_{u^j}^2-t) \cr
 & - & (m_{l^i}^2+m_{\tilde \chi_a^0}^2-s) (m_{d^k}^2+m_{u^j}^2-s) ]
\bigg ] \cr
 2 Re [M_s M_u^* (\tilde \chi_a^0 \bar l_i)] & = &  {   {\l'_{ijk}}^2
  \over 6  (s-m_{\tilde l^i_L}^2) (u-m_{\tilde d^k_R}^2) }
 \bigg [ -{g m_{l^i}^2 m_{\tilde \chi_a^0} \over m_W \cos \beta}
 \bigg ( -{e Re(N'_{a1}N'_{a3}) \over 3} +{g \sin^2 \theta_W
Re(N'_{a2}N'_{a3})
\over 3  \cos \theta_W} \bigg ) \cr
  & & ( s-m_{d^k}^2-m_{u^j}^2 )
 - {g m_{d^k}^2 m_{\tilde \chi_a^0}  \over m_W \cos \beta}
 \bigg ( -e Re(N'_{a1}N'_{a3})+{g Re(N'_{a2}N'_{a3}) \over \cos \theta_W}
(\sin
\theta_W^2-{1 \over 2}) \bigg ) \cr
  & & ( m_{l^i}^2+m_{u^j}^2-u )
 + {g^2 m_{l^i}^2 m_{d^k}^2 \vert N'_{a3} \vert ^2
  \over 2  m_W^2 \cos^2 \beta} (m_{\tilde \chi_a^0}^2+m_{u^j}^2-t)
 + \bigg ( {e^2 \vert N'_{a1} \vert ^2 \over 3 } \cr
 & - & {e g Re(N^*_{a1}N'_{a2}) \over 3  \cos \theta_W}
( 2 \sin \theta_W^2-{1 \over 2})
 + {g^2 \vert N'_{a2} \vert ^2 \sin^2 \theta_W  \over 3 \cos^2 \theta_W}
  (\sin^2 \theta_W-{1 \over 2}) \bigg ) \cr
& &  [  (m_{l^i}^2+m_{u^j}^2-u) (m_{\tilde \chi_a^0}^2+m_{d^k}^2-u)
 -(m_{l^i}^2+m_{d^k}^2-t) (m_{\tilde \chi_a^0}^2+m_{u^j}^2-t) \cr
& + & (m_{l^i}^2+m_{\tilde \chi_a^0}^2-s) (m_{d^k}^2+m_{u^j}^2-s) ],
\bigg ]
\label{fnel}
\end{eqnarray}

where, $s=(p(u^j)-p(\bar d_k))^2$, $t=(p(u^j)-p(\tilde \chi_a^0))^2$
and $u=(p(u^j)-p(\bar l_i))^2$.

\begin{eqnarray}
      \vert M_s (d_j \bar d_k \to \tilde \chi_a^- \bar l_i) \vert ^2 & = &
{ g^2 {\l'_{ijk}}^2
      \over 6  (s-m_{\tilde \nu^i_L}^2)^2 }
     (s-m_{d^j}^2-m_{d^k}^2)
      \bigg [ ( {m_{l^i}^2 \vert U_{a2} \vert^2 \over
     4  m_W^2 \cos^2 \beta}+ { \vert V_{a1} \vert ^2 \over 2} )
      ( s-m_{\tilde \chi_a^+}^2-m_{l^i}^2 ) \cr
       & + & {\sqrt{2} Re(V_{a1}U_{a2}) m_{l^i}^2 m_{\tilde \chi_a^+}
     \over m_W \cos \beta} \bigg ] \cr
      \vert M_t (d_j \bar d_k \to \tilde \chi_a^- \bar l_i) \vert ^2 & = &
    {     g^2 {\l'_{ijk}}^2
      \over 3  (t-m_{\tilde u^j_L}^2)^2 }
      ( t-m_{d^k}^2-m_{l^i}^2 ) \bigg [ ( t-m_{\tilde \chi_a^+}^2-m_{d^j}^2
)
       ( {\vert V_{a1} \vert ^2 \over 4} + {m_{d^j}^2 \vert U_{a2} \vert^2
     \over 8  M^2_W \cos^2 \beta} ) \cr
     & + & {Re(V_{a1}U_{a2}) m_{\tilde \chi_a^+} m^2_{d^j}
     \over \sqrt{2} m_W \cos \beta}  \bigg ] \cr
      \vert M_u (\bar u_k u_j \to \tilde \chi_a^- \bar l_i) \vert ^2 & = & {
g^2 {\l'_{ijk}}^2
      \over 24  (u-m_{\tilde d^k_R}^2)^2  }
      (m_{\tilde \chi_a^+}^2+m_{u^k}^2-u)
      (m_{l^i}^2+m_{u^j}^2-u)
     { \vert U_{a2} \vert^2  m_{d^k}^2 \over m_W^2 \cos^2 \beta} \cr
       2 Re [M_s M_t^* (\tilde \chi_a^- \bar l_i)] & = & {  g^2
{\l'_{ijk}}^2
      \over 12  (s-m_{\tilde \nu^i_L}^2) (t-m_{\tilde u^j_L}^2) }
     \bigg [  \vert V_{a1} \vert ^2
      [ -(m_{l^i}^2+m_{d^j}^2-u) (m_{\tilde \chi_a^+}^2+m_{d^k}^2-u) \cr
         & + & (m_{l^i}^2+m_{d^k}^2-t) (m_{\tilde \chi_a^+}^2+m_{d^j}^2-t)
         +(m_{l^i}^2+m_{\tilde \chi_a^+}^2-s) (m_{d^k}^2+m_{d^j}^2-s) ] \cr
 & + & {Re(V_{a1}U_{a2}) m_{\tilde \chi_a^+} \sqrt{2} \over m_W \cos \beta }
 [m^2_{l^i} (s-m_{d^j}^2 - m_{d^k}^2)
- m^2_{d^j} (m_{l^i}^2+m_{d^k}^2-t)] \cr
 & - &  { \vert U_{a2} \vert^2  m_{l^i}^2  m^2_{d^j} \over m_W^2 \cos^2
\beta}
(m_{\tilde \chi_a^+}^2+m_{d^k}^2-u) \bigg ],
\label{fchal}
\end{eqnarray}

where, $s=(p(d^j)-p(\bar d_k))^2$, $t=(p(d^j)-p(\tilde \chi_a^-))^2$
and $u=(p(u^j)-p(\bar l_i))^2$.}

\end{document}